\documentclass[preprint,12pt,sort&compress]{article}
\usepackage[utf8]{inputenc}
\usepackage{caption}
\usepackage{graphicx}

\usepackage{amssymb}
\usepackage{amsthm,amsmath}
\newcommand{\figref}[1]{\figurename~\ref{#1}}
\theoremstyle{definition}
\newtheorem{exmp}{Example}[section]
\usepackage{subcaption}
\usepackage[T1]{fontenc}
\usepackage{authblk}
\title{QuickCent: a fast and frugal heuristic for harmonic centrality estimation on scale-free networks}
\author[1]{Francisco Plana\footnote{franciscoplana@gmail.com}}
\author[1,2]{Andr\'es Abeliuk}
\author[3]{Jorge P\'erez}
\affil[1]{Department of Computer Science, Universidad de Chile,Beauchef 851,Santiago,Chile.}
\affil[2]{National Center for Artificial Intelligence (CENIA), Vicu\~na Mackenna 4860, Macul, Chile.}
\affil[3]{cero.ai,https://www.cero.ai/,Chile.}
\date{}

\begin{document}
\maketitle

This preprint has not undergone peer review or any post-submission improvements or corrections. The Version of Record of this article is published in the Springer Journal \textit{Computing}, and is available online at 

https://doi.org/10.1007/s00607-024-01303-z

View-only: https://rdcu.be/dKh8v





\begin{abstract}
We present a simple and quick method to approximate network centrality indexes.
Our approach, called \emph{QuickCent}, is inspired by so-called \emph{fast and frugal} heuristics, which are 
heuristics initially proposed to model some human decision and inference processes.
The centrality index that we estimate is the \emph{harmonic} centrality, which is 
a measure based on shortest-path distances, so infeasible to compute on large networks.
We compare \textit{QuickCent} with known machine learning algorithms on synthetic data generated with preferential attachment, and some empirical networks. 
Our experiments show that \textit{QuickCent} is able to make estimates that are competitive in accuracy with the best alternative
methods tested, either on synthetic scale-free networks or empirical networks. QuickCent has the feature of achieving low error variance estimates, even with a small training set. 
Moreover, \textit{QuickCent} is comparable in efficiency --accuracy and time cost-- to 
more complex methods.
We discuss and provide some insight into how QuickCent exploits the fact that in some networks, such as those generated by preferential attachment, local density measures such as the in-degree, can be a proxy for the size of the network region to which a node has access, opening up the possibility of approximating centrality indices based on size such as the harmonic centrality. Our initial results show that simple heuristics and biologically inspired computational methods 
are a promising line of research 
in the context of network measure estimations.

\end{abstract}










\section{Introduction}
\paragraph{Heuristics are proposed as a model of cognitive processes} Some models based on heuristics have been proposed to account for cognitive mechanisms \cite{
tversky1975judgment}, which 
assume that, though these heuristics are used at a lesser computational cost, they sacrifice accuracy and lead to systematic errors. 
This viewpoint has been challenged by the so-called 
\textit{Fast and frugal} heuristics \cite{simpleheu}, which are simple heuristics initially proposed
to model some human decision and inference processes.
They have shown that very simple human-inspired methods, by relying on statistical patterns of the data, 
can reach accurate results, in some cases even better than 
methods based on more information or complex computations \cite{katsikopoulos2010robust,simpleheu}. Due to these features, 
{fast and frugal} heuristics have been applied in problems different from their original motivation, 
including medical decision-making \cite{backlund2009improving}, predicting the outcomes of sport matches \cite{scheibehenne2007predicting} and geographic profiling~\cite{snook2005complexity}. 

\paragraph{The problem of centrality computation} In this paper, we provide an example of the usefulness of one of these simple heuristics for estimating the centrality index in a network. Roughly speaking, the centrality index is a measure of the importance of a node in a network. 
We chose to estimate the \textit{harmonic centrality} index \cite{marchiori2000harmony} since it satisfies a set of necessary axioms that any centrality should meet 
\cite{axiomcent}, namely that nodes belonging to large groups are important (\textit{size} axiom); that nodes with a denser neighborhood, i.e. with more connections, are more important (\textit{density} axiom); and that the importance increases with the addition of an arc (\textit{score-monotonicity} axiom). Consider a directed graph $G=(V,A)$, with $V$ the
set of nodes and $A$ the set of arcs or edges. 
Formally, let $d_G(y,x)$
be the length of the shortest path from node $y$ to $x$ in the digraph $G$.
The harmonic centrality of $x$ is computed as
\[
H_G(x)=\sum_{y\in V, y\neq x}\frac{1}{d_G(y,x)},
\]
which has the nice property of managing unreachable nodes in a clean way.

Besides its good properties, to compute the harmonic centrality for 
all nodes in a network we need first to solve the all-pairs shortest-path problem. 
Notice that by the total number of pairs of nodes, there is an intrinsic 
lower bound of $|V|^2$ for computing this centrality, and $O(|V|^2)$ is 
already a huge constraint for modern networks.
There has been a lot of work on optimizing the
computation of all-pairs shortest-paths for weighted networks~\cite{pettie2002computing,pettie2002comparison,planken2012computing} 
but even under strict constraints on the structure of the networks~\cite{planken2012computing} 
this computation is unfeasible for networks with a large number of nodes,
usually needing time $O(|A|\cdot |V|)$.
Thus, in order to use harmonic centrality in practice we need ways of estimating or approximating
it. 

Though there are few centrality indexes satisfying the three axioms \cite{axiomcent}, some simple measures can be built that do satisfy them. One way of doing this, is by taking the simple product of a density measure, such as the in-degree, with a size measure, such as the number of weakly reachable nodes \cite{axiomcent}. While the in-degree is cheap to compute, many times stored as an attribute so accessible in constant time, size measures have a higher time complexity. For example, the number of reachable nodes, for each node, can be computed from the condensation digraph of strongly connected components, which may give, in the worst case, a total time complexity of $O(|A|\cdot |V| + |V|^2)$. In this paper, we explore whether expensive indexes, sensitive to either density and size, such as the harmonic centrality, may be approximated by cheap local density measures such as the in-degree. 



\paragraph{Our proposal} Our proposed method, called \textit{QuickCent}, is a modification of QuickEst \cite{hertwigquick}, a heuristic proposed to represent the processes underlying human quantitative estimation. QuickCent can be considered as a generalization of QuickEst, in the sense that, albeit in this work we focus on centrality approximation, it proposes a general procedure to regress a variable on a predictor when some assumptions are met.
QuickCent is a very simple heuristic based on sequences of \emph{binary clues} associated with nodes in a network;
the value of a clue is an indicator of the presence or absence of an attribute 
signal of greater centrality for a node.
The method simply finds the first clue with value $0$ (absence), and 
it outputs an estimate according to this clue. All the clues used in QuickCent are based on the in-degree of the node, thus QuickCent can be seen as a
method to regress a variable (harmonic centrality) that correlates with a predictor variable (in-degree) that is cheaper to compute. Another key characteristic of QuickCent 
is that it is designed to estimate magnitudes 
distributed according to a power-law~\cite{newman2005power}, which can model a wide range of natural and human-made phenomena. This paper extends previous work by some of the authors, mainly by adding the study of networks defying the heuristics assumptions and the performance over empirical networks \cite{plana2018quickcent}.

\paragraph{Results and future work} Our method is able to generate accurate estimates 
even if trained with a small proportion --$10\%$-- of the dataset.
We compare QuickCent with three standard machine learning algorithms trained with the same predictor variable
over synthetic data and some empirical networks.
Our results show that QuickCent is comparable in accuracy to the best-competing methods tested, and has the lowest error variance.
Moreover, the time cost of QuickCent is in the middle range compared to the other methods, even though we developed a naive version of QuickCent. We also discuss how QuickCent exploits the fact that in some networks, where higher degree nodes are more likely to be found because more paths lead to them, local density measures such as in-degree can be a good proxy for the size of the network region to which a node has access, opening up the possibility of approximating centrality indices based on size, such as harmonic centrality. This insight supports the conjecture that QuickCent may be better suited to information networks, such as the Internet, citation, or scientific collaboration, which can be well approximated by the preferential attachment growth mechanism \cite{barabasi1999emergence,jeong2003measuring,vazquez2003growing}, than to more purely social networks \cite{jackson2007meeting,broido2019scale}, which is an interesting question for future work. Also, working in the future with more general notions of local density \cite{wang2018improved,curado2022anew} may serve to extend the validity of the heuristics for more general networks. 
The results of this paper are a proof of concept to illustrate the potential of using methods based on simple heuristics 
to estimate some network measures. Whether or not these heuristics provide a realistic model of human cognition, is a wide problem \cite{broder2008challenging} which is out of the scope of this work. 


\paragraph{Structure of the paper} The rest of this paper is structured as follows. 
We begin in Section \ref{QC} by 
introducing the general mechanism of QuickCent, while Section \ref{QC2} presents our concrete implementation. 
In Section \ref{res}, we present the results of our proposal, including the comparison with other machine learning methods on either synthetic or empirical networks.
Section \ref{disc} gives a final discussion of the results including directions for possible future work.

\section{The QuickCent Heuristic}\label{QC}

In this section, we give a general abstract overview of our proposal, which we call QuickCent. 
The setting for QuickCent is as follows: the input is a network $G=(V,A)$
and we want to get an accurate estimate of the value of a centrality function 
$f_C:V \longrightarrow \mathbb{R}$. That is,  
for every $v \in V$, we want to compute a value $\tilde{f}_v$ that is an estimation of $f_C(v)$.
In our abstract formulation, it does not matter which particular centrality function we are estimating, and the details of the implementation of the heuristic 
for the particular case of harmonic centrality are given in the next section.
We next explain the general abstract idea of the components of QuickCent. 

Analogously to QuickEst \cite{hertwigquick}, our QuickCent method relies on vectors of \emph{$n$ binary clues}.
We associate to every node $v\in V$ a vector $\vec{x}_v=(x_v^1,x_v^2,\ldots,x_v^n) \in \{0,1\}^n$.
The intuition is that the value of the $i-$th component (clue) $x_v^i$
is an indicator of the presence ($x_v^i=1$) or absence ($x_v^i=0$) of an attribute signal of greater centrality for node $v$.
Our method also considers the following $n+1$ sets of nodes:
\[
\begin{array}{ccl}
S_1 & = & \{v\in V\mid x_v^1 = 0\} \\
S_i & = & \{v\in V\mid x_v^i = 0\text{ and }x_v^{i-1}=1\} \;\;\; (2\leq i\leq n)\\
S_{n+1} & = & \{v\in V\mid x_v^n= 1\} \\
\end{array}
\]

That is, $S_i$ corresponds to nodes that do not have the $i-$th attribute while 
having the previous one.
For each one of the sets $S_i$, with $1\leq i\leq n+1$, our method needs a quantity 
$\bar{f}_i$ which is 
a summary statistic of the centrality distribution 
of the nodes in set $S_i$.
{QuickCent} must ensure that successive clues are associated with higher centrality values, 
thus we will have that 
\begin{equation}\label{promedios}
\bar{f}_1 <\bar{f}_2<\cdots <\bar{f}_n <\bar{f}_{n+1}.
\end{equation} 
With the previous ingredients, the general estimation procedure corresponds to the following simple rule.
\medskip

\noindent
{\bf General QuickCent heuristic:} \textit{For node $v$, we iterate over the $n$ clues, considering every value $x_v^i$. When we find the first $i$ verifying that ${x}_v^i=0$, 
we stop and output the value $\bar{f}_i$. 
If node $v$ is such that $x_v^i=1$ for every $i \in \{1,\ldots,n\}$, we output $\bar{f}_{n+1}$}.
\medskip

\begin{exmp}\label{the-exmpl}
This is only a very simple example to exhibit the working of QuickCent, where we assume complete knowledge of the centrality values of all nodes. Let us consider the following network in \figref{exmpl_nw} of size $25$ obtained as a random instance of linear preferential attachment, defined in \ref{lin-PA}. Table \ref{exampl_nw_stat} displays only the non-zero values of in-degree and harmonic centrality in this network. A reasonable way to aggregate these values is to consider four sets $S_i, i=1,2,3,4$, with the following binary clues, ${x}_v^i=1$ ($i=1,2,3$) if and only if $\deg^{\text{in}}(v)>d_i$, with $d_1=0$, $d_2=3$ and $d_3=4$. With this choice, for simple centrality approximation it is natural to take, for example, the median of harmonic centrality on every set $S_i$ as summary statistics, $\bar{f}_1=0$, $\bar{f}_2=1$, $\bar{f}_3=4.666$ and $\bar{f}_4=15.75$.  

\begin{figure}[t]
\centering
  \includegraphics[width=.4\linewidth]{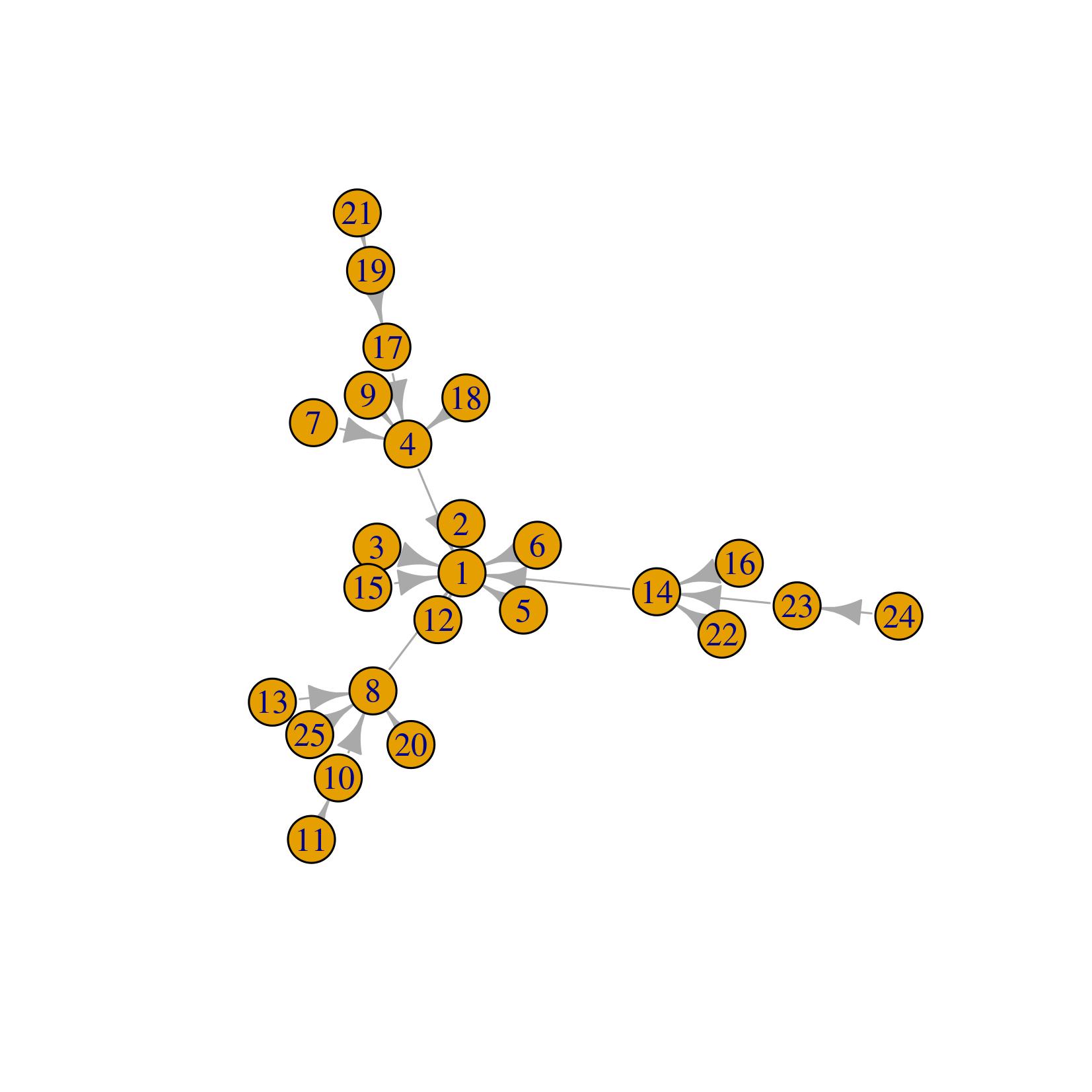}  
\caption{{\bf A network randomly generated with linear preferential attachment.}}
\label{exmpl_nw}
\end{figure}
\begin{table}
{\small
\begin{center}
  \begin{tabular}{c|cccccccc}
\textbf{Node}  & $1$ & $4$ & $8$ & $10$ & $14$ & $17$ & $19$ & $23$\\
\hline
\textbf{In-degree} & $9$ & $4$ & $4$ & $1$ & $3$ & $1$ & $1$ & $1$\\
\textbf{Harmonic}  & $15.750$ & $4.833$ & $4.500$ & $1.000$ & $3.500$ & $1.500$ & $1.000$ & $1.000$ \\
\textbf{QC100} & $13.429$ & $2.973$ & $2.973$ & $1.309$ & $1.309$ & $1.309$ & $1.309$ & $1.309$ \\
\textbf{QC70} & $6.531$ & $2.197$ & $2.197$ & $1.214$ & $1.214$& $1.214$& $1.214$& $1.214$\\
\end{tabular}
\caption{\textbf{In-degree and harmonic centrality values for each node of the network from \figref{exmpl_nw}.} Nodes that do not appear here have a zero in-degree and centrality. The last two rows correspond to QuickCent models described in Example \ref{exmpl_3}. The number of decimal places is truncated to three with respect to the source.}
\label{exampl_nw_stat}
\end{center}}
\end{table}

\end{exmp}

QuickCent provides a simple stopping rule:
for each node, the search is finalized when the first clue with value $0$ is found. 
Therefore, if our input is a network in which the vast majority of nodes is having similar and small centrality values --as it would be the case if the centrality were distributed according to a power law-- the procedure 
is likely to stop the search early and give an estimate quickly.
In this sense, the heuristic is \emph{frugal}, 
given that in many cases it can output an estimate without 
passing over all the clues, or without using all the available information. 

Up to this point, QuickCent remains similar to QuickEst. The reader can review the details of QuickEst in the book chapter by Hertwig et al (1999) \cite{hertwigquick}. The most critical aspects 
that distinguish QuickCent from QuickEst, as well as a specification of each part of the heuristic, are presented in the next section.

\section{A QuickCent implementation}\label{QC2}
In this section, we propose an instantiation of our general QuickCent method, including a way to compute the clues $x_v^i$ for every node $v$ based on its in-degree in Section \ref{using_indeg}, and an efficient way to compute the summary statistic $\bar{f}_i$ of the centrality for every set $S_i$ in Section \ref{all-pieces}. Section \ref{QC_assumpt} makes explicit the assumptions on the structure of graphs that QuickCent requires to be a \textit{ecologically rational heuristic}~\cite{hertwigquick}, i.e.  the proper problem conditions that ensure a successful application of the heuristic, including that the centrality has a power-law distribution.
Necessary concepts of the  power-law distribution are introduced in \ref{interl_pl}.

\subsection{Using the in-degree for the clues}\label{using_indeg}
Our approach to compute the binary clues is to employ a proxy variable related to the centrality by means of a monotonic function which ensures that Equation \eqref{promedios} holds. 
The idea is to use a proxy which should be far cheaper to compute than computing the actual centrality value.
The proxy variable we chose is the in-degree of the node, that is, the number of neighbors of the node given by incoming arcs of the network. The intuition for this proxy is that greater in-degree will likely be associated with shorter distances, which likely increases the harmonic centrality.
The in-degree is one of the most elementary properties of a node, 
and in many data structures it is stored as an attribute of the node 
(thus accessible in $O(1)$ time).
The in-degree can itself be considered as a centrality measure~\cite{axiomcent}. 
For a node $v$ we denote by $\deg^{\text{in}}(v)$ its in-degree.

Now, starting from a set of proportions $\{p_i\}_{i=1}^n$, 
where $0\leq\cdots\leq p_i\leq p_{i+1}\leq \cdots\leq 1$, 
we can get the respective \emph{quantile degree values} $\{d_i\}_{i=1}^n$. 
That is, if $F$ is the cumulative distribution function (CDF) for the in-degree, 
then $d_i = F^{-1}(p_i)$ for each $i=1,\ldots,n$. 
Then, we define the $i-$th clue for node $v$ as
\begin{equation}
{x}_v^i=1 \text{\; if and only if\; } \deg^{\text{in}}(v)>d_i. \label{clues}
\end{equation} 
With this definition, the sets $S_i$ are
\begin{equation*}
\begin{array}{ccl}
S_1 & = &\{v\in V \mid \deg^{\text{in}}(v) \leq d_1 \}, \\
S_i & = & \{v\in V \mid d_{i-1}< \deg^{\text{in}}(v) \leq d_i \} \;\;\; (2\leq i\leq n),\\
S_{n+1} & = &\{v\in V \mid d_{n}< \deg^{\text{in}}(v) \}.
\end{array}
\end{equation*}

\begin{exmp}\label{exmpl_2}
This type of clues was already used in Example \ref{the-exmpl}. In fact, the quantile degree values $\{0,3,4\}$ used there can be obtained via the inverse of the in-degree CDF applied to the set of proportions $\{0.68, 0.84, 0.96\}$. 
\end{exmp}

The final piece to apply QuickCent is to show how to compute
the summary statistic $\bar{f}_i$ for every set $S_i$.
We propose 
to compute $\bar{f}_i$ analytically as the median of each $S_i$
based on estimating the parameters of a power-law distribution. This idea is developed in the next subsections and the required background on this distribution is on \ref{interl_pl}. 

\subsection{Computing the summary statistic via a power-law distribution assumption}\label{QC_assumpt}

Our first assumption is the existence of a non-decreasing 
function $g$ relating the in-degree and the centrality\footnote{It is required an additional assumption --left continuity-- which can be consulted at Hosseini (2010) \cite{hosseini2010quantiles}.}. 
If there exists a function $g$ satisfying this condition, then the quantiles in the centrality side are equivalent to the application of $g$ on the same degree quantiles~\cite{hosseini2010quantiles}. With this result, the quantile proportions can be specified according to characteristics of the centrality distribution, as it is explained in Section \ref{all-pieces}.
In practice, and even more so considering that the in-degree is a discrete variable while the centrality is continuous, the object $g$ is a relation rather than a function. More formally, let $\{C_i\}_{i=1}^n$ be the set of quantile centrality values associated 
to the proportions $\{p_i\}_{i=1}^n$ that were used to compute the {quantile degree values} $\{d_i\}_{i=1}^n$ (see Equation \eqref{clues}). 
Given the above assumption about $g$, we can rewrite the sets $S_i$ as follows:
\begin{equation*}
\begin{array}{ccl}
S_1 & = &\{v\in V \mid g(\deg^{\text{in}}(v)) \leq C_1  \} \\
S_i & = &\{v\in V\mid  C_{i-1}< g(\deg^{\text{in}}(v)) \leq C_i \} \;\;\; (2\leq i\leq n)\\
S_{n+1} & = &\{v\in V \mid C_{n}< g(\deg^{\text{in}}(v)) \}
\end{array}
\end{equation*}

Our second assumption is that the centrality index that we want to estimate
follows a power-law distribution. 
We add this assumption motivated by 
the argument that QuickEst would have a \textit{negative bias} \cite{hertwigquick}, in the sense that it is a negative clue (or absent attribute) that stops this heuristic. Thus, a distribution such as the power law where most values are small (with mostly negative clues) and only a few high values exist (with mostly positive clues), would provide an optimal context for the performance of QuickEst, which is consistent with the finding that this heuristic predicts well the estimation behavior by some people on this kind of data \cite{von2008mapping}.   
Moreover, power-laws have a pervasive presence in many natural phenomena and magnitudes produced by humans too \cite{newman2005power},
although there has been some recent controversy on this topic~\cite{broido2019scale,barabasi2018love}
.
As we next show, our assumption of power-law distribution will allow us to use some particular properties  
to approximate the values $\{C_i\}_{i=1}^n$ used in the rewriting above, and then use them to efficiently compute 
the statistics $\{\bar{f}_i\}_{i=1}^{n+1}$ for every set $S_i$. In Section \ref{nw_defy}, we show some experiments to argue that these two assumptions of the heuristic are key to ensure its competitive accuracy. 


\subsection{Putting all the pieces together}\label{all-pieces}

Let $D=(V,A)$ be our input network, 
and recall that we are assuming that the centrality that we want to estimate for $D$ follows a power-law distribution.
Let $\hat{\alpha}$ be the estimate of the exponent parameter of the distribution (given by Equation~\eqref{mlealph}), and $\hat{x}_{\min}$ be the estimate of the lower limit of the distribution (given by the minimization of the functional of ~\eqref{KS}), which have been computed by considering a set of $m$ nodes in $V$ and their (real) centrality values. 
With all these pieces, we can compute the values $\{C_i\}_{i=1}^n$ associated 
to the proportions $\{p_i\}_{i=1}^n$
easily by using the equation
\begin{equation*}
\int_{\hat{x}_{\min}}^{C_i}K x^{-\hat{\alpha}}dx=p_i
\end{equation*}
from which we get that 
\begin{equation*}\label{threxp}
C_i=\hat{x}_{\min}\cdot(1 - p_i)^{\frac{1}{1-\hat{\alpha}}}.
\end{equation*}

Now, in order to compute the summary statistics $\{\bar{f}_i\}_{i=1}^{n+1}$, 
we will use the median of every set $S_i$. This median can be computed as follows. 
Given that we rewrote $S_i$ as the set of centrality values $x$ such that $C_{i-1} \leq x \leq C_{i}$, 
then the median ${\it md}_i$ of $S_i$ must verify 
\begin{equation*}
\int_{{\it md}_i}^{C_{i}}K x^{-\hat{\alpha}} dx=\frac{1}{2}\int_{C_{i-1}}^{C_{i}}K x^{-\hat{\alpha}} dx
\end{equation*}
from which we obtain that
\begin{equation}
{\it md}_i=\left(\frac{(C_{i-1})^{1-\hat{\alpha}}+(C_{i})^{1-\hat{\alpha}}}{2}\right)^{\frac{1}{1-\hat{\alpha}}}=\bar{f}_i\;\;\; (2\leq i\leq n)
\end{equation}
Moreover, since the extreme points of the distribution are $x_{\min}$ (estimated as $\hat{x}_{\min}$) 
and $\infty$, the two remaining statistics $\bar{f}_1$ and $\bar{f}_{n+1}$ are computed as
\begin{equation}
\bar{f}_1  = \left(\frac{(C_1)^{1-\hat{\alpha}}+(\hat{x}_{\min})^{1-\hat{\alpha}}}{2}\right)^{\frac{1}{1-\hat{\alpha}}}
\end{equation}
and
\begin{equation}
\bar{f}_{n+1} = 2^{\frac{1}{\hat{\alpha} - 1}}\cdot C_n
\end{equation}
We stress that with these formulas we compute the summary statistic $\bar{f}_i$ for each set $S_i$ just by knowing the values $\{C_i\}_{i=1}^n$, which are computed by using only the 
values $\hat{\alpha}$, $\hat{x}_{\min}$, and the underlying vector of proportions $\{p_i\}_{i=1}^n$. This last element was chosen as the quantile probability values that produced equidistant points on the range of $\{\log(h(v))|v\in V,h(v)\geq \hat{x}_{\min}\}$, that is, the set of vertices where the power-law is well defined for the harmonic centrality. Logarithmic binning is chosen to gauge the tail of the power-law distribution with higher frequency. The length $n$ of the vector of proportions required to construct the clues (see Equation~\eqref{clues}) was chosen after pilot testing on each type of distribution. See \ref{subsecB}, and Sections \ref{not_pl} and \ref{real_case} for more details. The election of this vector is a way of adapting QuickCent to distinct centrality distributions. Research on possible improvements achievable by tuning this vector may be addressed in future work.

The last element we introduced in our procedure, is the use of an additional quantile centrality value $C_0=\hat{x}_{\min}$, with the goal of spanning the centrality values $h(v)< \hat{x}_{\min}$ with greater accuracy. 
Since for this range of the vertex set the power-law distribution is no longer valid, the representative statistic $\bar{f}_0$ we have used is simply the empirical median of the harmonic centrality in the set of nodes $v$ such that $\deg^{\text{in}}(v)\leq g^{-1}(\hat{x}_{\min})$. With this element, it turns out that, if we use a proportions vector $\{p_i\}_{i=1}^n$ of length $n$, the total number of medians $\{\bar{f}_i\}_{i=0}^{n+1}$ is $(n+2)$. In the code provided to produce the analyses of this paper \cite{QCrepo}, this element is optional (and activated by setting \textbf{rm}=True or \textbf{rms}=True). All the results in this paper were obtained with this centrality quantile and median.  

\begin{exmp}\label{exmpl_3}
We continue revisiting Example \ref{the-exmpl}. If we fix $x_{\min}=1$, the exponent $\hat{\alpha}(1)$ that fits the complete distribution of centrality values, by using Equation (\ref{mlealph}), is $2.067$. The set of proportions shown in Example \ref{exmpl_2} comes from evaluating the centrality CDF on the set of points $\{1, 2.506, 6.283\}$, which correspond to $x_{\min}$ and two points ($n=2$) that in logarithmic scale turn out to be equidistant to the minimum and maximum of the set $\{\log(h(v))|v\in V,h(v)\geq \hat{x}_{\min}\}$, the (log) centrality domain of the given network where the power law is valid. From these parameters and the expressions shown in this section, one can get the medians required by QuickCent to make estimates. These can be examined in Table \ref{exampl_nw_stat}, corresponding to the model \textbf{QC100}, which has a MAE (mean absolute error) over the whole digraph of $3.606e-01$. A more interesting case may be computed when $\hat{\alpha}(1)$ is derived from a random sample of the centrality distribution. For example, by taking a sample without replacement of size $70\%$ one may get an exponent estimate of $\hat{\alpha}(1)=2.477$, which has a MAE of $6.948e-01$ and QuickCent estimates that can be examined in the model \textbf{QC70} in Table \ref{exampl_nw_stat}.
\end{exmp}

This completes all the ingredients for our instantiation of {QuickCent}, as we have the
values for the clues $(x_v^1,x_v^2,\ldots,x_v^n)$ computed from the in-degree of the node $v$, 
plus the values $\{d_i\}_{i=1}^n$ as shown in Equation~\eqref{clues}, and also the summary statistics 
$\{\bar{f}_i\}_{i=0}^{n+1}$ for each set $S_i$, which are the two pieces needed to apply the heuristic.

\section{Results}\label{res}
In the present section, we show the results of applying {\it QuickCent} either on synthetic data or on some empirical networks,
and we compare it with alternative procedures for centrality estimation.
We first show 
the comparison of QuickCent with other methods when applied on synthetic networks, considering accuracy and time measurements
in Sections 
\ref{subsecC} and~\ref{subsecD}. The synthetic network model corresponds to the preferential attachment (PA) growth model introduced in \ref{lin-PA}. Section \ref{nw_defy} reviews the output of QuickCent on null network models where its accuracy is not as good relative to other methods, with the aim of showing that the two assumptions of QuickCent (Section \ref{QC_assumpt}) are jointly required as a necessary condition for the competitive performance of this heuristics. The same benchmark presented for the synthetic case was applied to the empirical datasets, and the results are shown in Section \ref{real_case}. The experiments to check the fulfillment of QuickCent assumptions by the different networks are shown in \ref{assump_valid_p_val}, \ref{assump-rand}, \ref{assump-ER} and \ref{assump-empir}.
In all our experiments we consider harmonic centrality as the target to estimate. The number of nodes chosen for the synthetic networks experiments is $10,000$ and $1000$ for the null models, with the aim of accelerating the bootstrap computations to check the assumptions of QuickCent on each network. Similar sizes were searched for when choosing the tested empirical networks.
These are not really big numbers compared with modern networks.
We select these numbers as we need to be able to compute the exact value of the harmonic centrality
for all nodes in the graph, in order to compare our estimations with the real value, and regard it to be enough for a first assessment of the heuristic.

\subsubsection{Experiments specifications}

The norm that we employed to summarize the error committed on each node is the mean absolute error (MAE). This measure is preferable to other error norms, such as the Root mean squared error, because the units of the MAE are the same as the quantity under consideration, in this case, the harmonic centrality. On the other hand, since the MAE can be understood as the \textit{Minkowski} loss with $\mathcal{L}_1$ norm for the regression of the variable of interest, and in this case, it is known that the solution is given by the conditional median \cite{bishop2006}, it is reasonable to use the MAE when the summary statistic chosen is the median of each centrality interval. Finally, all the experiments were performed on the R language \cite{Rref} with igraph library \cite{igraph} for graph manipulation, and ggplot2 library \cite{ggplot} to produce the plots. 




\subsection{Comparison with other methods}\label{compar}\label{subsecC}
In this section, we compare the performance of known existing regression methods with QuickCent. 
This exercise allows us to evaluate the potential uses and applications of our proposal.
Specifically, whether it can deliver reasonable estimates, in relation to alternative solutions for the same task. 
This is not a trivial matter, considering that QuickCent is designed to do little computational work of parameter estimation and output production, possibly with limited training data, while common alternative machine learning (ML) methods generally perform more complex computations.
For a fair comparison, all other methods use only the in-degree as an explanatory variable. In rigor, QuickCent is able to produce the estimates only from the binary clues, without using the in-degree.

The competing methods considered are linear regression (denoted by L in plots), a regression tree (T) \cite{
quinlan1992learning,wek} and a neural network (NN) \cite{rumelhart1988parallel}, which are 
representatives of some of the most known machine learning algorithms.
We used \textit{Weka} \cite{wek} and the \textit{RWeka} R interface \cite{rwek}
to implement T and NN using default parameters.
In the literature, there is previous work specifically tailored to centrality estimation using ML methods, but for other centrality indices beyond harmonic centrality.
In particular Brandes and Pich study specific estimations for \emph{closeness} and \emph{betweenness} centrality~\cite{brandes2007centrality}.
It would be interesting to compare our method with the one proposed by Brandes and Pich~\cite{brandes2007centrality}, but this would amount to changing and adapting their method to harmonic centrality. We leave this adaptation and further comparison as future work.

\begin{figure}
\begin{subfigure}{.5\textwidth}
  \centering
  \includegraphics[width=\linewidth]{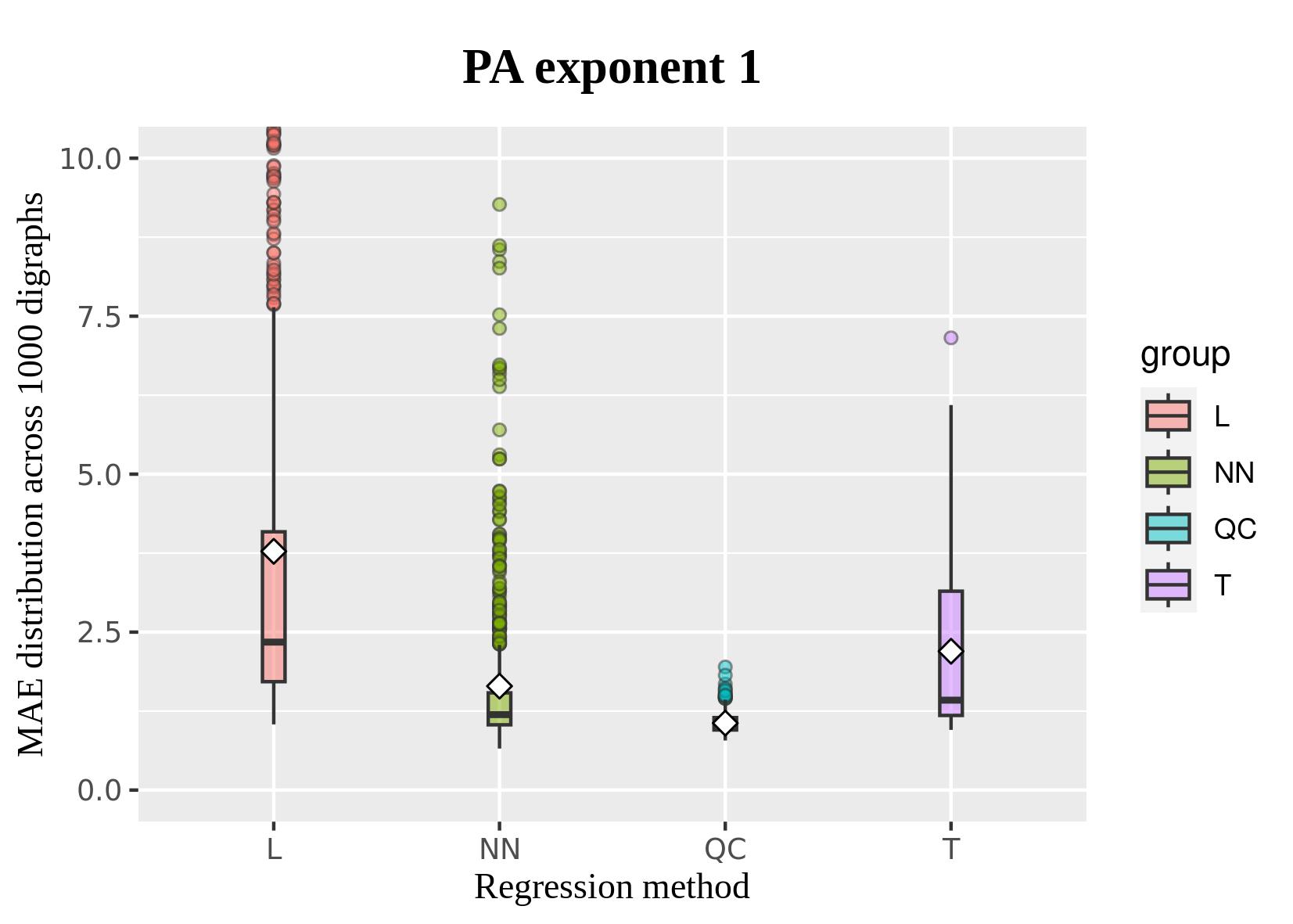}  
  \label{compar_10:sfig1}
\end{subfigure}
\begin{subfigure}{.5\textwidth}
  \centering
  \includegraphics[width=\linewidth]{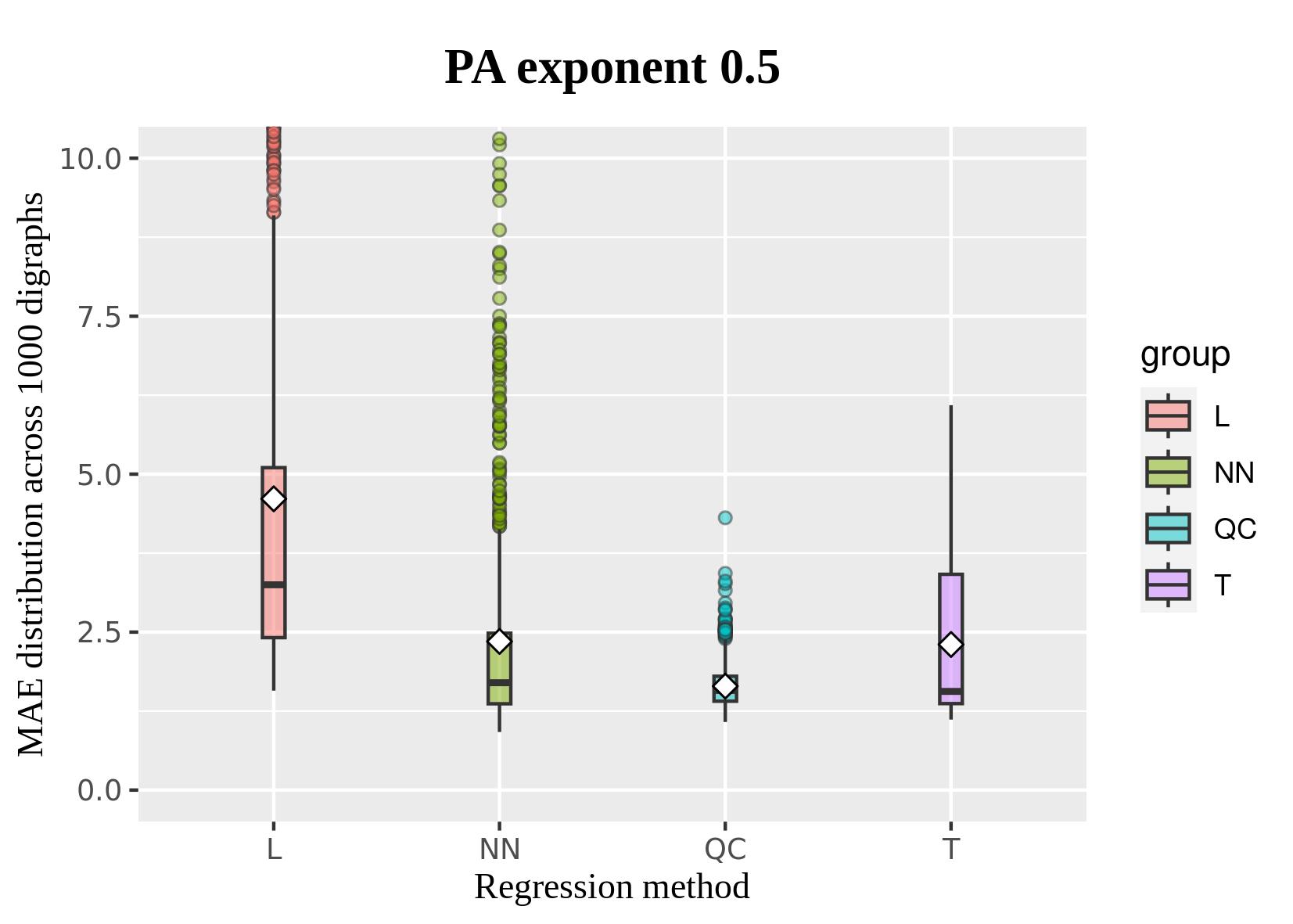}  
  \label{compar_10:sfig2}
\end{subfigure}
\newline

\begin{subfigure}{\textwidth}
  \centering
  \includegraphics[width=.5\linewidth]{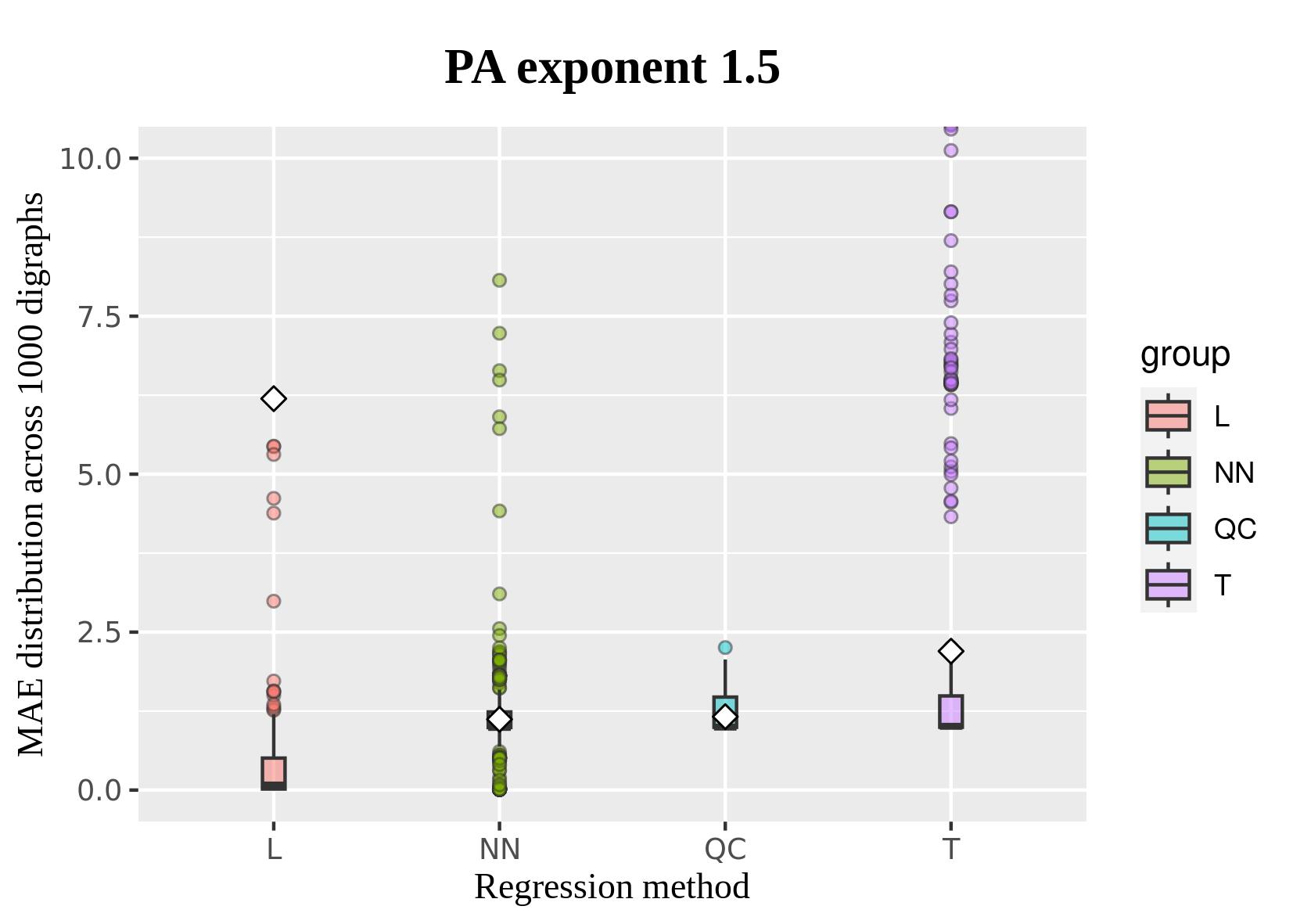}  
  \label{compar_10:sfig3}
\end{subfigure}

\caption{{\bf Benchmark with other ML methods for different exponents of PA digraph instances and 10 $\%$ of training size.} For each regression method, there is a boxplot showing the MAE distribution. Each boxplot goes from the $25-$th percentile to the $75-$th percentile, with a length known as the \textit{inter-quartile range} (IQR). The line inside the box indicates the median, and the rhombus indicates the mean. 
The whiskers start from the edge of the box and cover until the furthest point within $1.5$ times the IQR. Any data point beyond the whisker ends is considered an outlier, and it is drawn as a dot. For display purposes, the vertical limit of the plots has been set to $10$, since the highest MAE outliers of NN or L, depending on the PA exponent, blur the details of the model performance.}
\label{compar_10}
\end{figure}
\begin{figure}
\begin{subfigure}{.5\textwidth}
  \centering
  \includegraphics[width=\linewidth]{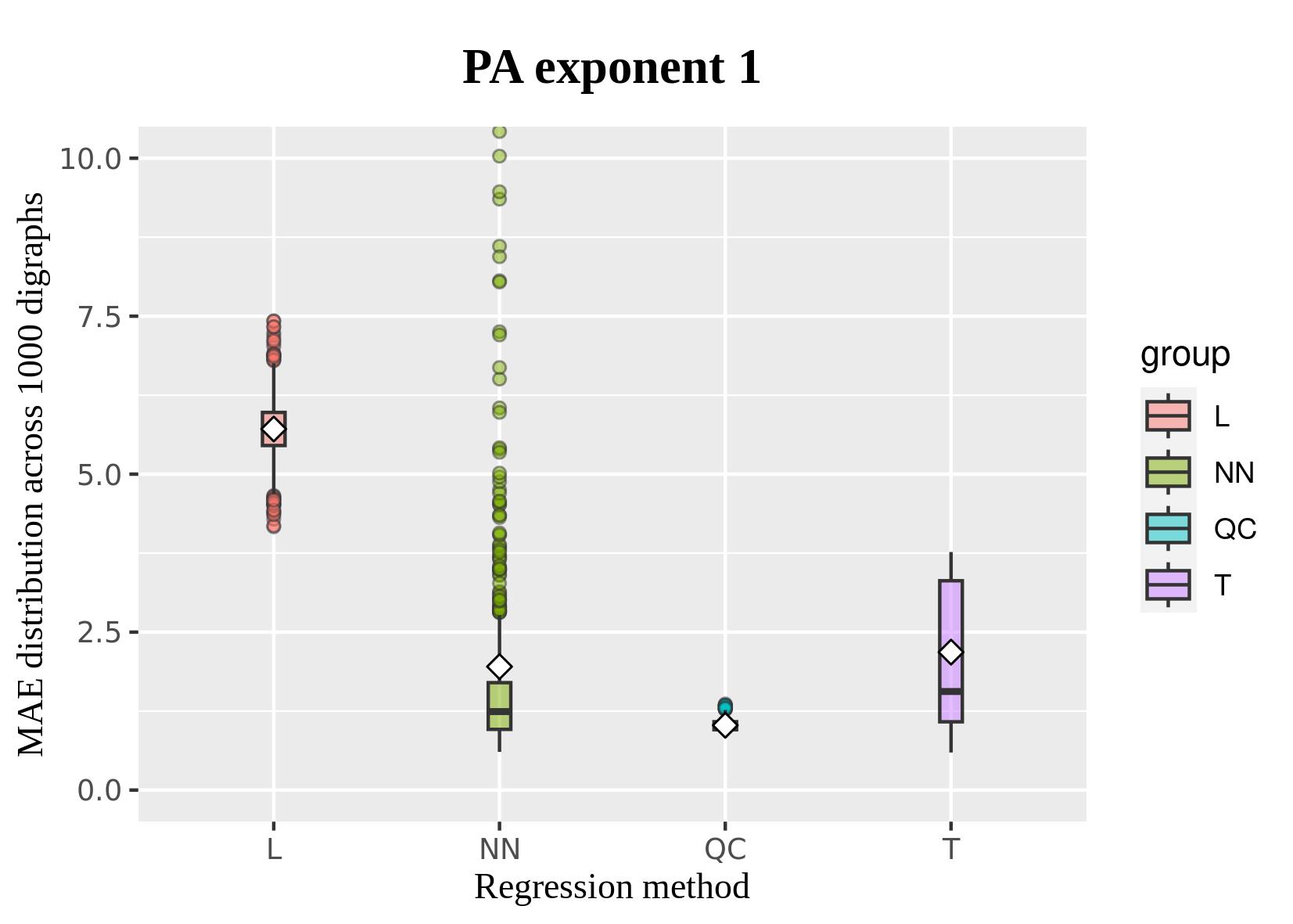}  
  \label{compar_100:sfig1}
\end{subfigure}
\begin{subfigure}{.5\textwidth}
  \centering
  \includegraphics[width=\linewidth]{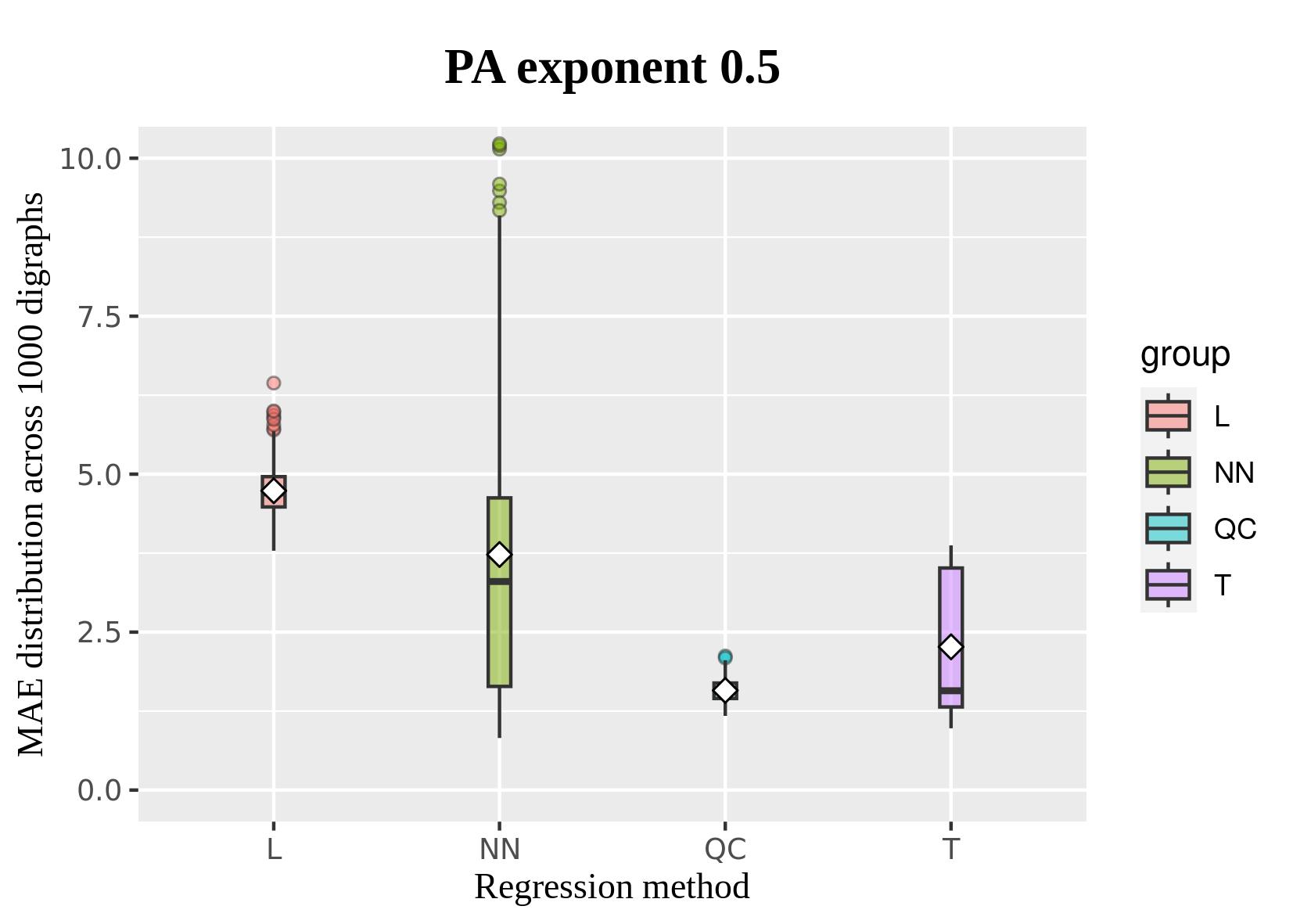}  
  \label{compar_100:sfig2}
\end{subfigure}
\newline

\begin{subfigure}{\textwidth}
  \centering
  \includegraphics[width=.5\linewidth]{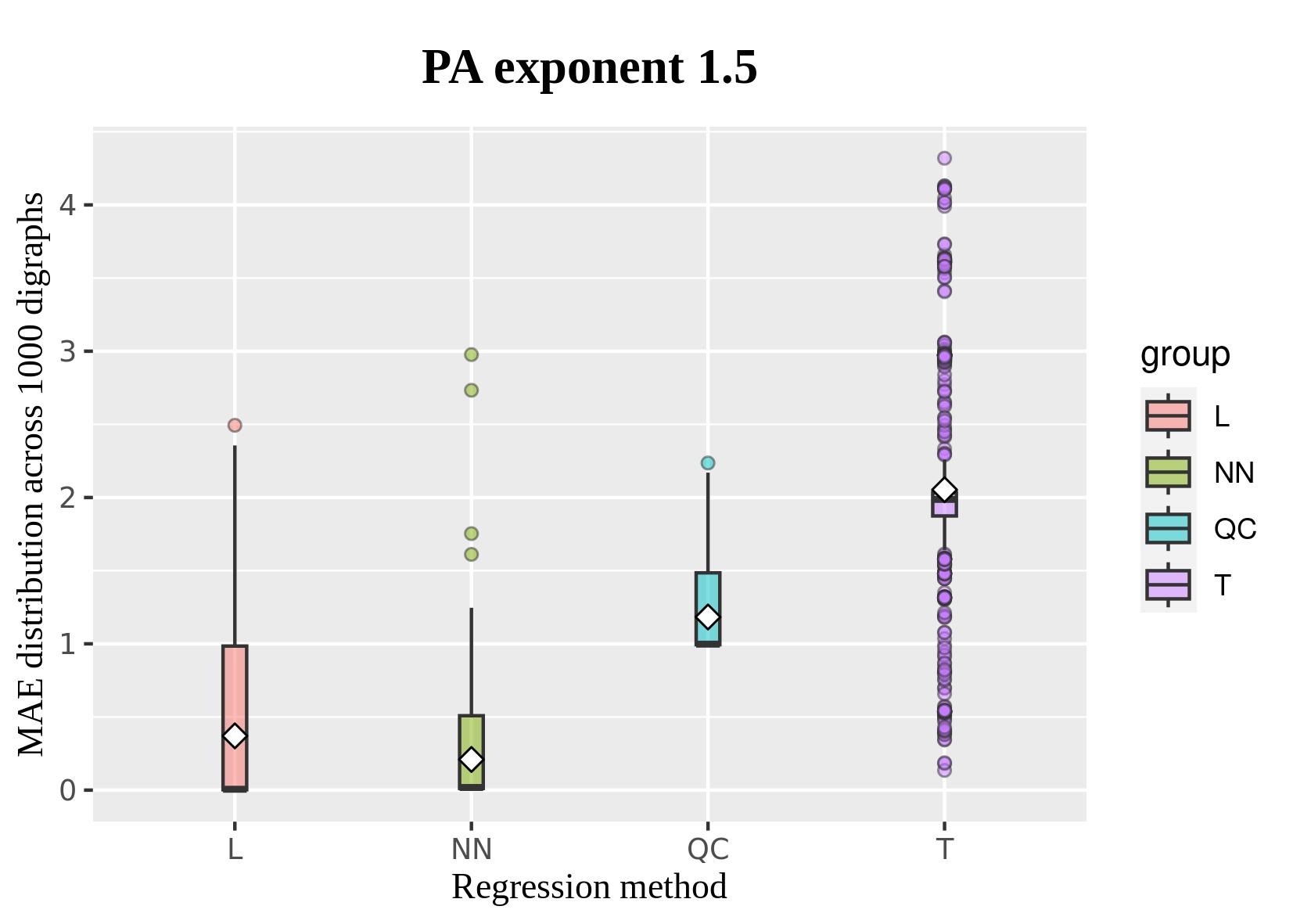}  
  \label{compar_100:sfig3}
\end{subfigure}

\caption{{\bf Benchmark with other ML methods for different exponents of PA digraph instances and 100 $\%$ of training size.} For each regression method there is a boxplot showing the MAE distribution. Each boxplot goes from the $25-$th percentile to the $75-$th percentile, with a length known as the \textit{inter-quartile range} (IQR). The line inside the box indicates the median, and the rhombus indicates the mean. 
The whiskers start from the edge of the box and cover until the furthest point within $1.5$ times the IQR. Any data point beyond the whisker ends is considered an outlier, and it is drawn as a dot. For display reasons, the vertical limit of the two first plots was set to $10$, since the highest MAE outliers of NN or L, depending on the PA exponent, blur the details of the model performance.}
\label{compar_100}
\end{figure}

The results of this experiment are shown in Figure \ref{compar_10} with a training size of 10 $\%$ and Figure \ref{compar_100} with a training size of 100 $\%$, where the test set is always the entire digraph. The two training sizes are studied with the goal of assessing the impact of scarce data on the distinct estimation methods, by contrasting a full versus a scarce data scenario. In the figures, it can be seen that QuickCent (QC) produces the lowest MAE errors of all the methods, either in terms of the IQR length, or the mean and outliers, for PA exponents 1 and 0.5. As noted in \ref{assump_valid_p_val}, these are the cases where the centrality distribution has a better fit by a power-law model, but it is anyway a remarkable result considering the error committed by fixing ${x}_{\min}=1$, see \ref{subsecB}. Examining the MAE medians of these simulations in Table \ref{mae_median}, one sees that the QC median is at the level of the most competitive ML methods in the simulations, NN and T, sometimes being the best of the three depending on the experiment. However, for exponents 1 and 0.5, where the power-law is present, the good thing about QC is that the upper quantiles and even the outliers remain low compared to other methods (see \figref{compar_10} and \figref{compar_100}). 

\begin{table}
{\small
\begin{center}
  \begin{tabular}{c|cccccccc}
 \textbf{PA $\beta$} &\textbf{L10} & \textbf{NN10} & \textbf{QC10} & \textbf{T10}& \textbf{L100} & \textbf{NN100} & \textbf{QC100} & \textbf{T100}\\
\hline
1&2.341& 1.194& 1.040& 1.422&5.711& 1.242& 1.009& 1.560\\
0.5& 3.249& 1.699&1.576& 1.561&4.704& 3.300&1.578& 1.571\\
1.5&0.079& 0.991& 0.996& 1.009&0.006& 0.018& 0.997&1.986\\
\end{tabular}
\end{center}
\caption{\textbf{Medians of the MAE distribution across 1000 digraphs.} These estimates are computed from the same simulationes displayed in \figref{compar_10} and \figref{compar_100}. The suffix of each method abbreviation corresponds to the size of the training size used. The exponent $\beta$ corresponds to the exponent of the preferential attachment growth (\ref{lin-PA}). The number of decimal places is truncated to three with respect to the source.}
\label{mae_median}
}
\end{table} 

Thus, the main takeaway is that QC, when its assumptions are fulfilled, is able to produce estimates at the same level as much more complex ML methods, with likely lower variance. This fact is consistent with the argument given by Brighton and Gigerenzer \cite{Brighton20151772} claiming that the benefits of simple heuristics are largely due to their low variance. The argument relies on the decomposition of the (mean squared) error into \textit{bias}, the difference between the average prediction over all data sets and the desired regression function, and \textit{variance}, the extent to which the estimates for individual datasets vary around their average \cite{bishop2006}. Thus, along the range of the bias-variance trade-off of models, simple heuristics are relatively rigid models with high bias and low variance, avoiding the potential overfitting of more complex models.

By examining the contrast of the outliers between \figref{compar_10} and \figref{compar_100}, it can be noticed that QC suffers the least impact from scarce data. In the case of L and NN, they show a similar pattern for power-law centralities (PA exponents 1 and 0.5). They have medians that are lower for the 10 $\%$ training size than those obtained with the whole network. Since there are only a few large values in the entire graph, when the training sample gets smaller, the sample values have a better linear fit, in comparison to larger samples. Therefore, a linear model adjusted to some small sample provides a good fit to the small-to-moderate centrality size nodes, which is the case for most of the nodes. This also explains the presence of higher outliers in the 10 $\%$ training size. On the other hand, the behavior of the regression tree is more similar to that of QuickCent.

\subsection{Time measurements} \label{subsecD}

Elapsed time was the other distinct aspect of the method's performance. The time cost is a critical feature of any approximation method because it measures the tradeoff between accuracy and cost. 
Elapsed time measurements were taken in the experiments shown in Section \ref{subsecC}, and the results are displayed in Table \ref{time}.
These times consider the training and inference time for each method, without including any centrality computation. In the case of QuickCent, the computation of the proportions vector is not considered for the elapsed times. The reason is that this computation is not directly part of the heuristics, and there is also the possibility of using a prototypical vector for a given type of centrality distribution. 

\begin{table}
{\small
\begin{center}
  \begin{tabular}{c|cccccccc}
 & \textbf{L10} & \textbf{QC10} & \textbf{T10} & \textbf{NN10}& \textbf{L100} & \textbf{QC100} & \textbf{T100} & \textbf{NN100}\\
\hline
\textbf{Mean} & 2&76&23&98 &3.9&77.5&28.0&760.9\\
\textbf{St. Dev.} & 0.25&2.95&2.98&2.16&0.35&3.27&4.66&9.88\\
\end{tabular}
\end{center}
\caption{\textbf{Mean and standard deviation of elapsed time in milliseconds over $1000$ digraphs with PA exponent 1.} The suffix of each method abbreviation corresponds to the size of the training set used.}\label{time}}
\end{table} 

From the table, we can see that the elapsed time of QC is in the middle range of the compared methods. The linear regression has the lowest times, around one order of magnitude faster than QuickCent and the regression tree, and the neural network has the highest elapsed time. 
Note also that there is no significant time difference between the $10\%$-case and the $100\%$-case 
for QC. This can be explained by the fact that the differences in sample sizes only affect the number of terms in the sum in Equation~\ref{mlealph} when estimating the exponent $\hat{\alpha}$, and summing a list of values is an extremely simple and quick procedure. 

Based on these results, we conjecture that QuickCent has the lowest time complexity among the tested methods. Among the computations that QuickCent performs, the most expensive ones correspond to the selection problem of finding the median of the lowest centrality values (Section \ref{all-pieces}), plus the quantile degree values (Section \ref{using_indeg}). The procedure used to compute the proportions vector (Section \ref{all-pieces}), which is not considered in the elapsed time measurements, also relies on solving the selection problem (of the maximum of the set of centrality values) and sorting of the centrality values set (to find the proportions). These problems may be solved in linear time \cite{cormen2022introduction} on the input size, that is, linear on the network size $\mathcal{O}(|V|)$. In contrast to the highly optimized R implementations for L, T, and NN, we considered only a naive implementation of QuickCent without, for example, architectural considerations. With these improvements such as using more appropriate data structures, these times could still be improved. We left as future work the construction of an optimized implementation for QuickCent.

\subsection{Networks defying QuickCent assumptions}\label{nw_defy}

Up to this point, we have mainly seen examples of networks where QuickCent exhibits quite good performance compared to competing regression methods. In order to give a full account of QuickCent capabilities and its \textit{ecological rationality} \cite{hertwigquick}, one should also have an idea of the networks where its accuracy deteriorates. To accomplish this, we will look at the two assumptions of QuickCent, namely, the power-law distribution of the centrality, and its monotonic map with the in-degree, to show that they are jointly required as a necessary condition for the competitive performance of the heuristic. Our approach here is to work with two null network models, each acting as a negation of the conjunction of the two assumptions, which provide strong evidence for this claim. 

\subsubsection{Response to the loss of the monotonic map}\label{not_mon}

\begin{figure}
  \includegraphics[width=\linewidth]{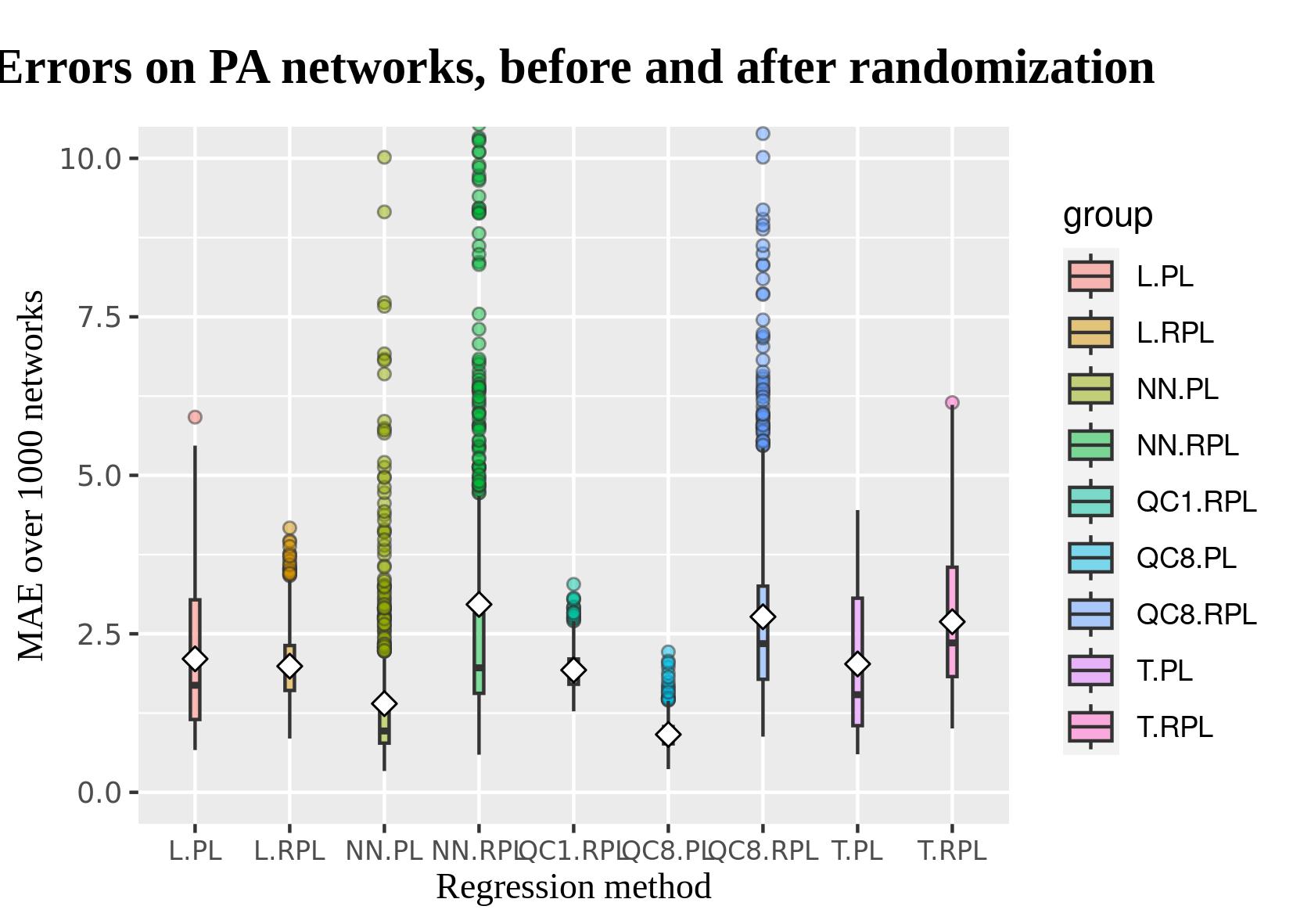}  

\caption{{\bf Effect of randomization on different ML methods using 30 $\%$ of the training size.} Each boxplot group is labeled with the name of the ML method, a dot, and the type of network on which the estimates are made (`PL' for the initial PA network, `RPL' for the network after randomization). QC8 corresponds to QuickCent with a proportion vector of length $8$, and analogously for QC1. For each regression method, there is a boxplot representing the MAE distribution. Each boxplot goes from the $25-$th percentile to the $75-$th percentile, with a length known as the \textit{inter-quartile range} (IQR). The line inside the box indicates the median, and the rhombus indicates the mean. 
The whiskers start from the edge of the box and extend to the furthest point within $1.5$ times the IQR. Any data point beyond the whisker ends is considered an outlier, and it is drawn as a dot. For display reasons, the vertical limit of the plots was set at $10$, since the highest MAE outliers of NN, make blur the details of the model performance.}
\label{compar_randomiz}
\end{figure}

Our first null model is a scale-free network built by preferential attachment, just as in the previous experiments, but after a \textit{degree-preserving randomization} \cite{maslov2002specificity} of the initial network, which is simply a random reshuffling of arcs that keeps the in- and out-degree of each node constant. The aim is, on the one hand, to break the structure of degree correlations found in preferential attachment networks \cite{li2004first,zhang2015exactly}, which may be a factor favoring a monotonic relationship between in-degree and harmonic centrality, and on the other hand, to maintain a power-law distribution for the harmonic centrality by preserving the degree sequence of nodes in the network. This last feature does not ensure that the centrality distribution is a power-law, since randomization also affects this property. \ref{assump-rand} displays the results of the experiments performed to check the assumptions of QuickCent on the randomized networks, showing that these networks satisfy them. Finally, \figref{compar_randomiz} shows the impact of randomization on each regression method. This is an experiment where 1000 PA networks (exponent 1) of size 1000 were created, and the four ML methods used in Section \ref{subsecC} were trained on each network with samples of size 30 $\%$ of the total node set, using only the in-degree as the predictor variable for the harmonic centrality. The same procedure was run on each network after applying degree-preserving randomization on 10000 pairs of arcs. The plot shows that the randomization has a similar impact on the performance loss of each method, which is an expected result due to the fact that the only source of information used by each method, the in-degree, becomes less reliable due to the weaker association with harmonic centrality thanks to the arc randomization. Since QuickCent was the most accurate of the methods tested on the initial PA networks, it appears to be also one of the methods most affected methods by the randomization. 

\subsubsection{Response to the loss of the power-law distribution of centrality}\label{not_pl}

Our second null model is the directed Erd{\"o}s-R{\'e}nyi graph model \cite{erdos1959random,bollobas1981degree,karp1990transitive}, and is chosen with the aim of gauging the impact of losing the power-law distribution of the centrality while maintaining the monotonic map from in-degree to centrality. This model is known to have a Poisson degree distribution \cite{bollobas1981degree}, a behavior very different from a heavy-tailed distribution, and according to our simulations, see \ref{assump-ER}, it turns out to be ideal for our purposes. We choose connection probabilities that ensure a unimodal distribution for centrality and a strong correlation with in-degree, i.e. with a mean in-degree greater than 1 \cite{karp1990transitive}.  In order to get a fair control on the performance of QuickCent, we have taken two empirical digraphs that satisfy the given condition of the mean in-degree, with node sets of size near $1000$, just to accelerate the bootstrap p-value computations. The networks are extracted from the KONECT database \cite{kunegis2013konect}\footnote{http://konect.cc/}, and their meta-data is shown in Table \ref{control_netw}. The fields $N$ and $\bar{\deg^{\text{in}}}$ given in this table, are used to determine the network size and the connection probability used to instantiate the respective ER digraphs from the identity $\bar{\deg^{\text{in}}}=p\cdot(N-1)$. 

\begin{table}
{\small
\begin{center}
  \begin{tabular}{c|cccccc}
\textbf{Name}  & \textbf{N} & \textbf{$\bar{\deg^{\text{in}}}$}&\textbf{Corr} & \textbf{Arc meaning}&\textbf{Ref.}\\
\hline
\textbf{moreno\_blogs} & 990 & 19.21 & 0.872 & Blog hyperlink & 
\cite{adamic2005political} \\
\textbf{subelj\_jung-j}  & 2208 & 62.81&0.808 & Software Class dependency & 
\cite{vsubelj2012software}\\
\end{tabular}
\end{center}
\caption{\textbf{General description of the two empirical control networks.} The fields in the table are the dataset name, the number of nodes with positive in-degree (N), the mean in-degree of nodes with positive in-degree ($\bar{\deg^{\text{in}}}$), the Spearman correlation between the positive values of in-degree and harmonic centrality (Corr), the meaning of the arcs, and the original reference. The name corresponds to the \textit{Internal name}field in the KONECT database. To access the site to download the dataset, append the internal name to the link \textit{http://konect.cc/networks/}.}
\label{control_netw}}
\end{table}

\begin{figure}
\begin{subfigure}{.5\textwidth}
  \centering
  \includegraphics[width=\linewidth]{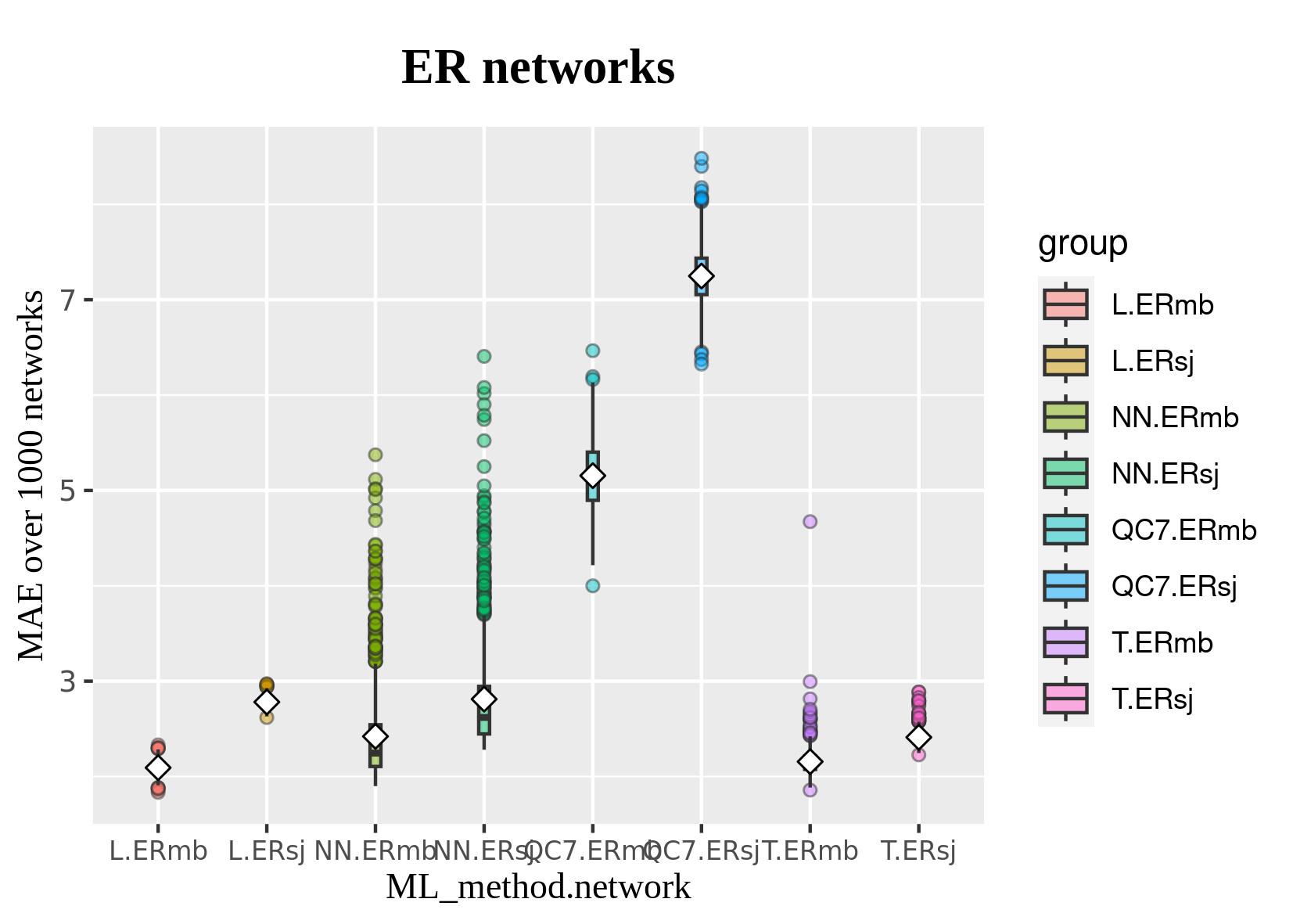}  
  \label{pl_vs_ER_sfig1}
\end{subfigure}
\begin{subfigure}{.5\textwidth}
  \centering
  \includegraphics[width=\linewidth]{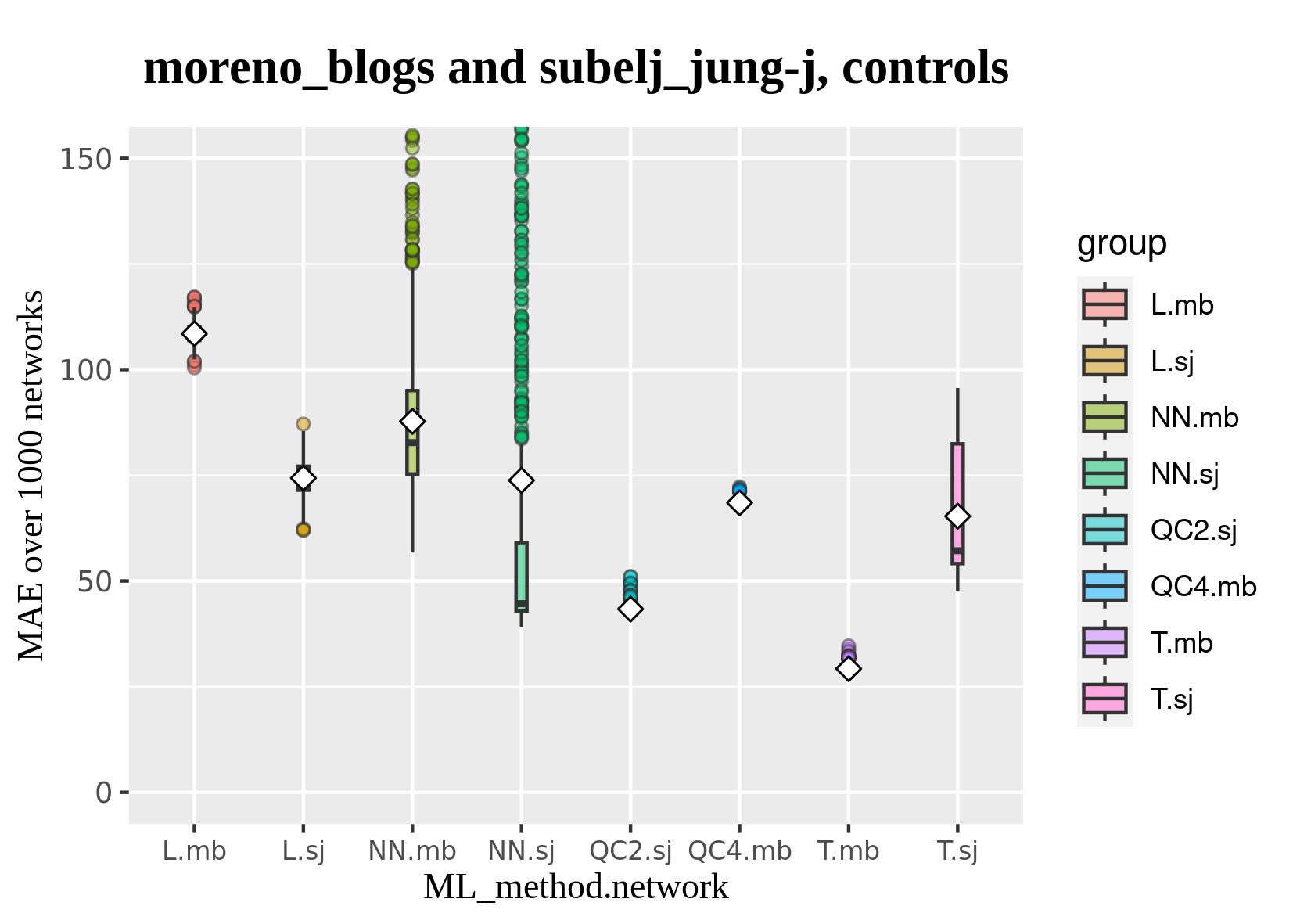}  
  \label{pl_vs_ER_sfig2}
\end{subfigure}

\caption{{\bf Effect of centrality distribution on different ML methods using 30 $\%$ of training size.} Each boxplot group is labeled with the name of the ML method, a dot, and the type of network on which the estimates are made (`mb' for moreno\_blogs, `sj' for subelj\_jung-j, `ERmb' for the ER digraph created with the parameters of moreno\_blogs, and analogously for `ERsj'). The number after `QC' is the length of the vector of proportions used by that method, corresponding to the best accuracy for the respective network. For each regression method, there is a boxplot representing the MAE distribution. Each boxplot goes from the $25-$th percentile to the $75-$th percentile, with a length known as the \textit{inter-quartile range} (IQR). The line inside the box indicates the median, and the rhombus indicates the mean. 
The whiskers start from the edge of the box and extend to the furthest point within $1.5$ times the IQR. Any data point beyond the whisker ends is considered an outlier, and it is drawn as a dot. For display reasons, the vertical limit of the control network plot has been set at $150$, as the highest MAE outliers of NN blur the details of the model performance.}
\label{pl_vs_ER}
\end{figure}

Finally, in \figref{pl_vs_ER} we can see the results of an experiment analogous to the one with the first null model, that is, there are 1000 iterations where the same four ML methods were trained on each network, two ER graphs with the two connection probabilities and sizes given by the two empirical/control networks, with random samples of size 30 $\%$ of the total node set, using only the in-degree. Since the unimodal distribution of ER digraphs is very different from a power-law, in this experiment we have taken the approach of using the parameter $\hat{x}_{\min}$ estimated by the method reviewed in \ref{interl_pl}, as well as the search space restriction explained there, instead of a fixed lower limit as in the previous experiments. Now, by comparing the two plots in \figref{pl_vs_ER}, one can observe a noticeable difference in the behavior of QuickCent in the two cases. While QuickCent achieves an average accuracy relative to other regression methods on the control networks with centrality distributions that are more or less close to heavy-tailed, on ER digraphs with similar characteristics to the controls, QuickCent consistently performs worse than other methods. The performance of QuickCent in these plots corresponds to the best possible for each network as a function of the length of the proportions vector, denoted by the number after `QC'. This output is also consistent with the difference in p-values of the power-law fit between the control networks and 1000 instances of the ER models, reported in Table \ref{verif_assump_ER_control}. These results reveal the critical importance of the centrality distribution of the data set for the proper functioning of QuickCent.   

On the other hand, all of the methods exhibit better performance on the ER digraphs than on the corresponding control network, probably due to less heterogeneity in the values to be predicted on the former. Finally, as a side note for working on empirical network datasets, for general networks it should be more accurate to use the fitted value of $\hat{x}_{\min}$ than a fixed value, although this depends on the variability range existing on the values less than $\hat{x}_{\min}$, which may introduce potentially large contributions to the estimation error. Observe that there is an additional computational overhead due to the calculation of $\hat{x}_{\min}$.

\begin{figure}
\begin{subfigure}{.5\textwidth}
  \centering
  \includegraphics[width=\linewidth]{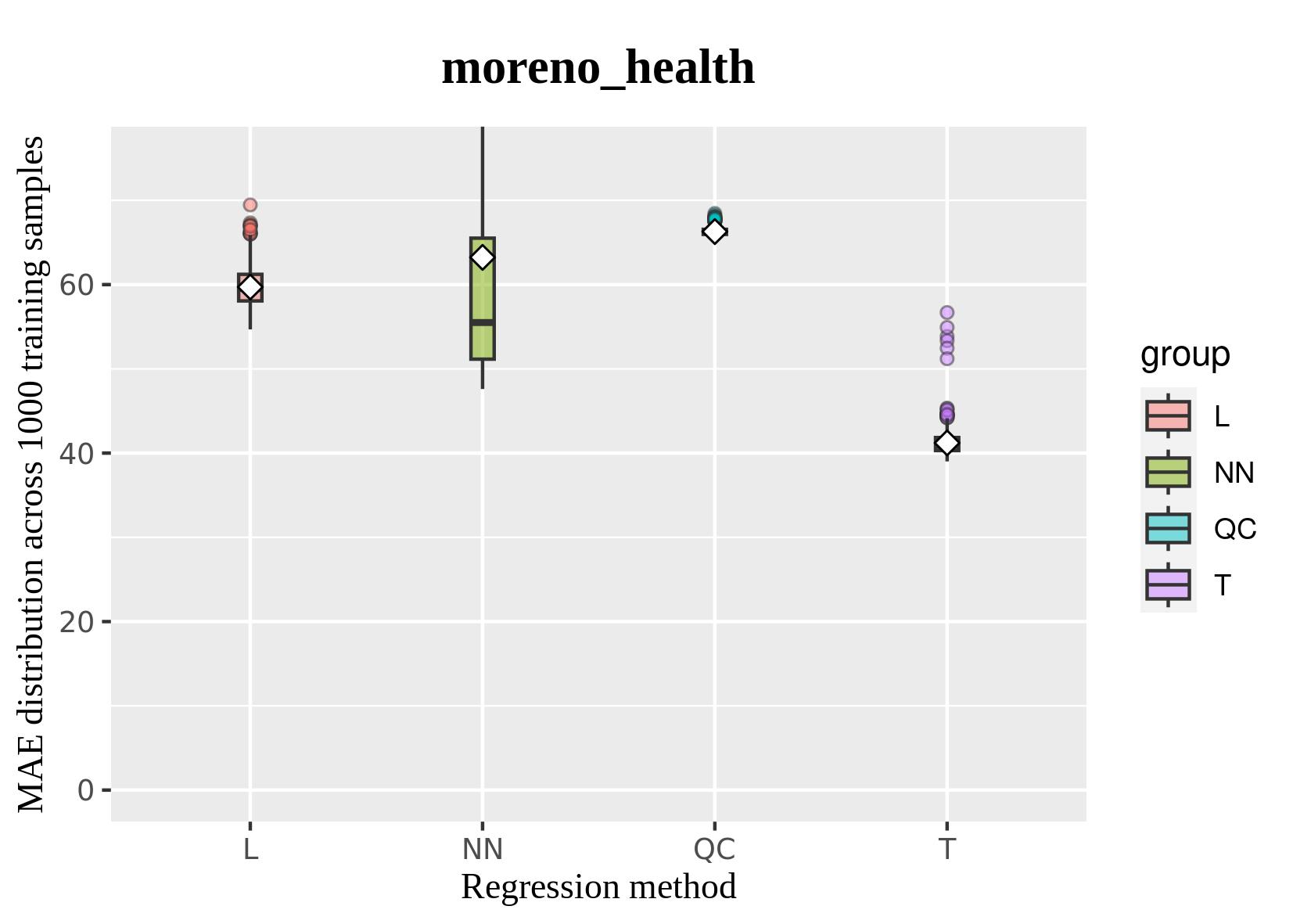}  
  \label{results_real:sfig1}
\end{subfigure}
\begin{subfigure}{.5\textwidth}
  \centering
  \includegraphics[width=\linewidth]{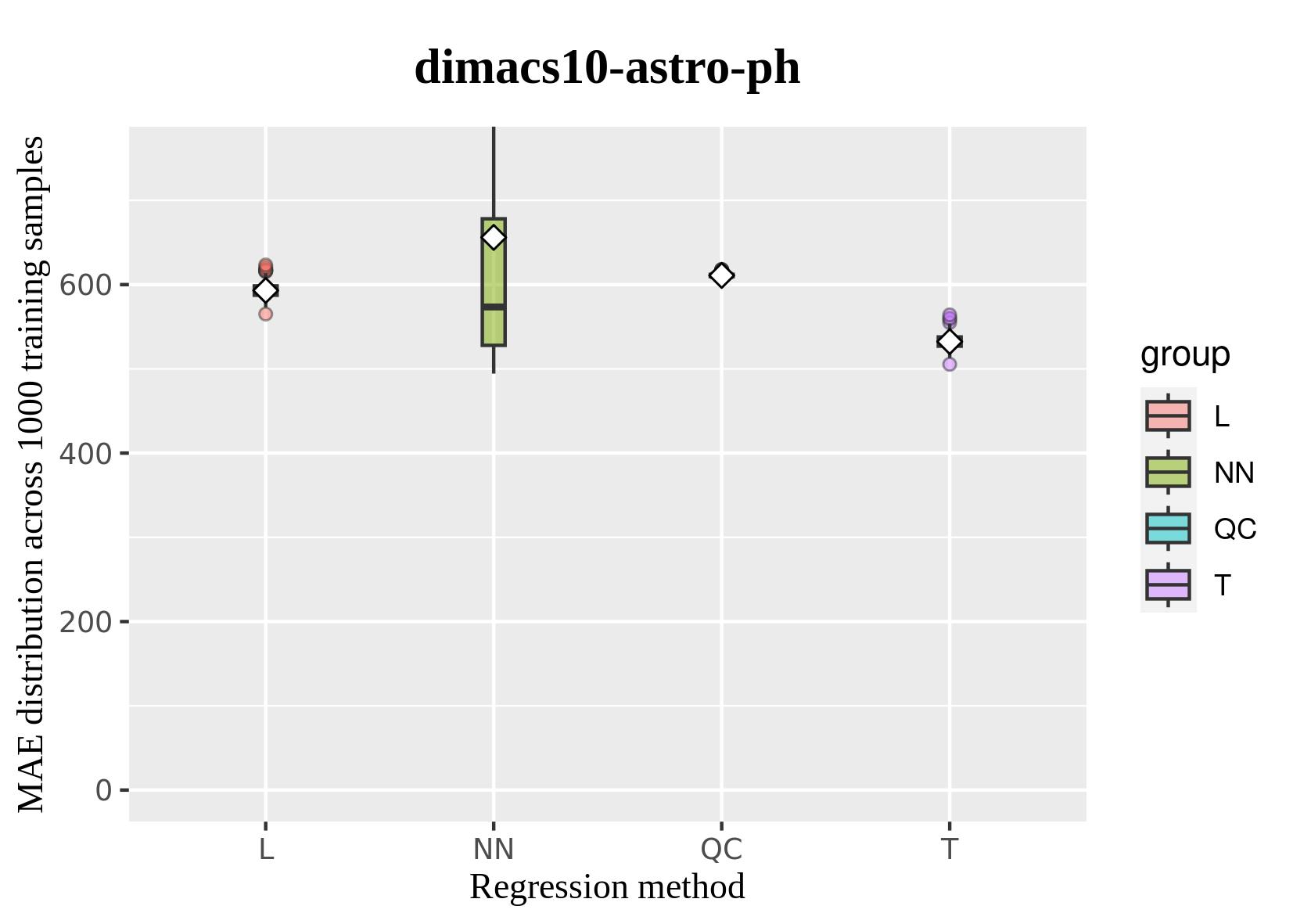}  
  \label{results_real:sfig2}
\end{subfigure}

\begin{subfigure}{.5\textwidth}
  \centering
  \includegraphics[width=\linewidth]{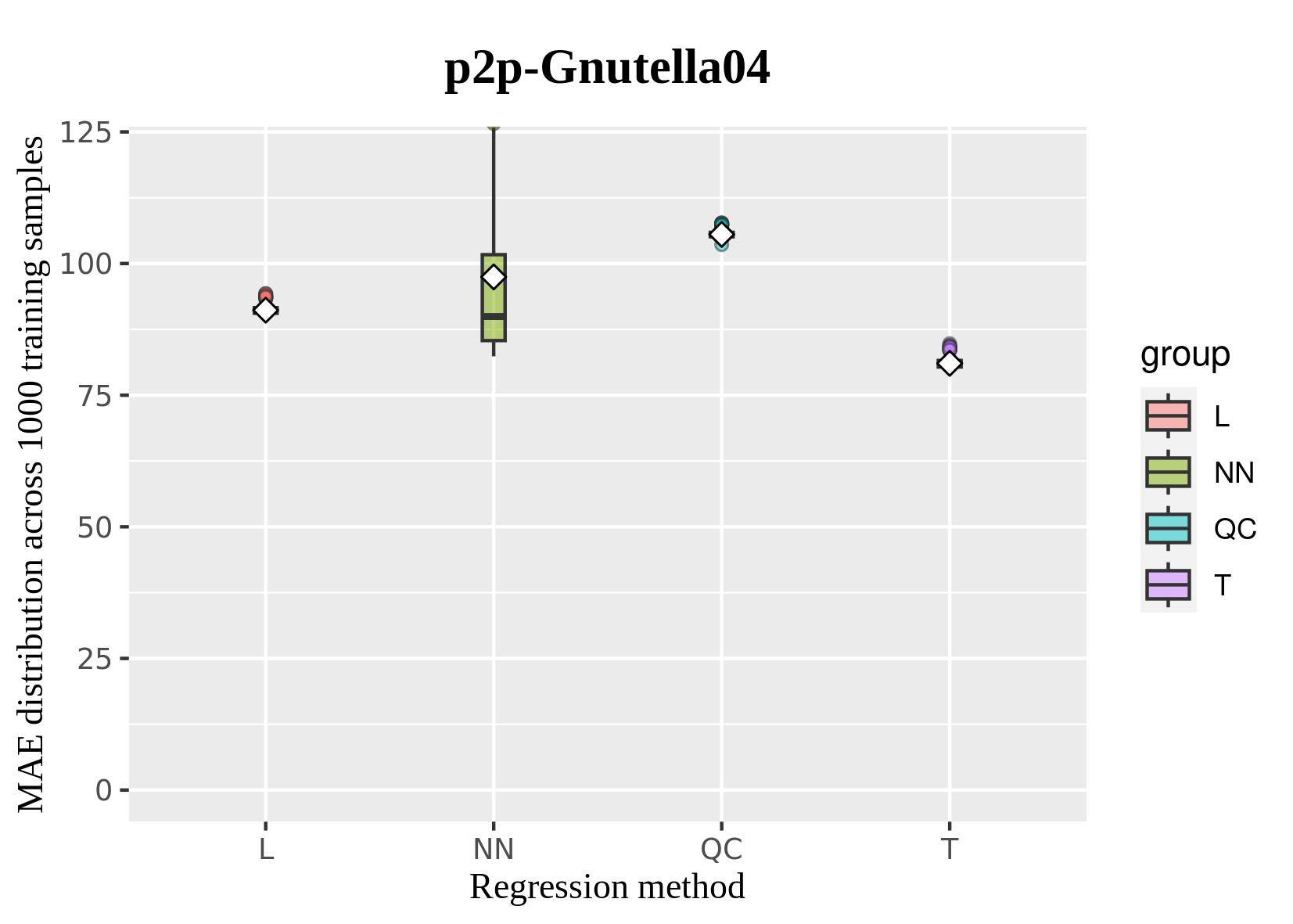}  
  \label{results_real:sfig4}
\end{subfigure}
\begin{subfigure}{.5\textwidth}
  \centering
  \includegraphics[width=\linewidth]{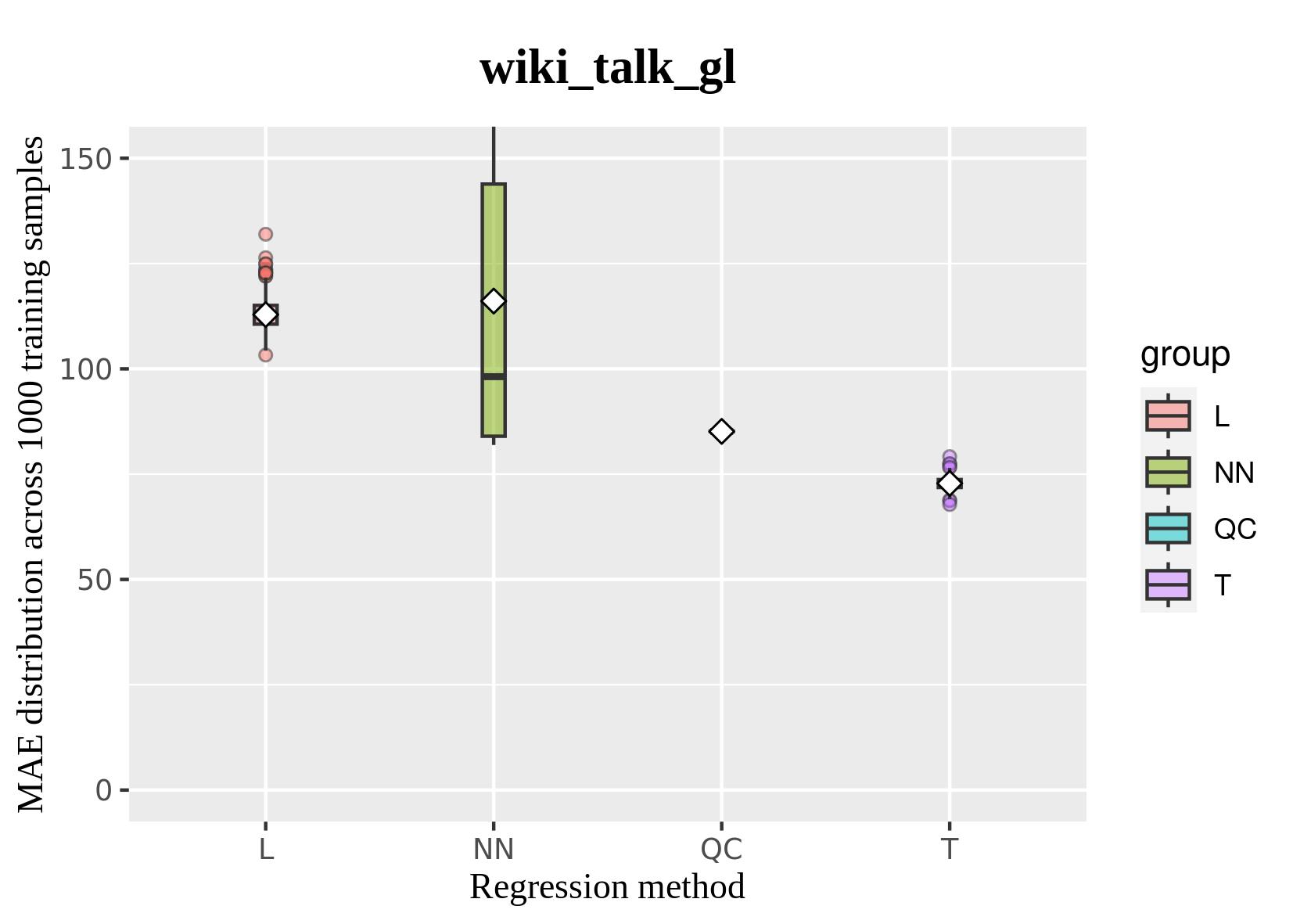}  
  \label{results_real:sfig5}
\end{subfigure}
\caption{{\bf Performance of QuickCent against  known ML algorithms on each dataset.} The competing algorithms are the same as in Section \ref{subsecC}, that is, a linear regression (L), a neural network (NN), and a regression tree (T), all with default parameters. Each point from each boxplot is the MAE of the respective model trained with a random sample of nodes of size 10 $\%$ of the total, and all the samples come from the same respective network. The white rhombus in each boxplot is the mean of the distribution.}
\label{results_real}
\end{figure}
\subsection{Experiments with empirical networks}\label{real_case}

In this section, we present the performance of QuickCent on some real network datasets of similar size to the synthetic networks already tested, also in comparison to other machine learning methods. 
These results are only a first glimpse of the challenges this heuristics may encounter when dealing with real datasets, and they should also be considered as a proof of concept. 

We selected five datasets, all of them extracted from the KONECT network database \cite{kunegis2013konect}\footnote{http://konect.cc/}, a public online database of more than a one thousand network datasets. The criteria for selecting the networks were that, besides being similar in size to our synthetic networks($10000$ nodes), each network had a distinct meaning, i.e., networks representing distinct systems from different contexts. General descriptors of these datasets are displayed on Table \ref{real_netw}. There, we can see that we have selected: a social network of friendships among students created from a survey \cite{moody2001peer}, a co-authorship network from the \textit{astrophysics} section of arXiv from 1995 to 1999 \cite{newman2001structure}, a citation network of publications from DBLP \cite{ley2002dblp}, a network of connected Gnutella hosts from 2002 \cite{ripeanu2002mapping}, and the communication network of messages among users from the Galician Wikipedia \cite{sun2016predicting}. See \ref{assump-empir} to review the experiments of assumption verification on these empirical datasets.

\begin{table}
{\small
\begin{center}
  \begin{tabular}{c|ccccc}
\textbf{Name}  & \textbf{Dir.} & \textbf{N} & \textbf{m} & \textbf{Edge meaning}&\textbf{Reference}\\
\hline
\textbf{moreno\_health}  & Y & 2539 & 12969 & Friendship & 
\cite{moody2001peer} \\
\textbf{dimacs10-astro-ph}  & N & 16046	 & 121251 & Co-authorship & 
\cite{newman2001structure} \\
\textbf{dblp-cite}  & Y & 12590 & 49759 & Citation & 
\cite{ley2002dblp} \\
\textbf{p2p-Gnutella04}  & Y & 10876 & 39994 & Host Connection & 
\cite{ripeanu2002mapping} \\
\textbf{wiki\_talk\_gl}  & Y & 8097	 & 63809 & Message & 
\cite{sun2016predicting} \\
\end{tabular}
\end{center}
\caption{\textbf{General description of the five empirical network datasets.} The fields in the table are the dataset name, whether the network is directed, the number of nodes (N), the number of edges (m), the meaning of the edges, and the original reference. The name corresponds to the \textit{Internal name} field in the KONECT database. To access the site to download the dataset, append the internal name to the link \textit{http://konect.cc/networks/}.}
\label{real_netw}}
\end{table}

We end this section with several plots in \figref{results_real} showing the performance of QuickCent compared to the same ML algorithms from Section \ref{subsecC}, all of them trained with samples of size equal to the 10 $\%$ of each dataset. The feature of QuickCent having the smallest error dispersion observed in the synthetic datasets is also observed in this case. The QC performance, although not as good as in the synthetic datasets, is competitive with the other ML methods, and even better than an important number of instances of the neural network for the dimacs10-astro-ph and wiki\_talk\_gl datasets. These results are obtained with a length of the proportions vector equal to $2$, which delivers the best performance found among several vector lengths tested, in contrast to the larger length of $8$ used in the synthetic case. These two differences with the synthetic case, support the hypothesis that the overall goodness of the power-law fit found by QC is better for the synthetic distributions than for the empirical ones. Finally, it is noteworthy that in these two datasets either QC or T, and in general the regression tree for all the datasets, obtain the best accuracy beating more flexible methods such as NN, considering that these methods provide a limited number of distinct output values.

\section{Discussion and future work}\label{disc}

In this section, we analyze the results presented in the last section. We start with a summary of the results, and then the discussion is mainly centered on the type of network patterns on which the performance of QuickCent is based. We end up with a series of ideas for future work and concluding remarks. 

\paragraph{Summary of results} The results presented in Section \ref{res} show that QuickCent can be a competent alternative to perform a regression on a power-law centrality variable. The method generates accurate and low variance estimates even when trained on a small -$10\%$- proportion of the dataset, comparable in precision to some more advanced machine learning algorithms. Its accuracy is available at a time cost that is significantly better than one of the machine learning methods tested, namely, the neural network. In this sense, QuickCent is an example of a simple heuristic based on exploiting regularities present in the data, which can be a competitive alternative to more computationally intensive methods. 

\paragraph{The patterns on which QuickCent relies} An interesting question is why our initial attempt to approximate an expensive centrality sensitive to size and density by a cheap density measure is successful, at least for the network cases tested. The same question framed in terms of the QuickCent method, would be why the two method assumptions, the power-law of harmonic centrality, and the strong correlation between harmonic centrality and in-degree, do hold for power-law, or more specifically for some preferential attachment networks. It was already mentioned in the text that while in-degree and PageRank centrality of a digraph obey a power-law with the same exponent \cite{litvak2007degree}, we have no knowledge of results describing the distribution of harmonic centrality on digraphs. A possible intuition for the scale-free behavior of harmonic centrality observed on PA networks (\figref{synth_distr}), may come from the motivation for the harmonic centrality given by Marchiori and Latora \cite{marchiori2000harmony,latora2001efficient}. The reciprocal shortest-path distances are used to informally define the efficiency of communication between nodes in a network. Therefore, it is reasonable to think that the scale-free degree distribution induces an analogous distribution in the efficiency to receive information sent.

\begin{figure}
\begin{subfigure}{.5\textwidth}
  \centering
  \includegraphics[width=\linewidth]{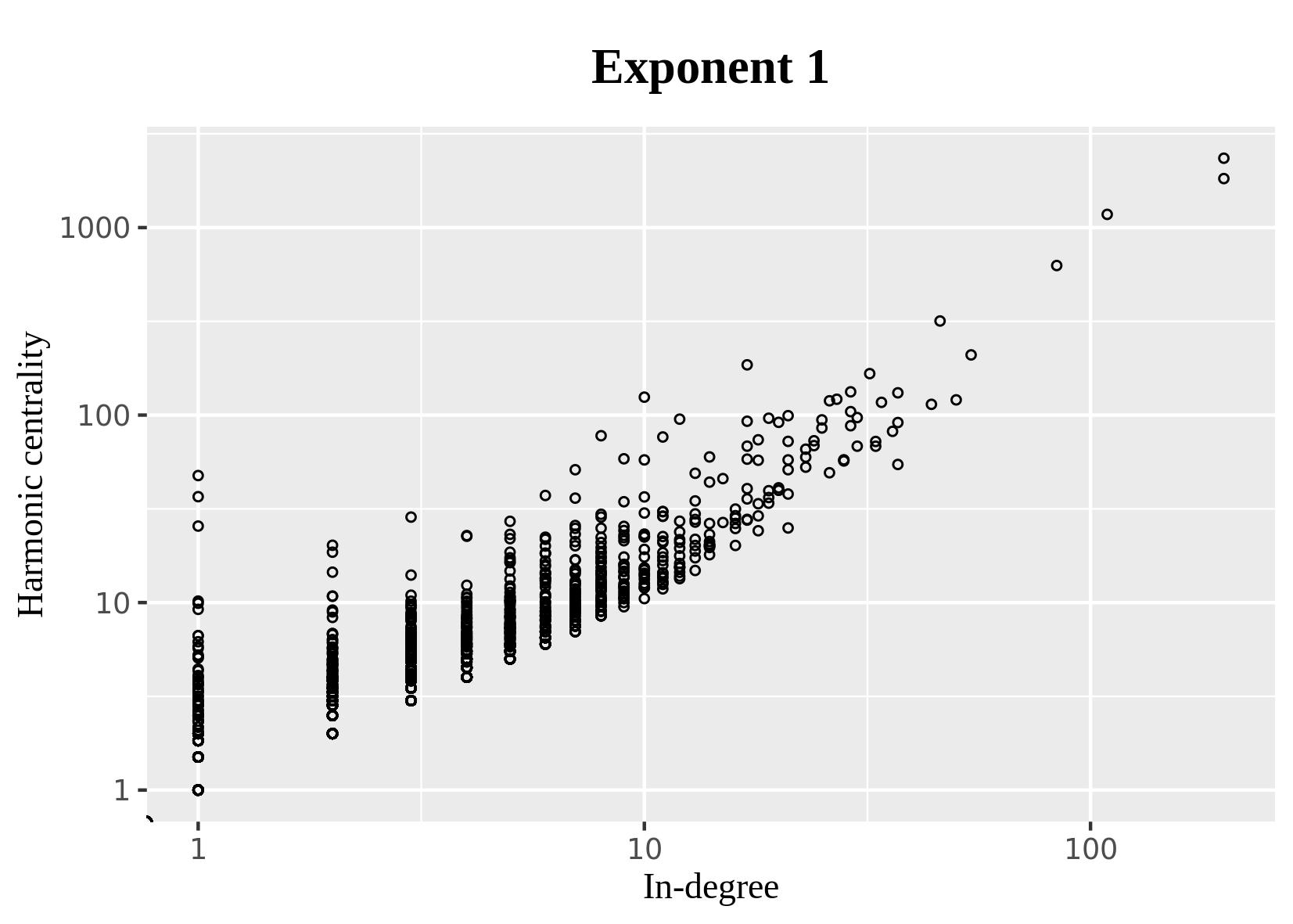}  
  \label{ind_harm_rel:sfig1}
\end{subfigure}
\begin{subfigure}{.5\textwidth}
  \centering
  \includegraphics[width=\linewidth]{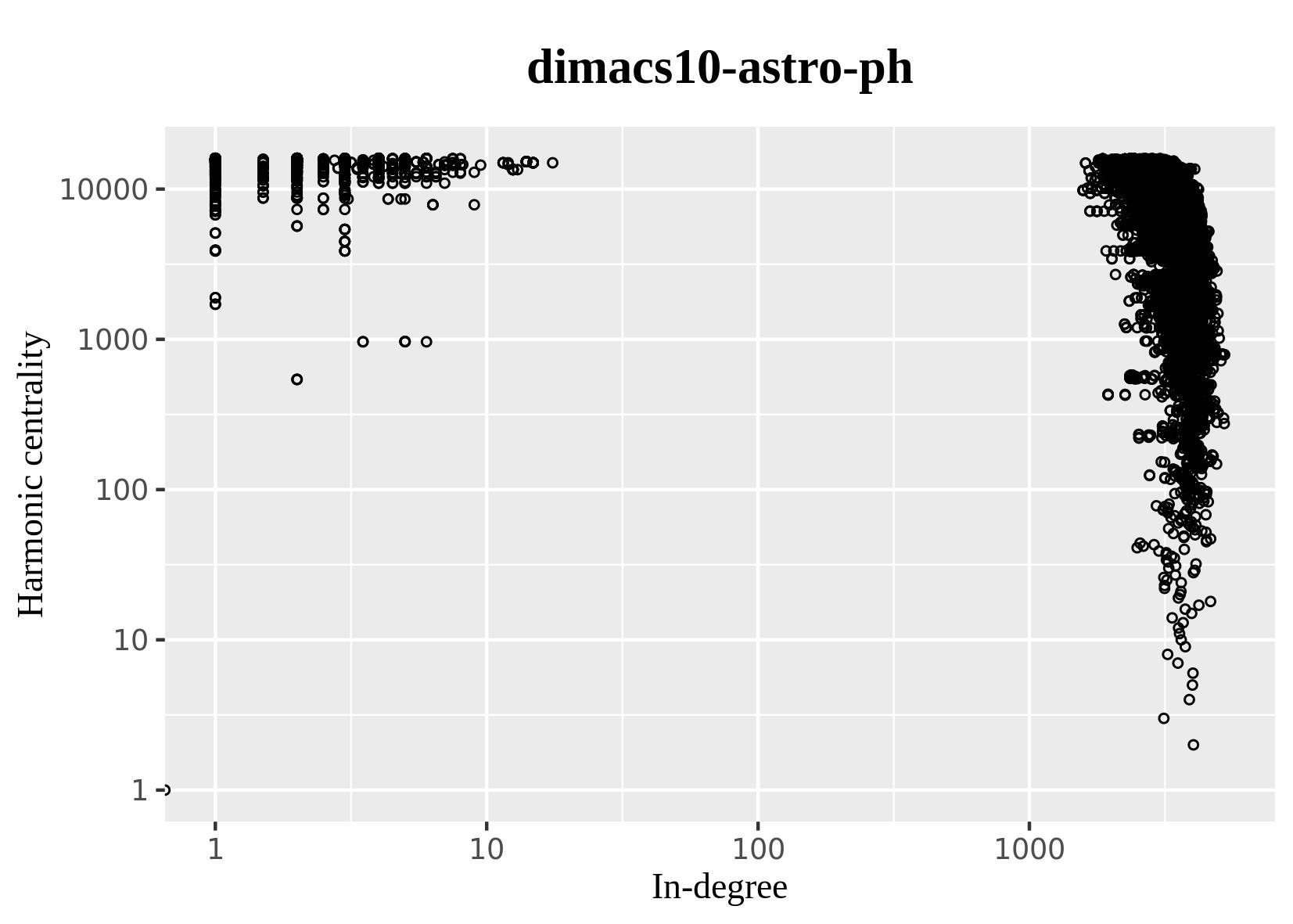}  
  \label{ind_harm_rel:sfig2}
\end{subfigure}

\caption{{\bf Scatterplot of in-degree versus harmonic centrality for a synthetic and an empirical network.} The plot on the left is obtained with a PA exponent of 1, and the plot on the right is that of the dimacs10-astro-ph dataset. The axes are in logarithmic (base 10) scale.}
\label{ind_harm_rel}
\end{figure}

The correlation and the monotonic relationship between harmonic centrality and in-degree, is strongly favored by the network generation mechanism. There is converging evidence showing that preferential attachment, which in its usual formulation requires global information about the current degree distribution, can be the outcome of link-creation processes guided by the local network structure, such as a random walk adding new links to neighbors of connected nodes, or in simple words, meeting friends of friends \cite{jackson2007meeting,vazquez2003growing}. The reason is that the mechanism of choosing a neighbor of a connected node makes those higher-degree nodes more likely to be chosen by the random walk, which in turn makes more paths lead to them. That is, the local density could indeed reflect the access to larger parts of the network. Of course, preferential attachment is not the only mechanism capable of producing scale-free networks \cite{krioukov2010hyperbolic,zhou2005maximal,doye2002network}, and the distinct generative mechanisms may engender or not, a stronger relationship between density and size in the resulting network. This insight may be the reason why the monotonic relationship between harmonic centrality and in-degree is more apparent in the preferential attachment model than in some of the empirical networks, as \figref{ind_harm_rel} shows.

The reasons explaining why and when fast and frugal heuristics work is an active research problem \cite{Brighton20151772,hogarth2005ignoring}. This question is not addressed in this paper, but some related results are mentioned next. It has been claimed and tested by simulations that the QuickEst estimation mechanism works well on power-law distributed data, but poorly on uniform data \cite{von2008mapping}. In the case of QuickCent, this dependence on the power-law distribution is reinforced by the parameter construction of this method (see Section \ref{QC2}). Another fact given by the definition of our clues that has been pointed out as a factor favoring simple strategies is the correlated information \cite{lee2012evaluating}, that is, information found early in the search is predictive of information found later. In the case of QuickCent, this simply means that since all clues are based on the in-degree, and this number can only belong to one of several disjoint real intervals, information from additional clues will not provide contradictory evidence once the heuristic has terminated its search. Other tasks where it is necessary to weigh the contribution of many possibly contradicting variables may present a more challenging context for these simple heuristics.

\paragraph{Future work} 

The insight described above about the monotonic map between the in-degree and the harmonic centrality on networks generated by preferential attachment nurtures the conjecture that QuickCent may be better suited to, for example, networks with an information component such as the Internet or citations, which can be well approximated by this growth mechanism \cite{barabasi1999emergence,jeong2003measuring,vazquez2003growing}, than to more pure social networks such as friendships \cite{jackson2007meeting,broido2019scale}. There is evidence that, if one assumes that some nodes to form links are found uniformly at random, while others are found by searching locally through the current structure of the network, it turns out that the more pure social networks appear to be governed largely through random meetings, while others like the World Wide Web and citation networks involve much more network-based
link formation \cite{jackson2007meeting}. Testing this hypothesis on a large corpus of diverse empirical network datasets is an interesting question for future work.

Applying QuickCent to other types of networks or centrality measures is not a direct task, since, depending on the type of network considered, degree and centrality may be strongly or weakly related. We plan to address these extensions in future work, where one possible line of research is to formulate the problem of finding the proportion quantiles as that of obtaining an optimal quantizer \cite{gray1998quantization}. There is some resemblance between our problem of finding the quantiles minimizing the error with respect to some distribution and that of finding the optimal thresholds of a piecewise constant function minimizing the distortion error of reproducing a continuous signal by a discrete set of points. On the other hand, QuickCent requires an explanatory variable that is correlated with the fitted variable to construct the clues. Future work should deal with extensions and flexibility of the clues employed, trying other clues or new ways to integrate different clues. The idea raised in our work of using a local density measure to approximate expensive size-based centrality indices could be generalized in order to be valid on more general networks, for example, by using a more general notion of local density than the in-degree, such as combined indicators of the spreading capability \cite{wang2018improved} or random-walk based indices of community density \cite{curado2022anew}.

\paragraph{Concluding remarks} The results of this paper are a proof of concept to illustrate the potential of using methods based on very simple heuristics to estimate some network centrality measures. Our results show that QuickCent is comparable in accuracy to the
best-competing methods tested, with the lowest error variance, even when trained on a small proportion of the dataset, and all this at intermediate time cost relative to the other methods using a naive implementation. We give some insight into how QuickCent exploits the fact that in some networks, such as those generated by preferential attachment, local density measures, such as the in-degree, can be a good proxy for the size of the network region to which a node has access, opening up the possibility of approximating centrality indices based on size such as the harmonic centrality. 

    



\bibliographystyle{acm} 
\bibliography{bibliografia}

\newpage
 \appendix

\section{Definitions and parameter specification of the power-law distribution}\label{interl_pl}

A random variable follows a power-law when its probability density function is given by an expression
of the form
$p(x)=K x^{-\alpha}$
where $\alpha$ is called the \textit{exponent} of the power-law, and $K$ is a \emph{normalization constant} 
depending on $\alpha$. Few real-world distributions follow a power-law on their whole range; many times the 
power-law behavior is observed only for higher values, in whose case it is said that the distribution ``has a power-law tail''. In analytic terms, since this density function diverges when $x$ goes to $0$, 
there has to be a lower limit $x_{\min}$ from which the power-law holds. 
That is, $x_{\min}$ is a value that satisfies
\begin{equation}
\int_{x_{\min}}^{\infty}K x^{-\alpha}dx=1. \label{xmin}
\end{equation} 
Moreover, from \eqref{xmin} it is simple to solve an expression for $K$ 
\begin{equation}
K=(\alpha -1)(x_{\min})^{\alpha -1}. \label{Kexp}
\end{equation}
The $\ell-$th moment of $x$ is given by

\begin{align*}
\langle x^{\ell} \rangle = \int_{x_{min}}^{\infty} y^{\ell}p(y)dy &= K\int_{x_{min}}^{\infty} y^{\ell -\alpha} dy \\ &= \frac{K}{\ell -\alpha + 1}\left(y^{\ell -\alpha + 1}\right)^{\infty}_{x_{min}}.
\end{align*}      

This expression is well defined for $\ell < \alpha - 1$, and in this case, by (\ref{Kexp}), its value is given by

\begin{equation*}
\langle x^{\ell} \rangle = \frac{\alpha -1}{\alpha - 1 -\ell}x_{min}^{\ell}.
\end{equation*} 

In particular, the second moment $\langle x^{2} \rangle$ diverges when the exponent $\alpha \leq 3$, and the first moment, the mean, diverges for exponents $\alpha \leq 2$. In our proposal, we need to estimate the $\alpha$ parameter of a power-law distribution.
A simple and reliable way to estimate $\alpha$ from a sample $\{x_i\}_{i=1}^m$ of $m$ observations from a power-law distribution is to employ the maximum likelihood estimator (MLE) which in this case is given by the formula \cite{newman2005power} 
\begin{equation}\label{mlealph}
\hat{\alpha} = 1+ m\left(\sum_{i=1}^m \ln\frac{x_i}{x_{\min}}\right)^{-1}.
\end{equation}
As it is apparent from the previous formula, there are at least two aspects that impact the quality of the estimate $\hat{\alpha}$: the number of observations
($m$ in the formula above), and, in case we do not know the exact value of $x_{\min}$, the
estimate $\hat{x}_{\min}$ that we use for it. 
It is clear how to improve in the first case: we just use more data points.
Estimating $x_{\min}$ is a bit more involved.
One possible way of computing an estimate for $x_{\min}$
is to visually inspect the log-log plot for the point where the CDF starts to look like a straight line.
However, this method is imprecise and highly sensitive to noise~\cite{clauset2009power}. A better method is the one proposed by Clauset et al. (2007) \cite{clauset2007frequency}, which selects the $\hat{x}_{\min}$ that makes the distributions of the empirical data and its fitted power-law model as similar as possible above $\hat{x}_{\min}$, that is, where the fit model is well defined. This similarity, or distance between two probability distributions, could be implemented through the Kolmogorov-Smirnov statistic (KS), whose expression, in this case, is the following
\begin{equation}\label{KS}
D(x_{\min})=\max_{x\geq x_{\min}} {|S(x) - P(x)|}
\end{equation}
where $S(x)$ is the CDF of the data for observations with a value greater than $x_{\min}$, and $P(x)$ is the CDF of the power-law model that best fits the data (for example, the MLE estimation (\ref{mlealph})) in the region $x\geq x_{\min}$. Finally, $\hat{x}_{\min}$ corresponds to the value $x_{\min}$ that minimizes $D(x_{\min})$. 

We estimated $x_{\min}$ with the bootstrap method implemented by the \textit{poweRlaw} R package \cite{powRlaw}, where several samples $x_{\min}$ are drawn and that minimizing $D(x_{\min})$ is selected. We noticed in our experiments that, with high frequency, this method selects $x_{\min}$ as a point with a high value, that is, a $x_{\min}$ value that discards a high portion of the distribution. In the case of empirical networks shown in Section \ref{real_case}, we have taken the heuristic approach of limiting the search space by an upper bound given by the percentile $20$ of the distribution of positive centrality values\footnote{This is the domain where the power-law fit can be computed.}, since we have seen for many datasets this is enough to span the point where the log-log plot of the complementary ECDF starts to behave like a straight line. Other authors giving implementations of this method have also noticed the difficulties when estimating $x_{\min}$\footnote{The commented code from https://github.com/keflavich/plfit says: \textit{...``The MLE for the power-law alpha is very easy to derive given knowledge of the lowest value at which a power law holds, but that point is difficult to derive and must be acquired iteratively.''}}. This method has a statistical consistency that has been proved only for some heavy-tailed models \cite{drees2020minimum}. There are alternative methods to optimize the KS statistic that perform, for example, a grid search over a predefined set of exponent values for each possible $x_{\min}$ that, however, have been claimed to present many drawbacks \cite{voitalov2019scale}.  

\section{Preferential Attachment growth}\label{lin-PA}

In this section, we review an early generalization of the preferential attachment hypothesis. The preferential attachment hypothesis states that the rate $\Pi(k)$ in which a $k-$degree node creates new links is an increasing linear function of $k$, and though Barab\'asi and Albert (1999) suggested this rate may follow a power-law, initial evidence from simulations showed that the scale-free property emerged only in the linear case \cite{barabasi1999emergence}. This intuition was confirmed by Krapivsky et al (2000) \cite{krapivsky2000connectivity}, where the following was found. Suppose the rate $\Pi(k)$ may have the following general form,
\begin{equation*}
\Pi(k_i)=\frac{k_i^\beta}{\sum_j k_j^\beta} = C(t) k_i^\beta.
\end{equation*}
Krapivsky et al (2000) prove that for $\beta=1$ this model reduces to the usual BA graph with exponent $3$. The original formulation by Krapivsky et al (2000) in fact is more general and allows tuning the exponent of the power-law degree distribution to every value larger than $2$. In the sublinear case, $\beta<1$, the degree distribution follows a stretched exponential, that is, the bias favoring more connected nodes is weaker, which produces an exponential cutoff that limits the size of hubs. On the other hand, for a superlinear attachment $\beta>1$, a single node becomes central and connects to nearly all other nodes. 

\section{Synthetic networks setting and assumptions verification}\label{assump_valid_p_val}

\begin{figure}
\begin{subfigure}{.5\textwidth}
  \centering
  \includegraphics[width=\linewidth]{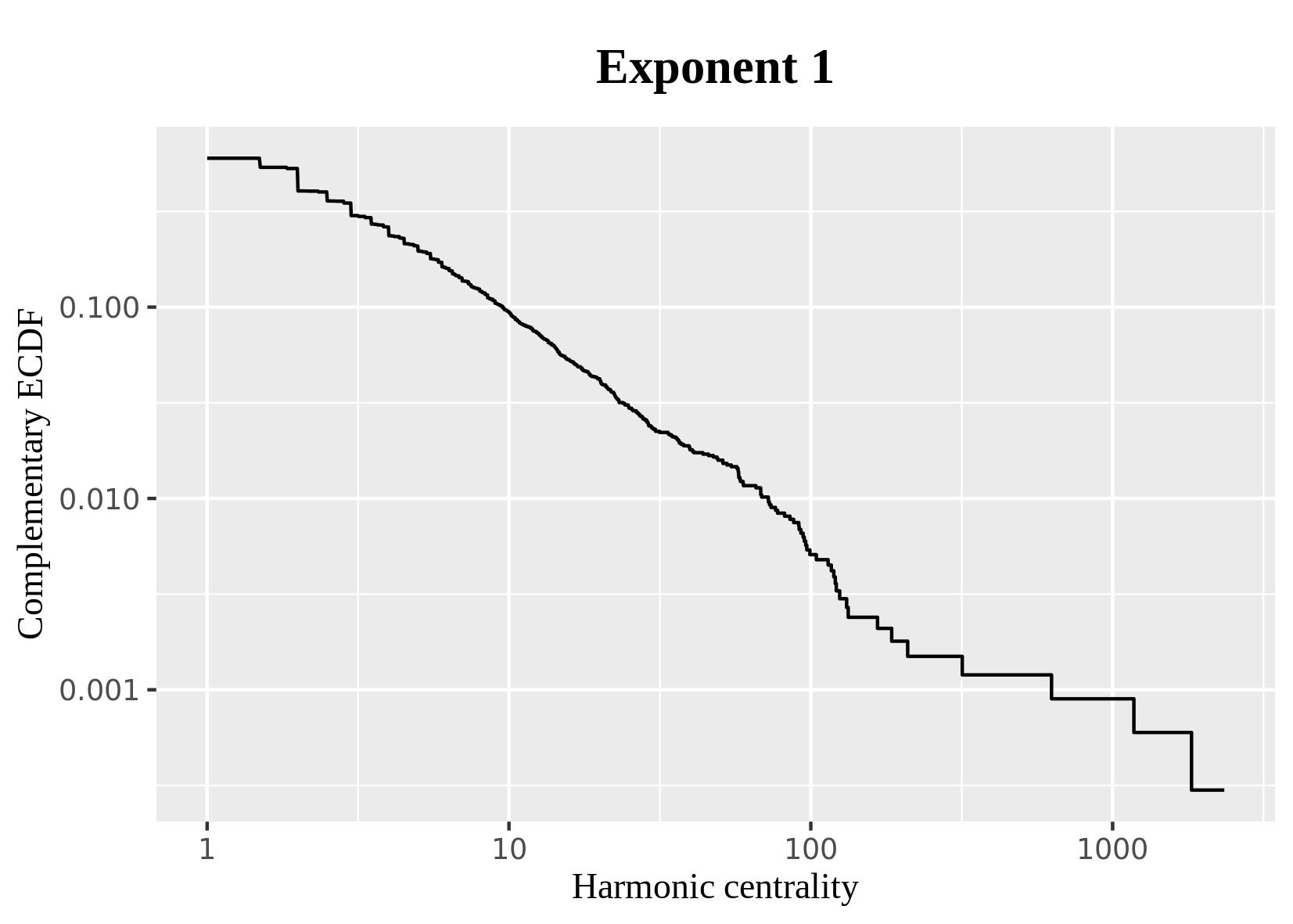}  
  \caption{Harmonic, PA exponent$=1$}
  \label{synth_distr:sfig1}
\end{subfigure}
\begin{subfigure}{.5\textwidth}
  \centering
  \includegraphics[width=\linewidth]{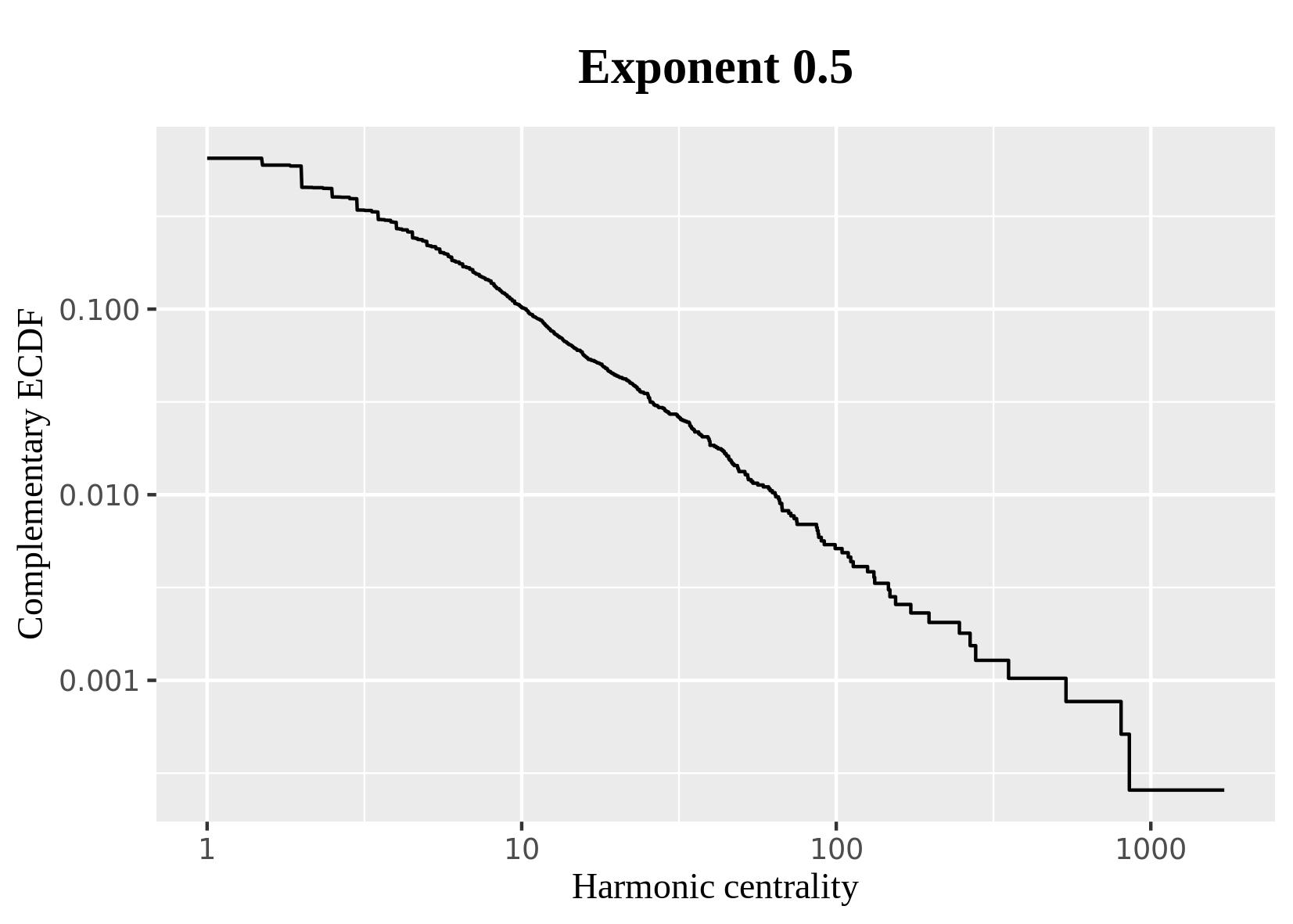}  
  \caption{Harmonic, PA exponent$=0.5$}
  \label{synth_distr:sfig2}
\end{subfigure}
\newline

\begin{subfigure}{.5\textwidth}
  \centering
  \includegraphics[width=\linewidth]{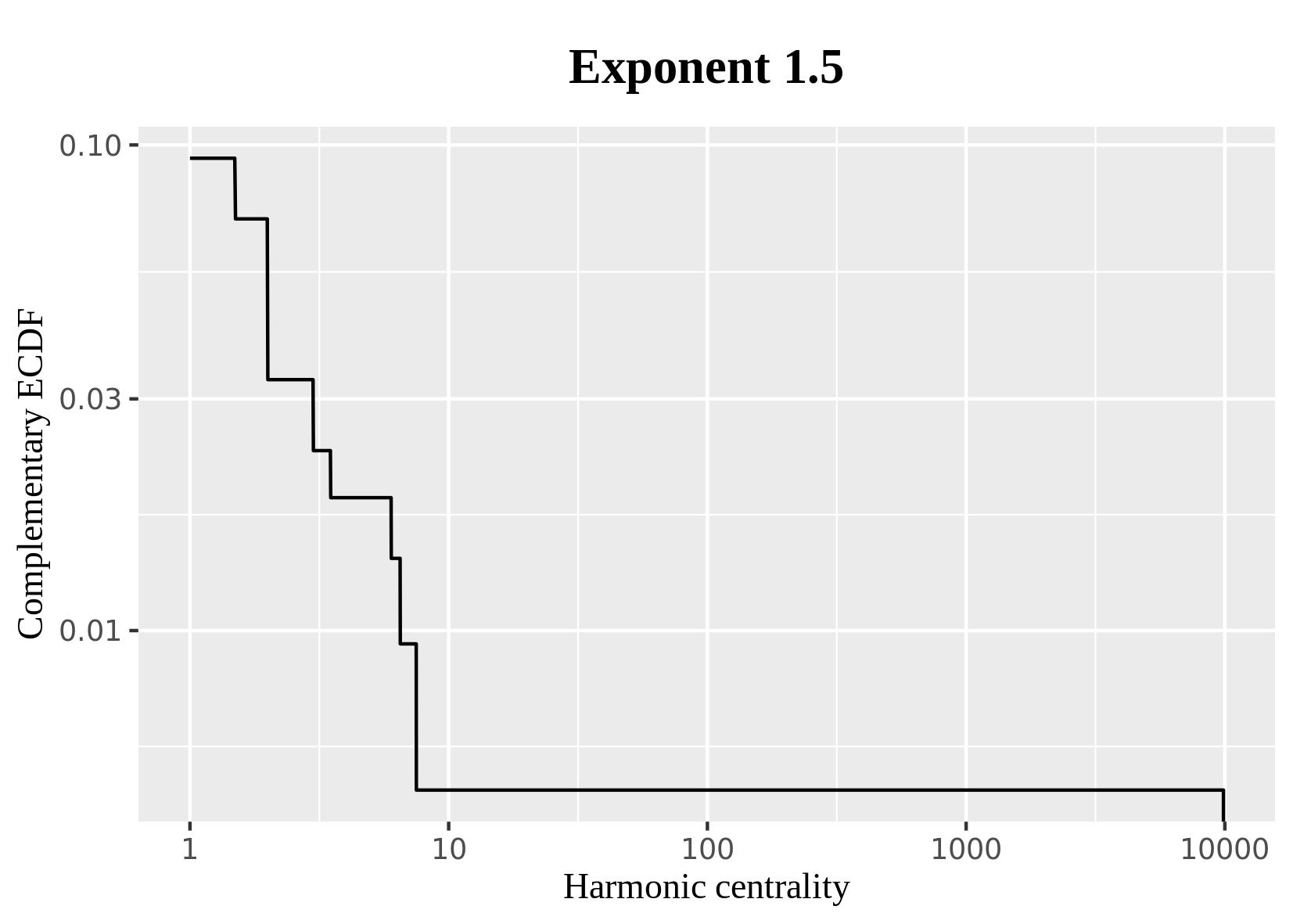}  
  \caption{Harmonic, PA exponent$=1.5$}
  \label{synth_distr:sfig3}
\end{subfigure}
\begin{subfigure}{.5\textwidth}
  \centering
  \includegraphics[width=\linewidth]{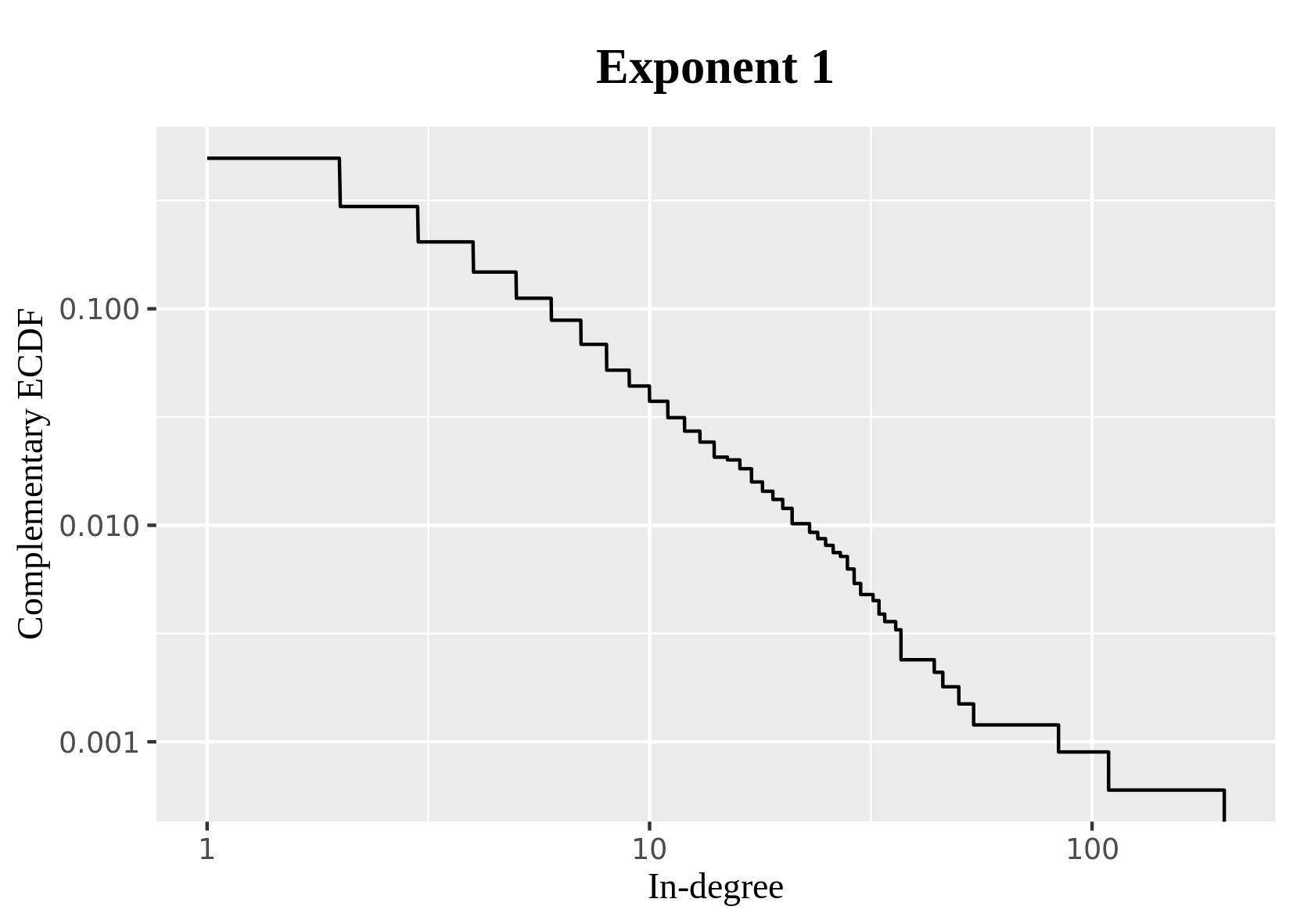}  
  \caption{In-degree, PA exponent$=1$}
  \label{synth_distr:sfig4}
\end{subfigure}
\newline

\begin{subfigure}{.5\textwidth}
  \centering
  \includegraphics[width=\linewidth]{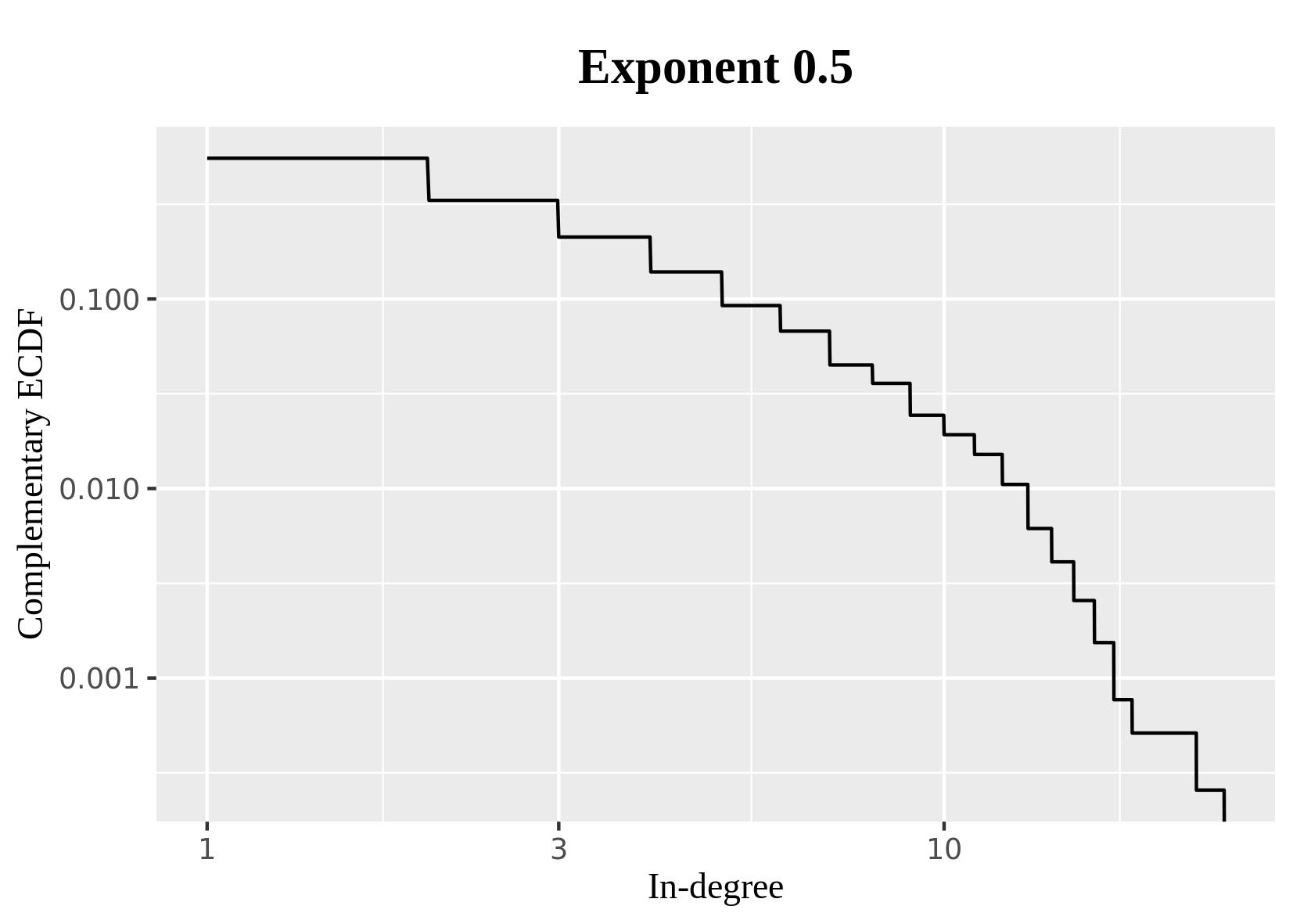}  
  \caption{In-degree, PA exponent$=0.5$}
  \label{synth_distr:sfig5}
\end{subfigure}
\begin{subfigure}{.5\textwidth}
  \centering
  \includegraphics[width=\linewidth]{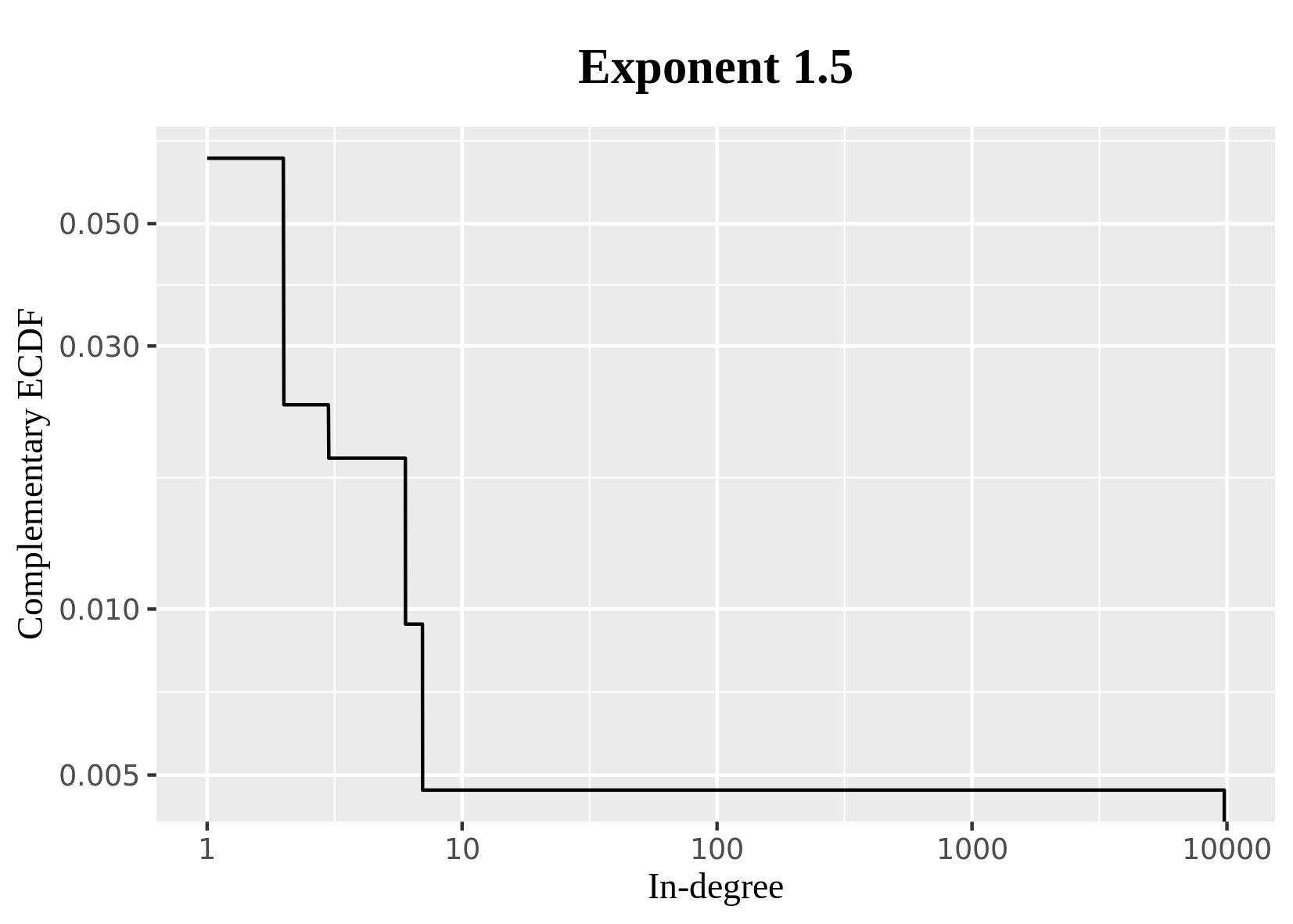}  
  \caption{In-degree, PA exponent$=1.5$}
  \label{synth_distr:sfig6}
\end{subfigure}
\caption{{\bf Complementary ECDF of harmonic centrality and in-degree for prototypical preferential attachment networks.} Each plot shows with logarithmic (base $10$) axes the complementary empirical cumulative density function of the harmonic centrality, and in-degree of randomly generated networks of each representative exponent of the PA model.}
\label{synth_distr}
\end{figure}

The synthetic network model we chose for our simulations is the preferential attachment (PA) growth reviewed in \ref{lin-PA}, including a small additive constant to the probability of acquiring links that allow isolated nodes to attract arcs\footnote{This setting corresponds to the default parameters of the method \textit{sample\_pa()} of package \textit{igraph} \cite{igraph} for the R language.} \cite{krapivsky2000connectivity,dorogovtsev2000structure,barabasi1999emergence}. This is a very simple model that reproduces the power-law degree distribution observed in real networks coming from a wide range of fields~\cite{barabasi1999emergence}, and provides a convenient setting for studying our heuristics on several prototypical distributions. With the goal of illustrating these distinct patterns, we have plotted in \figref{synth_distr} the harmonic centrality and in-degree distributions for randomly generated graphs of the model from \ref{lin-PA} for exponents $0.5,1,1.5$. 

In \figref{synth_distr:sfig4}, \figref{synth_distr:sfig5}, \figref{synth_distr:sfig6} we can visualize the respective patterns for the in-degree generated by PA described in \ref{lin-PA}. That is, a power-law distribution, a stretched exponential reflected in the scale on the horizontal axis, and the \textit{gelation} regime of a single gel connected to every node reflected in a heavy-tailed distribution even on a log-log plot. On the other hand, in \figref{synth_distr:sfig1}, \figref{synth_distr:sfig2}, \figref{synth_distr:sfig3} the respective patterns of the harmonic centrality are depicted, where the gelation pattern is also obtained for exponent $1.5$, and the scale-free behavior is observed for both exponents $1,0.5$, which is noteworthy. There are works in the literature showing that the in-degree and PageRank centrality of a digraph obey a power-law with the same exponent \cite{litvak2007degree}, but we have no knowledge of results describing the distribution of harmonic centrality on digraphs.

Before we can proceed with applying QuickCent, we need to ensure that its assumptions, 
namely the power-law distribution of the centrality and the monotonic relation between the degree and the centrality, are satisfied by the PA model.
We checked the assumption of the power-law distribution of the centrality by using the goodness-of-fit test proposed by Clauset et al.~\cite{clauset2009power}. This test is based on the KS distance $d$, see Equation \eqref{KS}, between the distribution of the empirical data and the fitted power-law model, and produces a p-value $p$, computed via a Monte Carlo procedure, which estimates the probability that the distance for any random sample is larger than $d$. Therefore, if $p$ is close to $1$, the fit is acceptable since the difference between the empirical data and the model fit can be explained by random fluctuations; otherwise, if $p$ is close to $0$, the model is not an appropriate fit to the data \cite{clauset2009power}.

The experiment executed to check the power-law assumption of harmonic centrality was to instantiate a PA network 
and compute the following magnitudes: an estimate $\hat{x}_{\min}$ of the distribution lower limit via minimization of Equation~(\ref{KS}), the respective exponent $\hat{\alpha}$ by using Equation (\ref{mlealph}), and the significance of this fit\footnote{This test requires the values tested to be strictly positive, so the zero harmonic values present in the sample were discarded for this test.}.
We also compute the value $\hat{\alpha}_{1}$ which is an estimation of the exponent $\alpha$ associated to value $x_{\min}=1$. The objective of computing $\hat{\alpha}_{1}$ was to verify the error obtained by working with a fixed value of $x_{\min}$ (equal to $1$ in this case), which may also tell about the goodness of the power-law fit. The described calculations were repeated over $1,000$ networks. Some descriptive statistics for the different magnitudes computed are shown in Table~\ref{verif_assump_exp1} (Q25 and Q75 are the $25$-th and $75$-th percentiles). Estimations were computed with the \textit{poweRlaw} R package \cite{powRlaw}.\par

\begin{table}
{\small
\begin{center}
  \begin{tabular}{c|cccccc}
&    $\hat{x}_{\min}$  &  $\hat{\alpha}$  &   \textbf{p-value}  & $\hat{\alpha}  -  \hat{\alpha}_{1}$& \textbf{Spearman} & \textbf{Spearman p-value}\\
\hline
\textbf{Q25}   & 6.333 & 2.135 & 0.333  & -0.027 & 0.923 & 0\\
\textbf{Median}  &   7.333 & 2.167 & 0.582  &   0.003 & 0.925 & 0\\
\textbf{Mean}   &    8.343 & 2.171 & 0.555   & 0.008 & 0.925 & 0\\
\textbf{Q75}   & 9.166 & 2.203 & 0.808 &  0.041 & 0.928 & 0\\
\end{tabular}
\end{center}
\caption{\textbf{Results of experiments testing QuickCent's assumptions for PA with exponent 1.} Fields correspond to: fitted lower limit and exponent of the power-law, KS-based p-value of this fit, difference between the fitted exponent and the exponent assuming $x_{\min}=1$, the Spearman correlation between the logarithms of centrality and in-degree, and its significance. The number of decimal places is truncated to three with respect to the source.}\label{verif_assump_exp1}}
\end{table}

\begin{table}
{\small
\begin{center}
  \begin{tabular}{c|cccccc}
&    $\hat{x}_{\min}$  &  $\hat{\alpha}$  &   \textbf{p-value}  & $\hat{\alpha}  -  \hat{\alpha}_{1}$& \textbf{Spearman} & \textbf{Sp. p-value}\\
\hline
\textbf{Q25} & 2.500 & 1.429 & 0.056 & -5.438 & 0.853 & 1.043e-82\\
\textbf{Median} & 3.500 & 1.521 & 0.274 & -4.518 & 0.882 & 1.645e-70\\
\textbf{Mean} & 93.351 & 1.670 & 0.346 & -4.610 & 0.876 & 5.647e-29\\
\textbf{Q75} & 4.208 & 1.774 & 0.620 & -3.650 & 0.903 & 3.977e-60\\
\end{tabular}
\end{center}
\caption{\textbf{Results of experiments testing QuickCent's assumptions for PA with exponent 1.5.} Fields correspond to: fitted lower limit and exponent of the power-law, KS-based p-value of this fit, difference between the fitted exponent and the exponent assuming $x_{\min}=1$, the Spearman correlation between the logarithms of centrality and in-degree, and its significance. The number of decimal places is truncated to three with respect to the source.}\label{verif_assump_exp15}}
\end{table}

\begin{table}
{\small
\begin{center}
  \begin{tabular}{c|cccccc}
&    $\hat{x}_{\min}$  &  $\hat{\alpha}$  &   \textbf{p-value}  & $\hat{\alpha}  -  \hat{\alpha}_{1}$& \textbf{Spearman} & \textbf{Sp. p-value}\\
\hline
\textbf{Q25} &  6.358 & 2.228 &  0.317& 0.188 & 0.923 & 0 \\
\textbf{Median} & 7.333 & 2.252 & 0.540 & 0.214 & 0.926 & 0 \\
\textbf{Mean} & 8.373 & 2.255 & 0.544 & 0.216 & 0.926 & 0 \\
\textbf{Q75} & 9.166 & 2.280 & 0.792 & 0.244 & 0.928 & 0 \\
\end{tabular}
\end{center}
\caption{\textbf{Results of experiments testing QuickCent's assumptions for PA with exponent 0.5.} Fields correspond to: fitted lower limit and exponent of the power-law, KS-based p-value of this fit, difference between the fitted exponent and the exponent assuming $x_{\min}=1$, the Spearman correlation between the logarithms of centrality and in-degree, and its significance. The number of decimal places is truncated to three with respect to the source.}\label{verif_assump_exp05}}
\end{table}

From this table, we can see that $75\%$ of the $1,000$ repetitions have a p-value greater than 0.333, which is greater than the rule-of-thumb threshold of 0.1 to rule out the power law~\cite{clauset2009power}. Hence, the fit given by  $\hat{x}_{\min}$ and $\hat{\alpha}$ is considered acceptable and the first assumption is satisfied. In the table, it can be appreciated that the estimated exponents $\hat{\alpha}$ are relatively stable for different values of $\hat{x}_{\min}$, and similar to $\hat{\alpha}_1$, in the sense that their difference is less than the critical amount of one integer which changes the behavior of the moments (see \ref{interl_pl}). Given this observation, in the experiments of Sections \ref{subsecC}, \ref{subsecD}, \ref{not_mon} where QuickCent is applied to estimate centrality on
synthetic PA networks, we consider a fixed value of $x_{\min}=1$. 
This further simplifies QuickCent training without incurring a big error in the parameter estimates. 

To check the assumption of monotonic relationship, in the same simulations performed before, we computed the Spearman correlation between the logarithm of the harmonic centrality and the logarithm of the in-degree\footnote{Here we also removed the zero harmonic values since $\log(0)=-\infty$.}. Some statistics for the distribution of correlation values, and the distribution of significant p-values can be reviewed in Table~\ref{verif_assump_exp1}. Here, we see that most of the correlations are almost perfect and significant. Since there is a linear relation between the logarithms of the two variables, we can conclude that there is a good exponential fit of the in-degree to the centrality, so the second assumption is satisfied. 

As an example for comparison, in Table \ref{verif_assump_exp15} the same statistics are reported for the PA model with exponent $1.5$. In this case, we see that while the p-values are not low in the central tendency, they are lower than those with exponent $1$ and there is an important proportion of them with values lower than the rule-of-thumb of 0.1. There is also an important difference between the general fitted $\alpha$ and that obtained with $x_{\min}=1$, which may be associated with the different slopes observed in the ECDF plot for lower and higher centrality values. This may also explain the greater variability observed in the values of $x_{\min}$, where the mean shows the presence of higher values. A similar pattern can be seen regarding the correlation between centrality and degree, in the sense of being lower than that of the previous case. The statistics for the exponent 0.5 of PA growth can be reviewed in Table \ref{verif_assump_exp05}, and are very similar to those obtained with exponent 1. However, in this case, there are higher differences between $\hat{\alpha}$, associated with $\hat{x}_{\min}$, and $\hat{\alpha}_{1}$, which in fact will have an effect on the robustness experiments reviewed in \ref{subsecB}.

\begin{figure}
\begin{subfigure}{.5\textwidth}
  \centering
  \includegraphics[width=\linewidth]{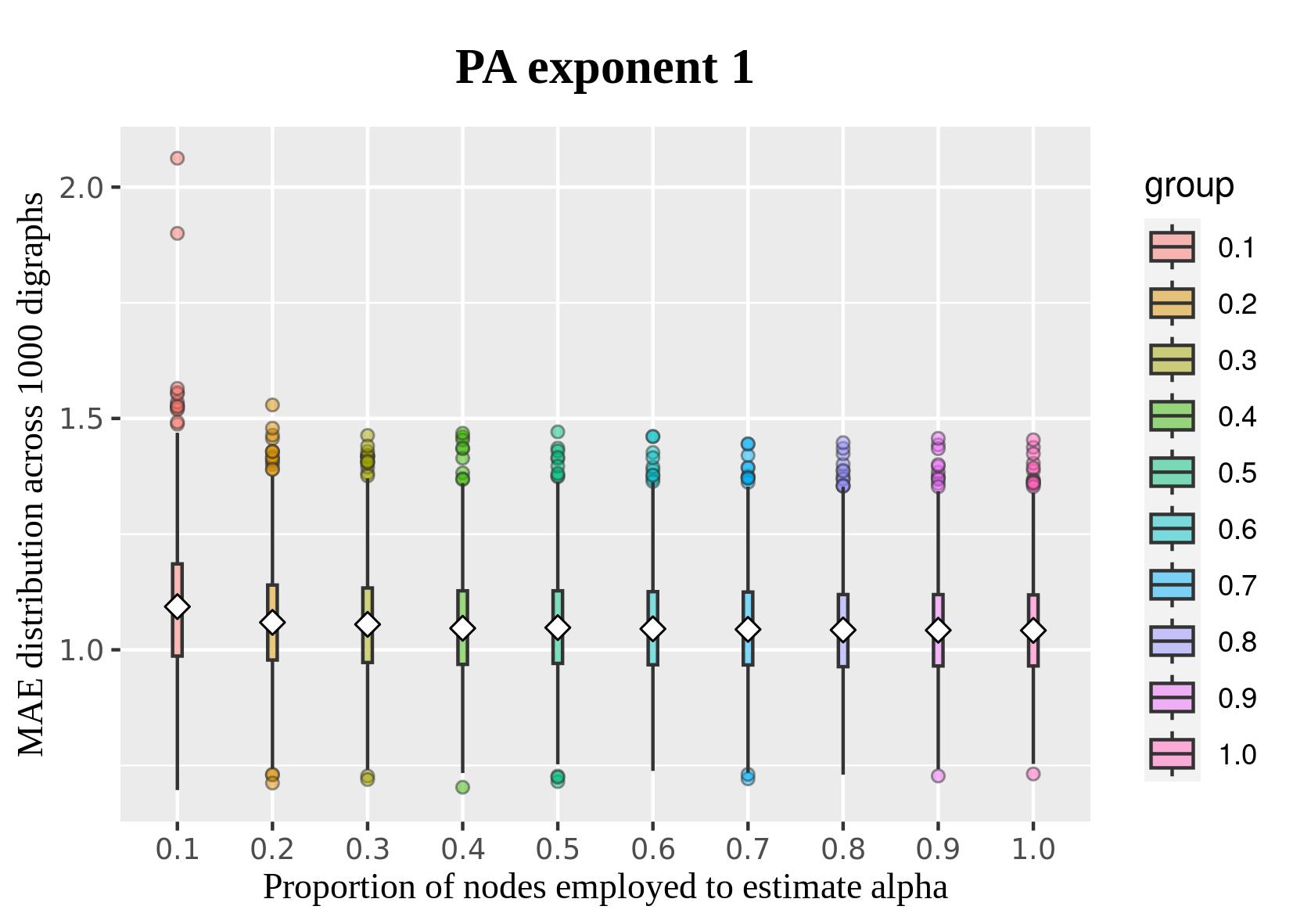}  
  \label{robust:sfig1}
\end{subfigure}
\begin{subfigure}{.5\textwidth}
  \centering
  \includegraphics[width=\linewidth]{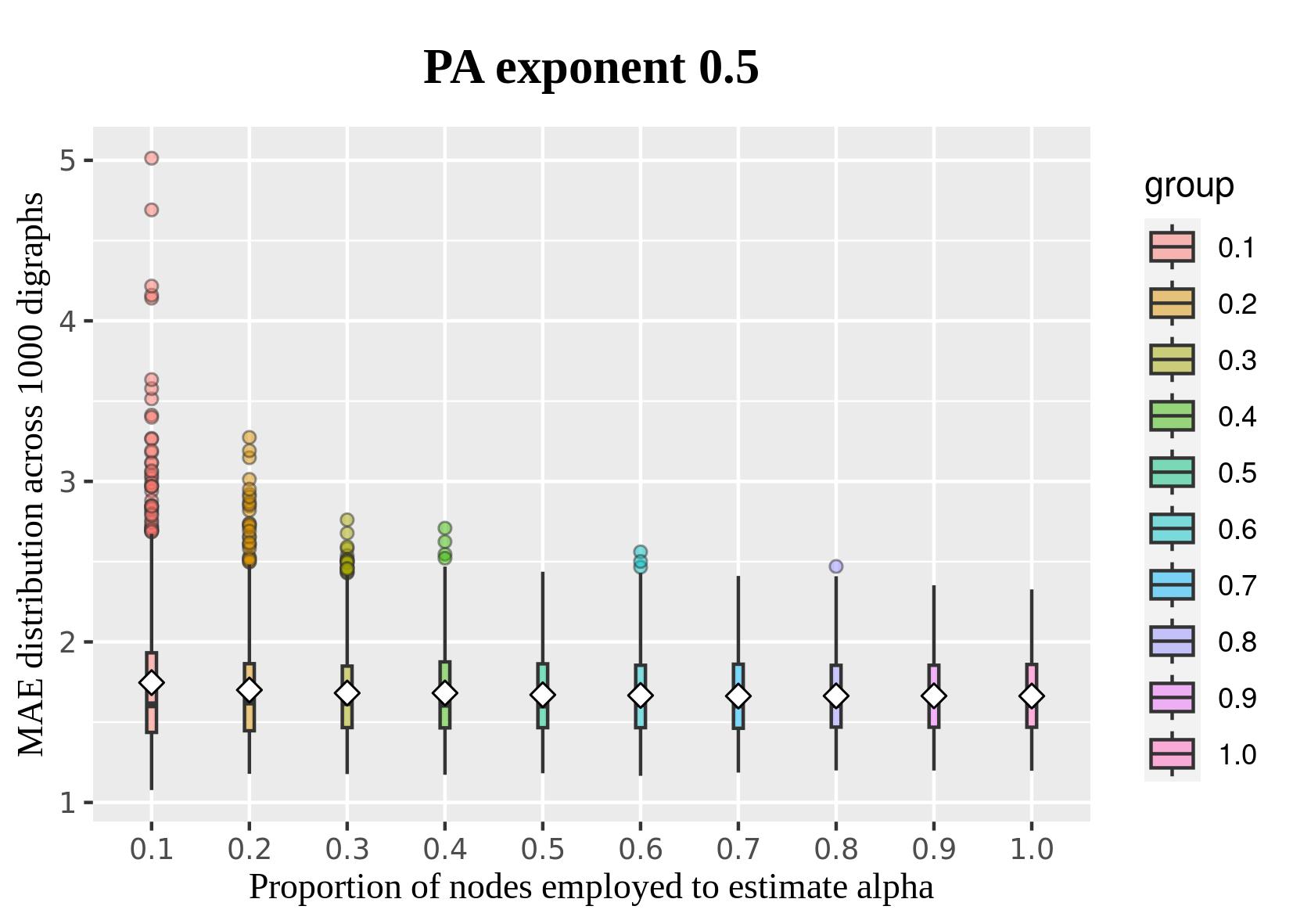}  
  \label{robust:sfig2}
\end{subfigure}
\newline

\begin{subfigure}{\textwidth}
  \centering
  \includegraphics[width=.5\linewidth]{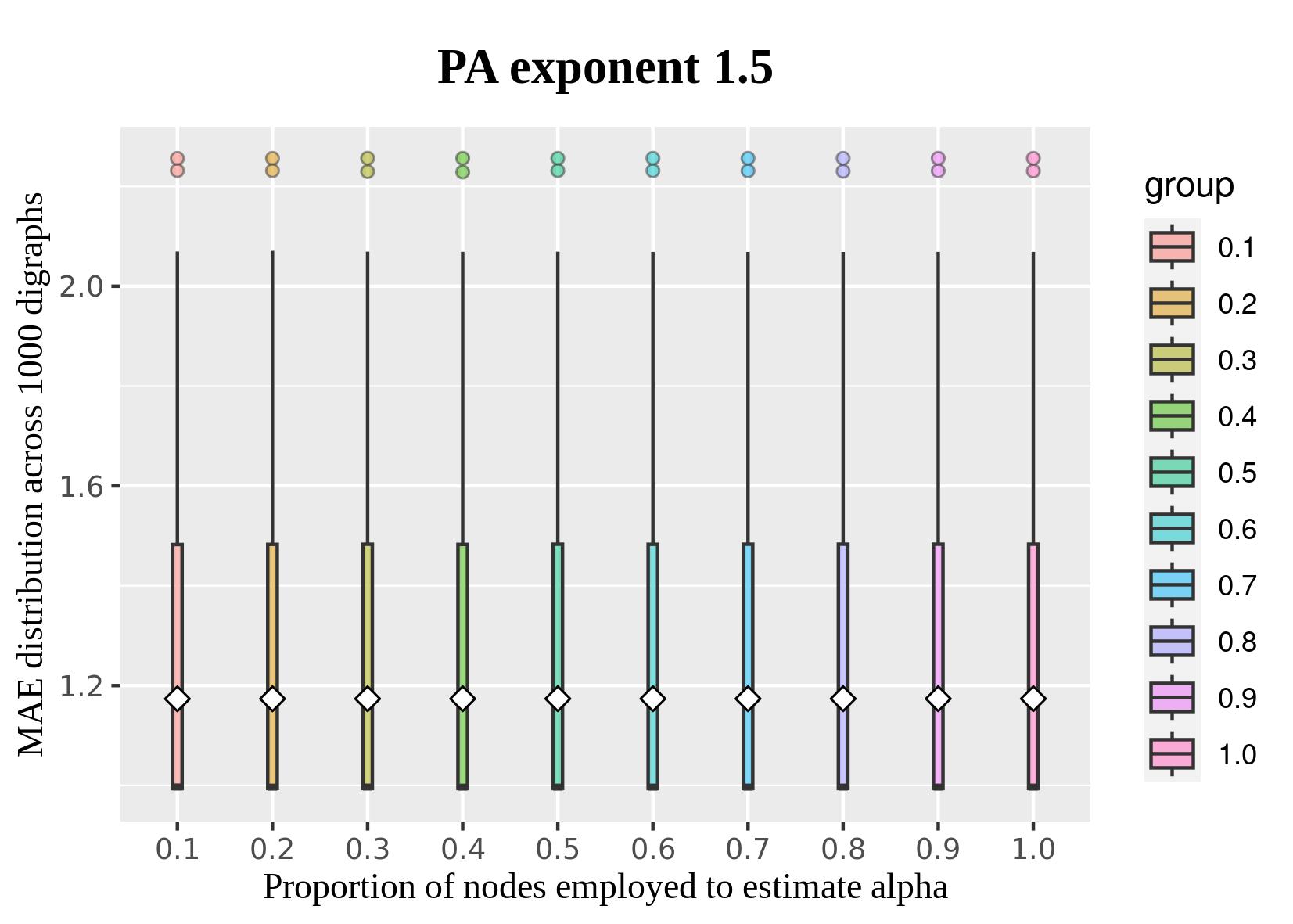}  
  \label{robust:sfig3}
\end{subfigure}

\caption{{\bf Robustness experiments for different exponents of PA digraph instances.} For each sample size there is a boxplot representing the MAE distribution. Each boxplot goes from the $25-$th percentile to the $75-$th percentile, with a length known as the \textit{inter-quartile range} (IQR). The line inside the box indicates the median, and the rhombus indicates the mean. 
The whiskers start from the edge of the box and extend to the furthest point within $1.5$ times the IQR. Any data point beyond the whisker ends is considered an outlier, and it is drawn as a dot.}
\label{robust}
\end{figure}

\section{Robustness of estimates}\label{subsecB}
We present here the results of the following robustness experiment. 
We consider an instance of a PA graph of $10,000$ nodes 
and created distinct {QuickCent} models with varying information amounts from the network. This information is extracted from uniform node samples and the formula from (\ref{mlealph}) to estimate the $\alpha$ exponent, 
with sample sizes (value $m$ in the formula) ranging from $10\%$ to $100\%$ of the total vertex set. Note that in this formula, there will be a number of nodes discarded from the sample depending on the value of $x_{\min}$. From the experiments shown in \ref{assump_valid_p_val}, we chose to use $x_{\min}=1$ for the experiments in Sections \ref{subsecC}, \ref{subsecD}, \ref{not_mon} using synthetic PA networks, since there is not a big penalty for doing so, at least with the exponents 0.5 and 1 with a centrality distribution nearer to a power-law. Then, for each model, we compute a centrality estimate on every vertex and the difference between this estimate and the real value. 
Finally, we computed from these differences the MAE on the entire digraph and examined the MAE distribution across $1,000$ different instances of PA digraphs. The idea is to assess how the availability of information from the network enhances or deteriorates the quality of the estimates.

Results are shown in Figure~\ref{robust}. The plots show a performance that is quite accurate and robust to the training size in the executed experiments. It is accurate because the central tendency and most of the distribution contained in the IQR of the MAE errors (see the caption in Figure~\ref{robust}), does not exceed the two units of difference, taking the average over all network nodes, from the real centrality value, at least for the exponents 1 and 1.5 of PA. Bigger MAE errors appear for the exponent $0.5$, however, the most extreme outliers even in this case do not surpass the five units of error. This case reveals that the in-degree does not need to have a scale-free distribution to be functional for QuickCent estimates. Actually, an implicit requirement for the correct working of QuickCent is that this distribution has an injective map from the proportion vector used to the degree quantile values, allowing to distinguish the distinct centrality intervals. For this reason, the length of the proportion vector in these experiments was fixed at 8, since while the accuracy increases with this parameter, for lengthier proportion vectors repeated degree quantile values appear. Thus, the bigger MAE errors for the PA exponent 0.5 come from the exponents $\hat{\alpha}_{1}$ used in the simulations, which are lower than those $\hat{\alpha}$ associated with the fitted $\hat{x}_{\min}$, see Table \ref{verif_assump_exp05}. Since they are lower, they predict centrality values that are higher than the real ones, particularly among the highest values.  

On the other hand, the performance of QuickCent is robust to the training size in the executed experiments. It is not highly deteriorated for the smaller training sizes of 10 $\%$. The case of exponent 1.5 is special since the performance seems to be quite independent of the training size. This has to do with the fact that there is a structural bias given by the maladjustment between the empirical centrality distribution and the assumed power-law model. However, given that in these simulations ${x}_{\min}=1$, the fitted power-law covers the bulk of the distribution composed of low centrality nodes, making the MAE error to be pretty accurate since it is an average over the network. This shows how this measure of the error is favored by low errors committed in a relevant mass of the distribution, as in the power-law case for PA exponents 0.5 and 1.



The feature of making estimates with relatively stable errors could be an advantage in relation to other regression methods, which can suffer potentially a greater impact from scarce data. This issue is addressed in Section \ref{subsecC}.

\section{Assumption verification experiments on randomized networks}\label{assump-rand}

Table \ref{verif_assump_exp1_rand} shows some summary statistics of several variables testing the assumptions to apply QuickCent, on a set of 1000 randomly generated PA networks (exponent 1), subject to degree-preserving randomization of arcs. The p-values show that, even by fixing $x_{\min}=1$, which is the approximation used by QuickCent, the harmonic centrality distribution is reasonably approximated by a power-law. The fitted parameters of the power-law distribution of randomized networks are in fact similar to those shown in Table \ref{verif_assump_exp1} corresponding to PA networks without randomization. The two columns with correlation values clearly show the impact of randomization over the map between in-degree and harmonic centrality.   

\begin{table}
{\small
\begin{center}
  \begin{tabular}{c|cccccc}
& $\hat{x}_{\min}$  &  $\hat{\alpha}$  & $\hat{\alpha}_{1}$ & \textbf{p-value} & \textbf{corr\_1} & \textbf{corr\_2}\\
\hline
\textbf{Q25}   & 2.775&  1.858&  1.859&    0.020&      0.918& 0.777\\
\textbf{Median}  &   3.833&  1.960&  1.894&      0.120&   0.927&   0.797\\
\textbf{Mean}   &    7.636&  2.183& 1.896&   0.269&       0.926&  0.796\\
\textbf{Q75}   & 7.500& 2.203& 1.932&  0.470& 0.935&       0.817\\
\end{tabular}
\end{center}
\caption{\textbf{Results of experiments testing the assumptions of QuickCent on 1000 PA networks (exponent 1) and degree-preserving randomization.} Fields correspond to: fitted lower limit ($\hat{x}_{\min}$) and power-law exponent($\hat{\alpha}$) of the harmonic centrality, power-law exponent ($\hat{\alpha}_{1}$) obtained by fixing $x_{\min}=1$ (the approximation used by QuickCent), the KS-based p-value of this last fit, the Spearman correlation between the positive values of harmonic centrality and in-degree, before (corr\_1) and after randomization (corr\_2). The number of decimal places is truncated to three with respect to the source. All fields except corr\_1 are computed on the network obtained after randomization by swapping the start and end of 10000 randomly selected arc pairs that do not change the in-degree network sequence.}
\label{verif_assump_exp1_rand}}
\end{table}

\section{Sensitivity to connection probability and assumptions verification on Erd{\"o}s-R{\'e}nyi digraphs and control networks}\label{assump-ER}

\begin{figure}
\begin{subfigure}{.45\textwidth}
  \centering
  \includegraphics[width=\linewidth]{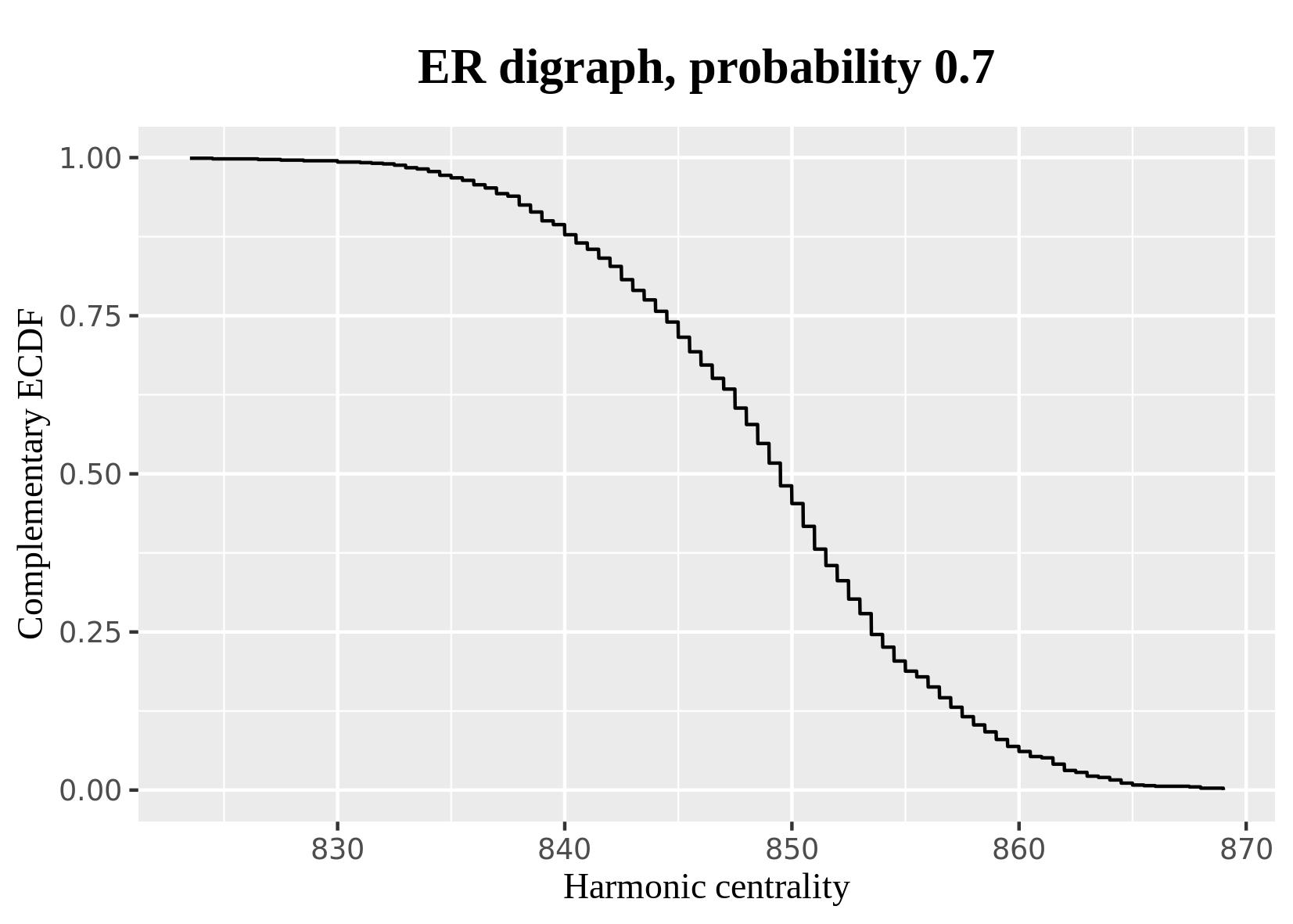}  
  \label{ER_patt:sfig1}
\end{subfigure}
\begin{subfigure}{.45\textwidth}
  \centering
  \includegraphics[width=\linewidth]{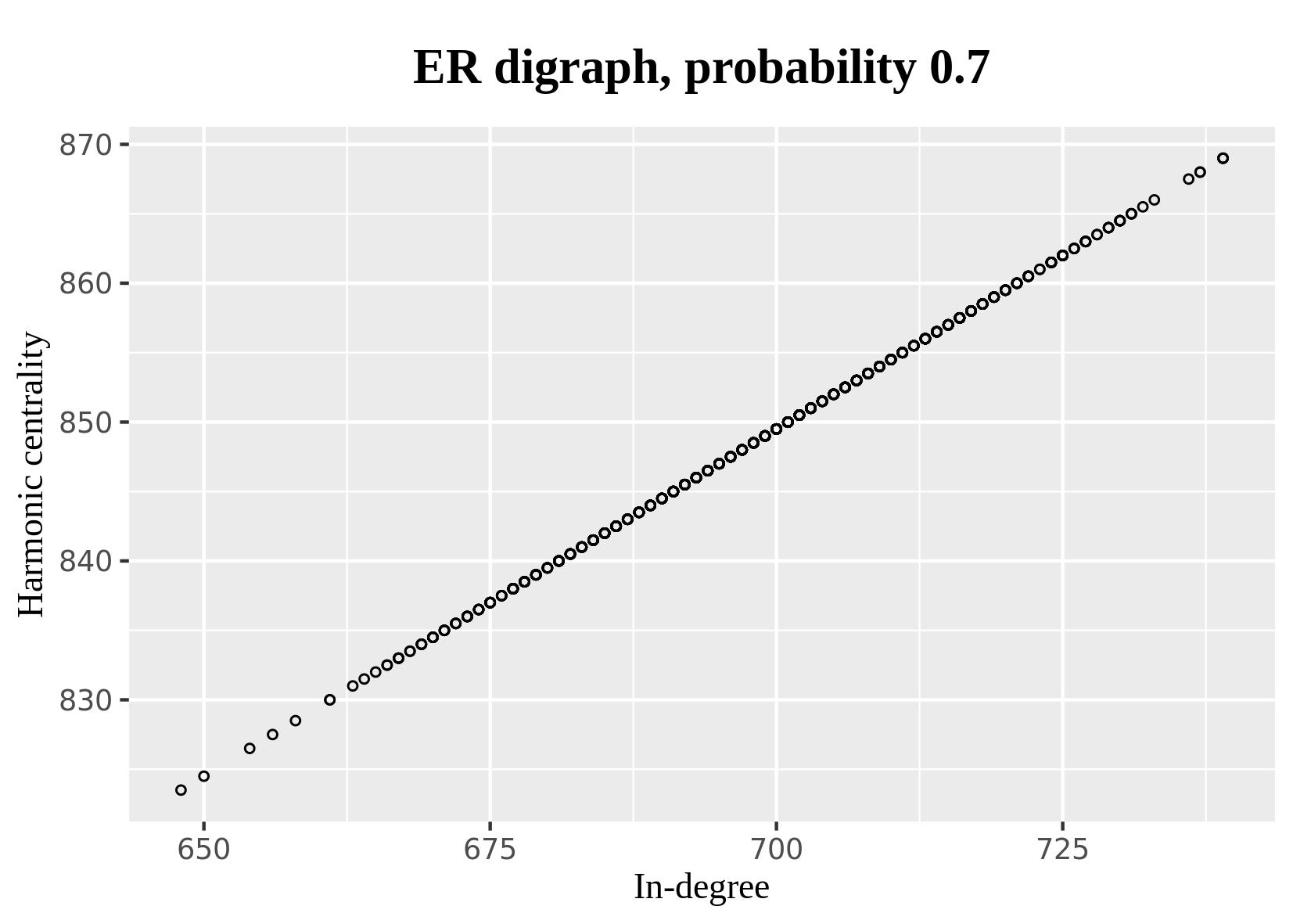}  
  \label{ER_patt:sfig2}
\end{subfigure}

\begin{subfigure}{.45\textwidth}
  \centering
  \includegraphics[width=\linewidth]{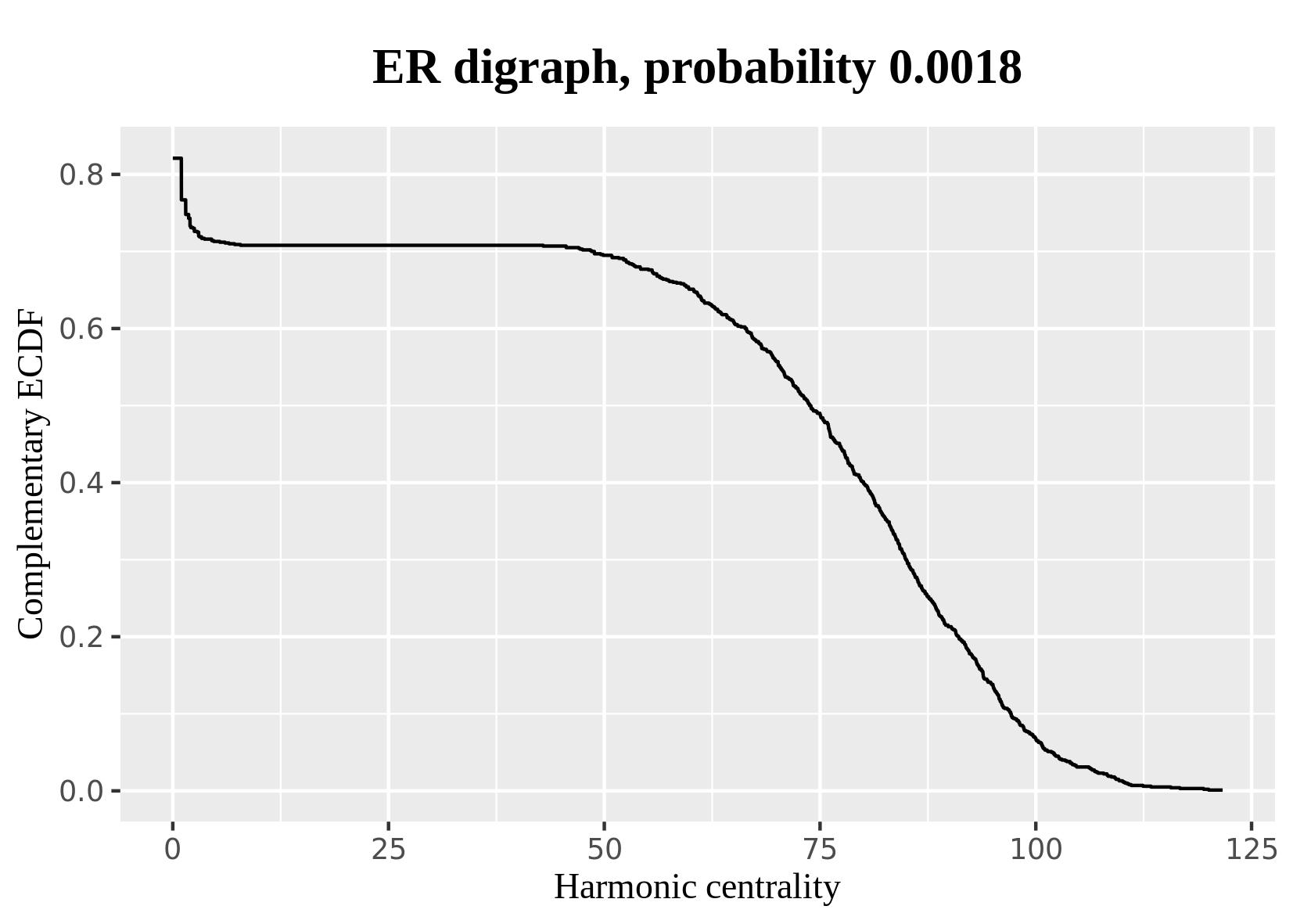}  
  \label{ER_patt:sfig3}
\end{subfigure}
\begin{subfigure}{.45\textwidth}
  \centering
  \includegraphics[width=\linewidth]{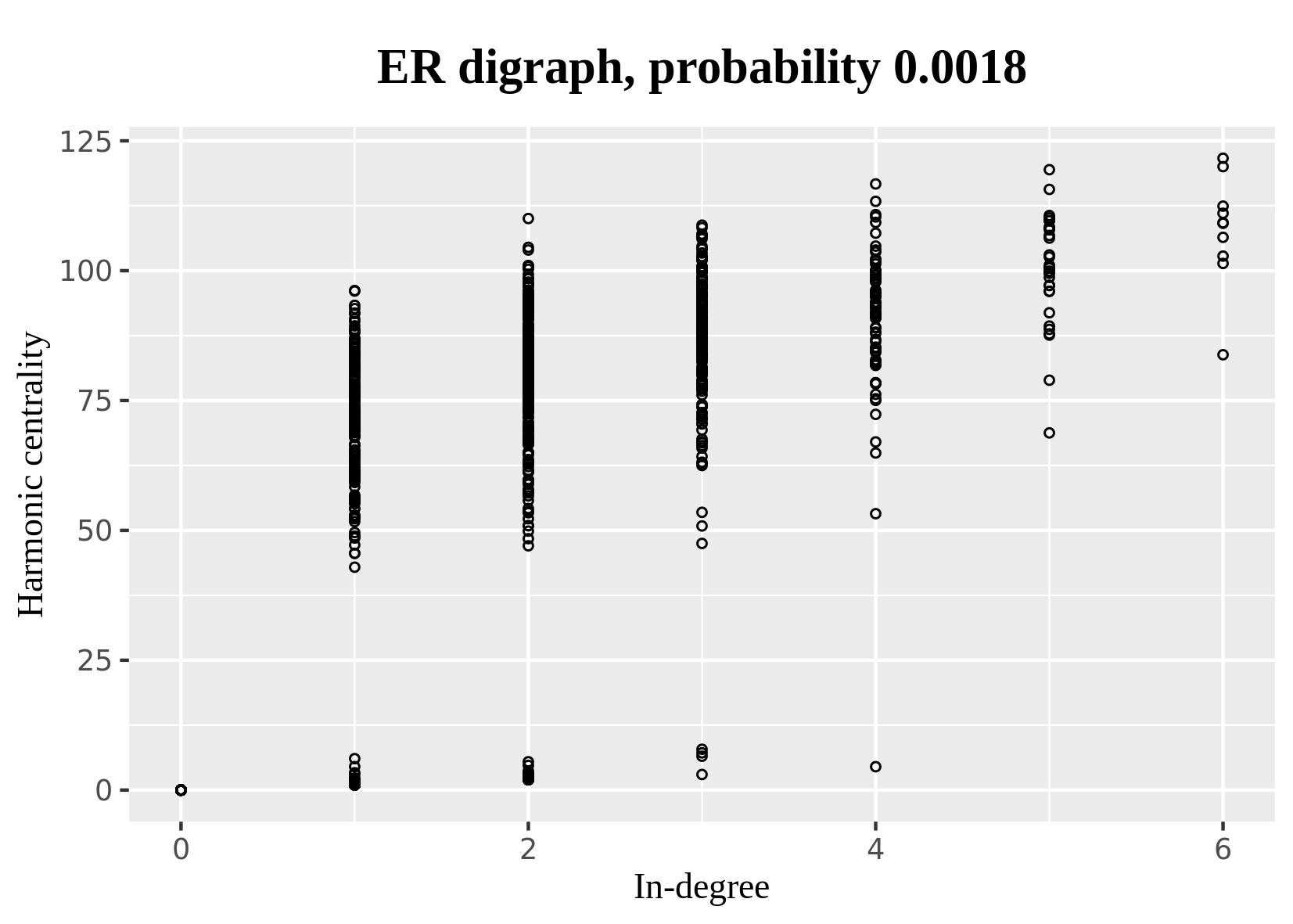}  
  \label{ER_patt:sfig4}
\end{subfigure}

\begin{subfigure}{.45\textwidth}
  \centering
  \includegraphics[width=\linewidth]{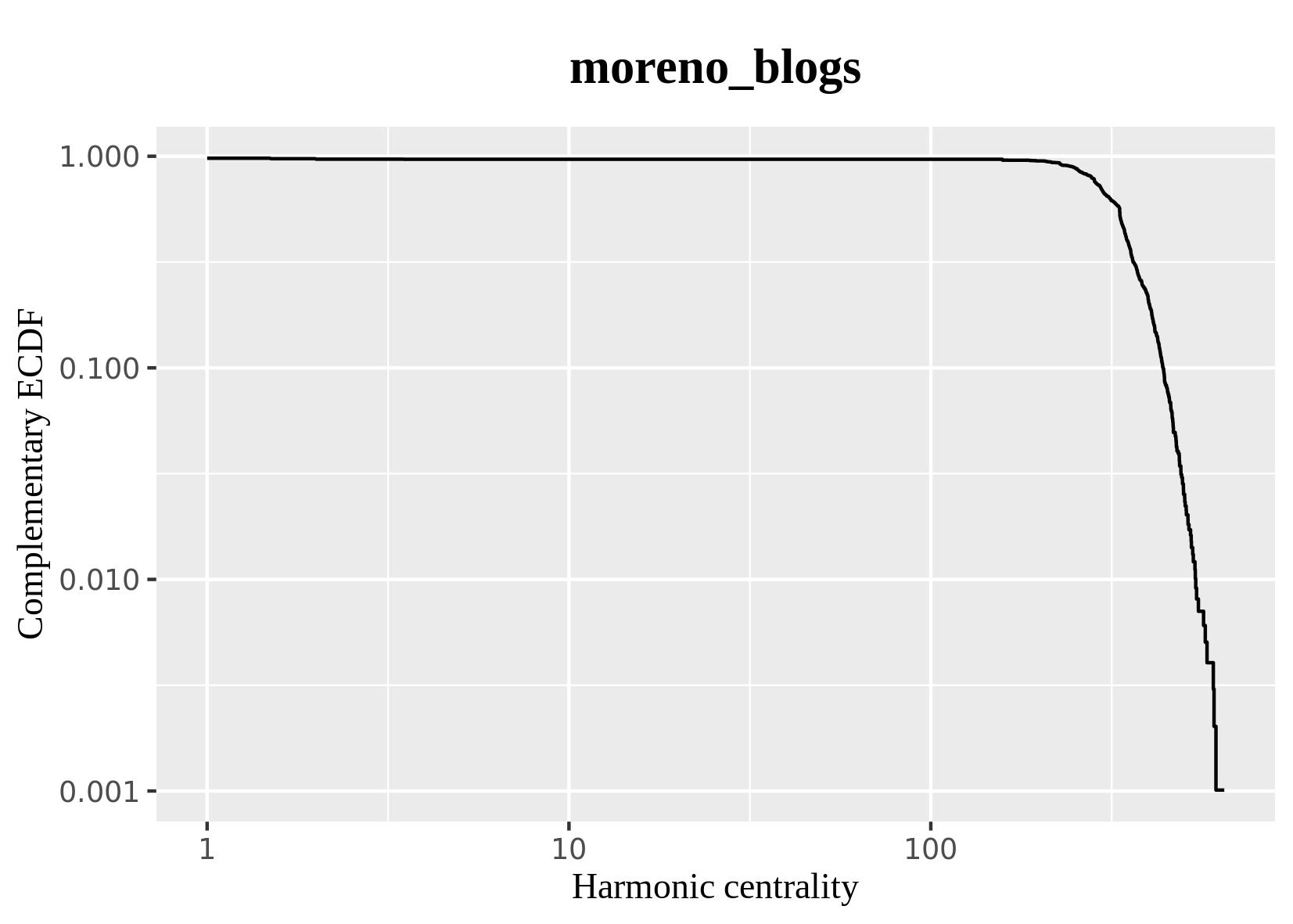}  
  \label{blogs_harm}
\end{subfigure}
\begin{subfigure}{.45\textwidth}
  \centering
  \includegraphics[width=\linewidth]{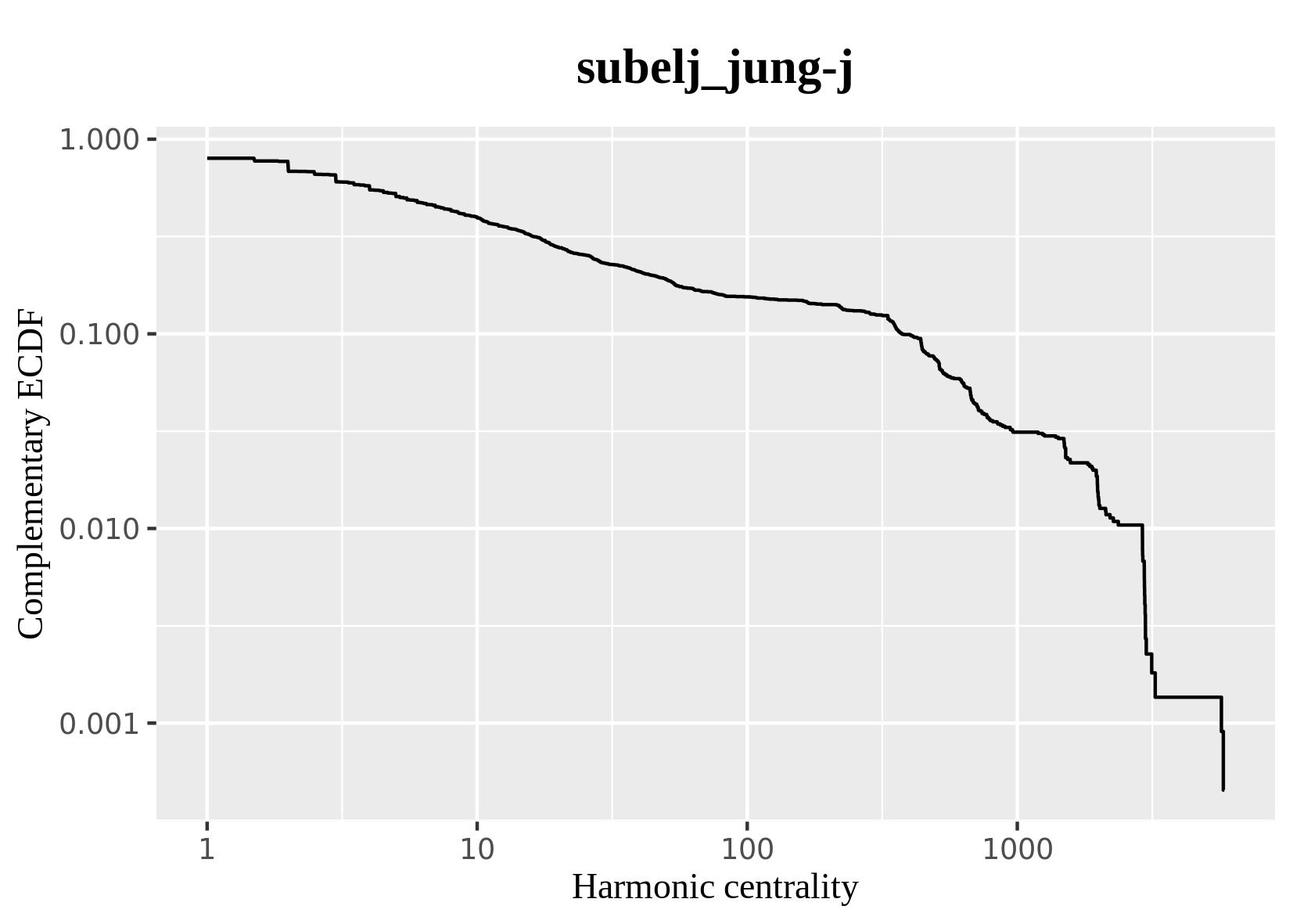}  
  \label{subelj_harm}
\end{subfigure}

\caption{{\bf Distribution of harmonic centrality (left) and map from in-degree to harmonic (right) on ER, and harmonic centrality in control digraphs.} The plots are generated from random instances of ER digraphs of size $N=1000$. The top plots represent networks where the probability of connection $p=0.7$ induces a mean in-degree far greater than $1$, the critical value for the existence of a unique giant strong component. This is reflected in a unimodal distribution of harmonic centrality, and a perfect correlation between in-degree and harmonic centrality. The middle plots represent a connection probability $p=0.0018$ near the transition point, where the correlation is now $0.799$. The bottom plots correspond to the harmonic centrality distribution of the two empirical networks used as controls.} 
\label{ER_patt}
\end{figure}

Erd{\"o}s-R{\'e}nyi (ER) digraphs seem to have a unimodal distribution of harmonic centrality, and an almost perfect correlation between this coefficient and the in-degree, at least for those values of the connection probability $p$ where the mean in-degree $\sim p\cdot N>1$, corresponding to the values where it is very likely that there is a unique large strongly connected component \cite{karp1990transitive}, see \figref{ER_patt}. For values of $p$ near the transition point, sets of reachable (or co-reachable) nodes from distinct nodes tend to be smaller, which produces greater variability in the relation mapping in-degree values to harmonic centrality, which lowers the strength of the correlation, see \figref{ER_patt}. Table \ref{verif_assump_ER_control} shows the fitted lower limits and p-values of the power-law fit either on the ER or control digraphs. Each one of the $1000$ iterations of fitting over each network is associated with a distinct random seed, impacting for example the bootstrap computations of p-values. In spite of this, for the control networks, the parameters are constant since we are working with the same network, in contrast to the ER digraphs which are newly instanced for each iteration.

\begin{table}
{\small
\begin{center}
  \begin{tabular}{c|cccccccc}
& $\hat{x}_{\min}^{\text{mb}}$  & $\textbf{p}^{\text{mb}}$ & $\hat{x}_{\min}^{\text{sj}}$ & $\textbf{p}^{\text{sj}}$&$\hat{x}_{\min}^{\text{ERmb}}$  &  $\textbf{p}^{\text{ERmb}}$ & $\hat{x}_{\min}^{\text{ERsj}}$ & $\textbf{p}^{\text{ERsj}}$\\
\hline
\textbf{Q25} &  277.416  &  1 &1&1&381.333& 0.010&1060.000&0.020\\
\textbf{Median} &  277.416  &  1&1&1&381.916&0.180&1060.667&0.380\\
\textbf{Mean} &  277.416  &  1&1&1&381.915&0.347&1060.621&0.431\\
\textbf{Q75}  &  277.416  & 1&1&1&382.500&0.650&1061.167&0.820\\
\end{tabular}
\caption{\textbf{Results of experiments testing the assumptions of QuickCent on ER and control digraphs.} Fields correspond to: fitted lower limit ($\hat{x}_{\min}$) and p-value ($\textbf{p}$) of this power-law fit to the harmonic centrality distribution of each network. The superscript of each parameter denotes the respective network, `mb' for moreno\_blogs, `sj' for subelj\_jung-j, `ERmb' for the ER digraph created with the parameters of moreno\_blogs, and analogously for `ERsj'. The number of decimal places is truncated to three with respect to the source.}\label{verif_assump_ER_control}
\end{center}
}
\end{table}

\section{Assumptions verification on empirical networks}\label{assump-empir}

\begin{figure}
\begin{subfigure}{.5\textwidth}
  \centering
  \includegraphics[width=\linewidth]{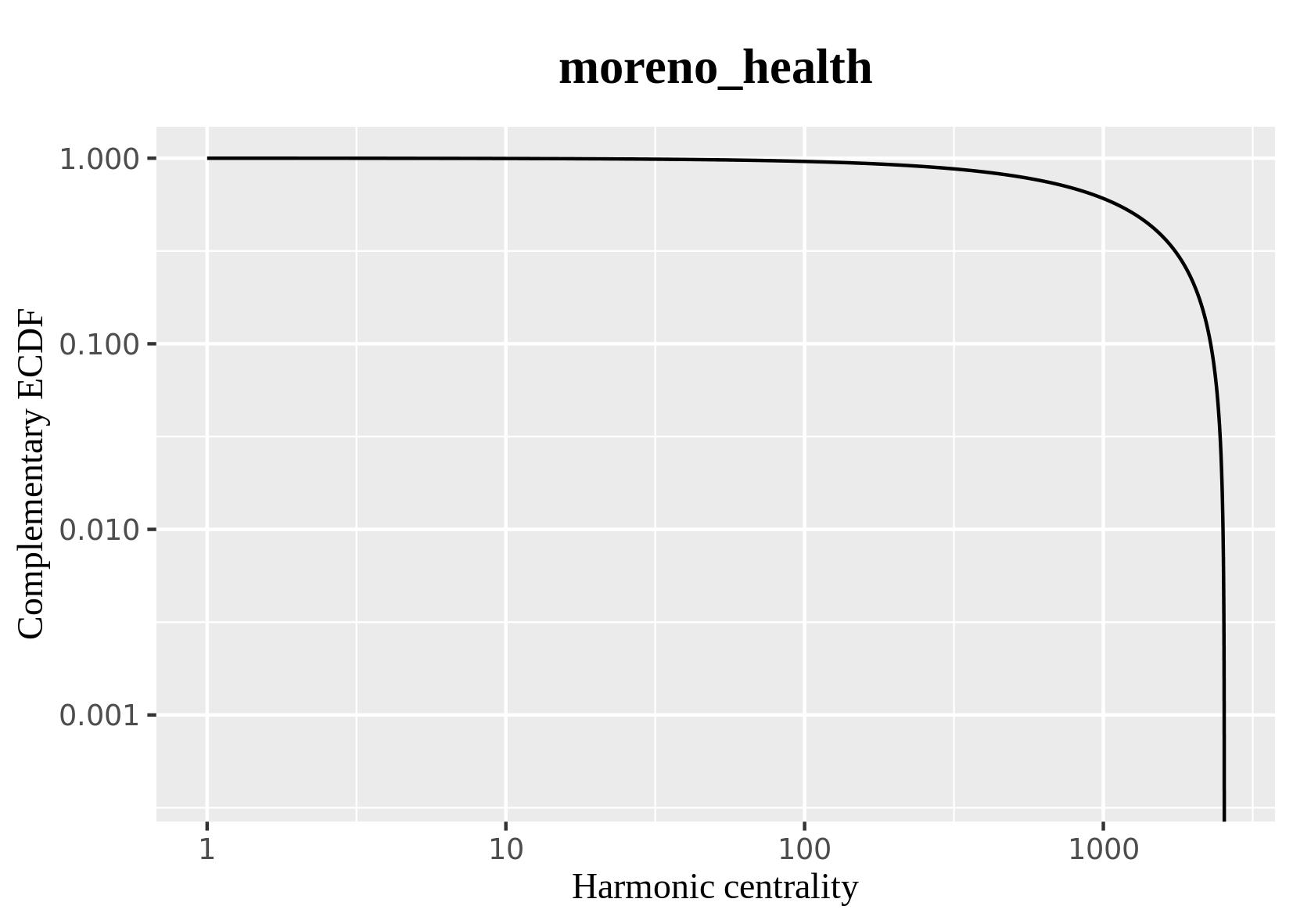}  
  \label{harm_ecdf:sfig1}
\end{subfigure}
\begin{subfigure}{.5\textwidth}
  \centering
  \includegraphics[width=\linewidth]{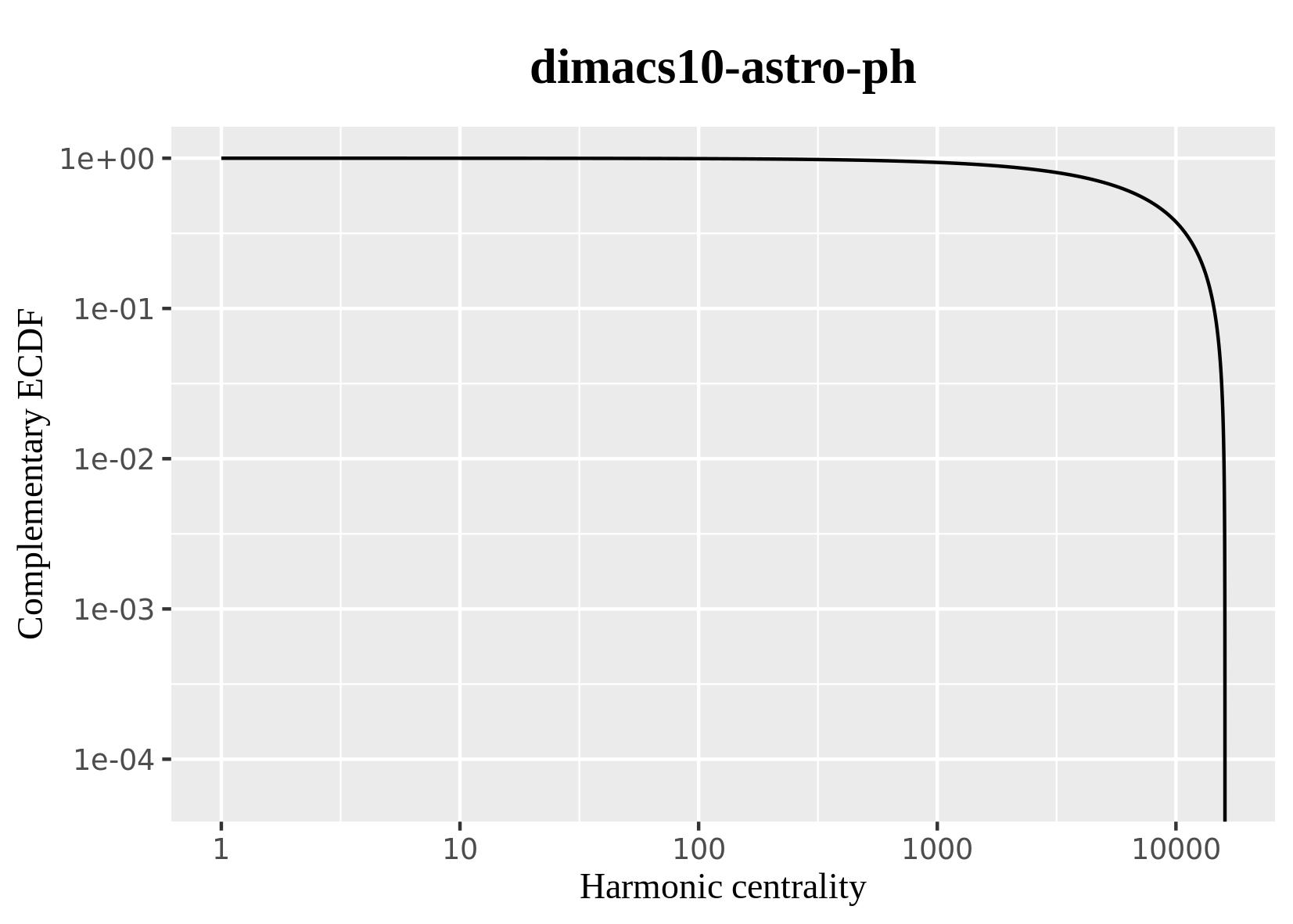}  
  \label{harm_ecdf:sfig2}
\end{subfigure}

\begin{subfigure}{.5\textwidth}
  \centering
  \includegraphics[width=\linewidth]{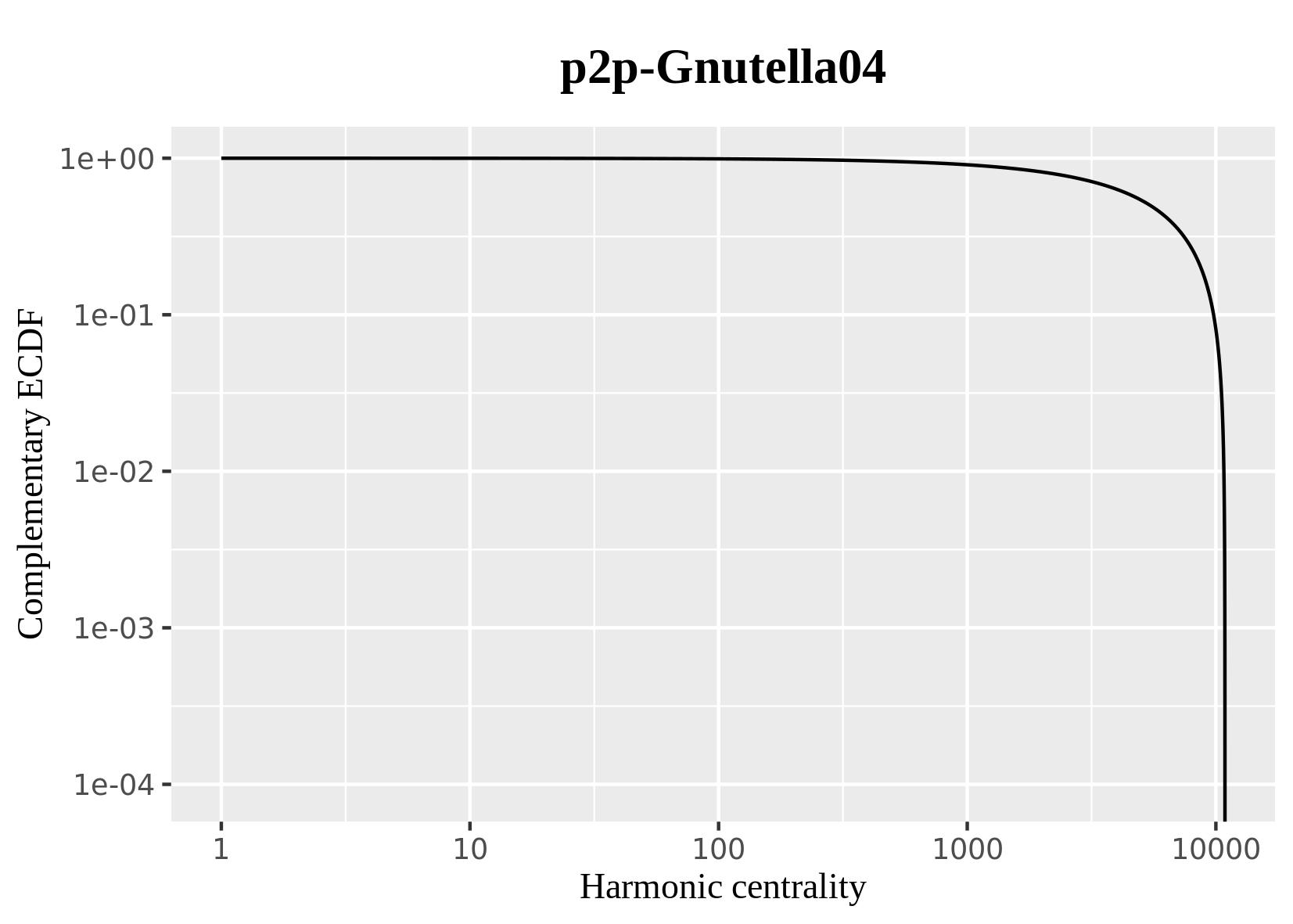}  
  \label{harm_ecdf:sfig4}
\end{subfigure}
\begin{subfigure}{.5\textwidth}
  \centering
  \includegraphics[width=\linewidth]{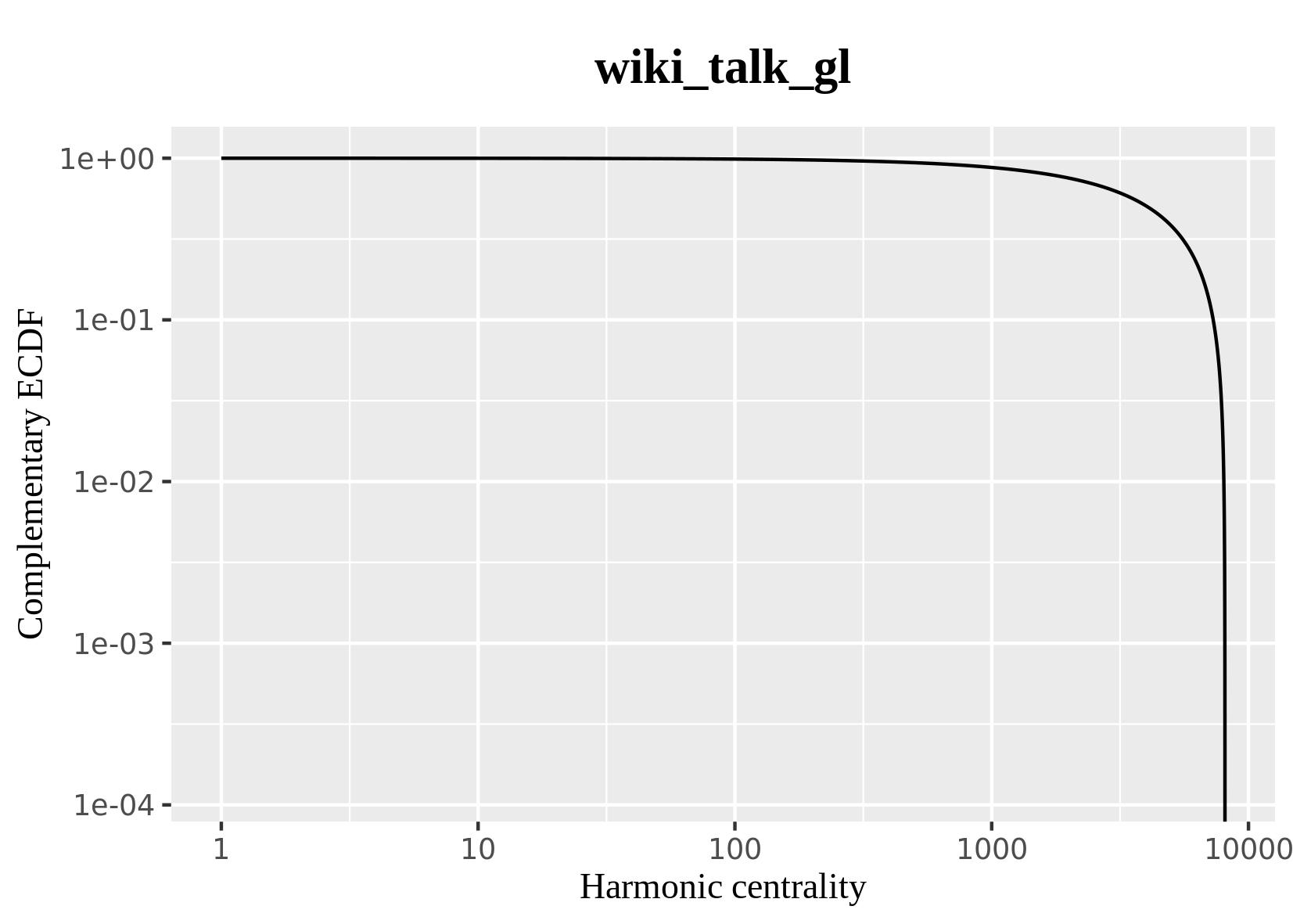}  
  \label{harm_ecdf:sfig5}
\end{subfigure}
\caption{{\bf Cumulative density function of the harmonic centrality on some empirical networks.} Each plot shows, with logarithmic (base $10$) axes, the complementary empirical cumulative density function of the harmonic centrality in each of the empirical datasets analyzed later.}
\label{ecdf_harm}
\end{figure}

In \figref{ecdf_harm} we can see the log-log plot of the complementary empirical cumulative density function of the harmonic centrality in each of the datasets. We can see that all the plots share the feature of having first a flat region of low centrality nodes with high probability, and then there is a turning point starting a new region more similar to a power-law, that however is restricted to a characteristic scale, excluding a possible scale-free behavior. This observation does not exclude this behavior for general datasets, since the observed pattern may well be a consequence of the size of the chosen datasets. In all the plotted datasets, the turning point that would correspond to $\hat{x}_{\min}$ in terms of the power-law distribution, can be visually placed around the percentile $10$ to $20$ of each distribution, which justifies the use of the upper bound given by the percentile $20$ for the search of $\hat{x}_{\min}$ explained in \ref{interl_pl}. The exception is the plot of the DBLP dataset which, in spite of being very similar to the plots displayed, therein $\hat{x}_{\min}$ may be visually placed around the percentile $60$. For this reason, we excluded this dataset in the later analyses. The interested reader can anyway run the provided code \cite{QCrepo} to review this dataset analogously to the others. 

We performed a similar experiment to that from \ref{assump_valid_p_val} to study whether the datasets fulfill the assumptions of QuickCent. 
A p-value to quantify the goodness of fit was computed with the bootstrap procedure from \ref{assump_valid_p_val}. We also computed the Spearman correlation between the logarithms of the in-degree and the centrality\footnote{As before, computed on non-zero in-degree and centrality.}. The results are displayed in Table \ref{assmpt_valid_real_netw}. There we can see that, in general, the datasets present a reasonable fulfillment of the assumptions, the only exception being perhaps the wiki\_talk\_gl dataset. These results should be taken with care since, from our experimentation, the method for p-value estimation is very sensitive to the value used to bound the search space. The overall exponents obtained agree with the discussion regarding the complementary ECDF plots given in the previous paragraph. 

\begin{table}
{\small
\begin{center}
  \begin{tabular}{c|ccccc}
\textbf{Name}  & \textbf{Corr.} & \textbf{p-val} & $\hat{x}_{\min}$ & $\hat{x}_{\min}$ p-value & $\hat{\alpha}(\hat{x}_{\min})$\\
\hline
\textbf{moreno\_health} &0.8&0& 339.2662&1& 5.519177 \\
\textbf{dimacs10-astro-ph}  & 0.75&0& 2719.642& 1
& 5.199691\\
\textbf{p2p-Gnutella04} &0.73&0& 582.6779 & 1& 4.991946\\
\textbf{wiki\_talk\_gl} & 0.22 & 3.5e-83 & 361.65& 0& 9.505685\\
\end{tabular}
\caption{\textbf{Indicators of the fulfillment of the assumptions by empirical data sets.} Fields in the table are the network name, its Spearman correlation between the logarithm of the in-degree and the logarithm of the harmonic centrality, the correlation p-value, the fitted lower limit, its corresponding p-value, and the power-law exponent.}
\label{assmpt_valid_real_netw}
\end{center}}
\end{table}






\end{document}